\documentclass[a4paper,12pt]{article}
\usepackage{jheppub}
\usepackage{amssymb}
\usepackage{amsmath}
\usepackage{graphicx}
\usepackage{multirow}
\usepackage{subfigure}
\usepackage{float}
\usepackage{graphics}
\usepackage{slashed}
\usepackage{booktabs}
\usepackage{color}
\usepackage[normalem]{ulem}
\usepackage{bm,psfrag}

\def\Slash#1{{#1\!\!\!\slash}}

\def\nslash{n\!\!\!\slash}
\def\bnslash{\bar n\!\!\!\slash}

\newcommand{\nn}{\nonumber}

\newcommand{\bn}{\bar n}

\newcommand{\mcdot}{\!\cdot\!}

\def\sech{{\rm sech}}
\def\kin{k_{\rm in}}
\def\kout{k_{\rm out}}
\def\dkin{\delta(k_{\rm in})}
\def\dkout{\delta(k_{\rm out})}
\def\kinp{\left(\frac{1}{k_{\rm in}}\right)_\star}
\def\koutp{\left(\frac{1}{k_{\rm out}}\right)_\star}
\def\logkinp{\left[\frac{1}{k_{\rm in}}\ln\left(\frac{k_{\rm in}}{\mu}2\cosh y_J\right)\right]_\star}
\def\logkoutp{\left[\frac{1}{k_{\rm out}}\ln\left(\frac{k_{\rm out}}{\mu}2\cosh y_J\right)\right]_\star}
\def\kain{\kappa_{\rm in}}
\def\kaout{\kappa_{\rm out}}
\def\muin{\mu_{s_{\rm in}}}
\def\muout{\mu_{s_{\rm out}}}
\def\etain{\eta_{\rm in}}
\def\etaout{\eta_{\rm out}}
\def\gE{\gamma_E}

\def\MXt{M_X^2}

\definecolor{darkred}{rgb}{0.7,0.0,0.0}
\definecolor{darkblue}{rgb}{0.0,0.0,0.9}
\definecolor{darkgreen}{rgb}{0.0,0.5,0.0}
\definecolor{brown}{rgb}{0.0,0.0,0.0}

\newcommand{\cA}{{\mathcal{A}}}
\newcommand{\cB}{{\mathcal{B}}}
\newcommand{\cC}{{\mathcal{C}}}
\newcommand{\cQ}{{\mathcal{Q}}}

\newcommand{\cI}{{\mathcal{I}}}
\newcommand{\cF}{{\mathcal{F}}}

\newcommand{\cM}{{\mathcal{M}}}

\newcommand{\cO}{{\mathcal{O}}}
\newcommand{\gcusp}{\gamma_{\mathrm{cusp}}}
\newcommand{\hcC}{\hat{{\mathcal{C}}}}
\newcommand{\hH}{\hat{H} }
\newcommand{\hws}{\hat{\widetilde{s}} }

\title{Resummation prediction on the jet mass spectrum in one-jet inclusive production at the LHC}
\author[a]{Ze Long Liu,}
\author[a,b]{Chong Sheng Li,}
\author[c]{Jian Wang}
\author[a]{and Yan Wang}

\affiliation[a]{School of Physics and State Key Laboratory of
Nuclear Physics and Technology, Peking University,\\Beijing 100871,China}
\affiliation[b]{Center for High Energy Physics,
Peking University,\\Beijing 100871,China}
\affiliation[c]{PRISMA Cluster of Excellence $\&$ Mainz Institute for Theoretical Physics,Johannes Gutenberg University,\\D-55099 Mainz, Germany}

\emailAdd{liuzelong@pku.edu.cn}
\emailAdd{csli@pku.edu.cn}
\emailAdd{jian.wang@uni-mainz.de}
\emailAdd{wangyanwww@pku.edu.cn}

\abstract{We study the factorization and resummation prediction on the jet mass spectrum in one-jet inclusive production at the LHC based on soft-collinear effective theory. The soft function with anti-$k_T$ algorithm is calculated at next-to-leading order and its validity is demonstrated by
checking the agreement between the expanded leading singular terms with the exact fixed-order result.
The large logarithms $\ln^{n} (m_J^2/p_T^2)$ and the global logarithms $\ln^{n} (s_4/p_T^2)$ in the process are resummed to all order
at next-to-leading logarithmic and next-to-next-to-leading logarithmic level, respectively.
The cross section is enhanced by about 23\% from the next-to-leading logarithmic level to next-to-next-to-leading logarithmic level.
Comparing our resummation predictions with those from Monte Carlo tool \texttt{PYTHIA} and ATLAS data at the 7 TeV LHC, we find that the peak positions of
the jet mass spectra agree with those from \texttt{PYTHIA} at parton level, and the predictions of the jet mass spectra with non-perturbative effects are in coincidence with the ATLAS data. We also show the predictions at the future 13 TeV LHC.}

\arxivnumber{1412.1337}
\preprint{MITP/14-094}

\begin{document}
\bibliographystyle{unsrt}
\maketitle
\flushbottom

\section{Introduction}\label{sec:intro}
The substructure of jets produced at the Large Hadron Collider (LHC) has become one of the hot topics for both theorists and experimentalists. The particles such as massive electroweak bosons, top quark and other possible new resonances produced with transverse momenta much greater than their masses, i.e., $p_T\gg m$, can decay to hadronic products, which are almost collinear and may be recombined into a single jet by jet algorithms. Therefore it is necessary to find a way to distinguish the interesting signal jets from the purely QCD backgrounds.

During the past few years, many studies on jet substructures have been performed~\cite{Butterworth:2002tt,Butterworth:2008iy,Kaplan:2008ie,Ellis:2009me,Thaler:2008ju,Krohn:2009th,Gallicchio:2010dq,Thaler:2010tr,Gallicchio:2010sw,Cui:2010km,Gallicchio:2011xq,Altheimer:2012mn,Ellis:2012sn},
in which new techniques and observables have been designed to analyze the events. The event generators such as \texttt{SHERPA}~\cite{Gleisberg:2003xi,Gleisberg:2008ta}, \texttt{PYTHIA}~\cite{Sjostrand:2006za,Sjostrand:2007gs} and \texttt{HERWIG++}~\cite{Bahr:2008pv,Gieseke:2011na},  can provide fully differential events, by which any observable can be predicted and compared with data.  However, the various event generators employ different models for parton shower and non-perturbative effects, such as the hadronization and multiparton interactions. As a consequence, they might provide very different predictions. For instance, the jet mass spectra from the \texttt{PYTHIA} and the \texttt{HERWIG++} do not agree with each other, as shown in ref.~\cite{ATLAS:2012am}.
Moreover, there is a type of color correlation between the initial and final colored particles that is not taken into account in these event generators.

In order to obtain more precise predictions and test the validity of the Monte Carlo tools, it is important to develop a theoretical framework to study the jet substructure. Recently, various jet substructure observables have been investigated analytically  based on soft-collinear effective theory (SCET)~\cite{Becher:2008cf,Cheung:2009sg,Ellis:2009wj,Ellis:2010rwa,Jouttenus:2009ns,Kelley:2011tj,Kelley:2011aa,Chien:2012ur,Chien:2014nsa} and the traditional perturbative QCD (pQCD) resummation formalism~\cite{Banfi:2010pa,Li:2011hy,Li:2012bw,Dasgupta:2012hg,Dasgupta:2013ihk}. For example, the factorization and resummation prediction of the jet angularity in the multijet production at $e^+e^-$ colliders have been studied in refs.~\cite{Ellis:2009wj,Ellis:2010rwa}, and the invariant mass and energy profile of jets at hadron colliders  have been explored  in refs.~\cite{Li:2011hy,Li:2012bw}.

The theoretical developments of prediction on jet mass spectrum at hadron colliders can be found in \cite{Li:2012bw,Dasgupta:2012hg,Chien:2012ur,Jouttenus:2013hs}.
In ref.~\cite{Li:2012bw}, the jet mass was investigated with the pQCD resummation formalism by focusing on the processes independent jet function,
where it was found that the nonperturbative effects are important at small jet mass.
The author of ref.~\cite{Dasgupta:2012hg} studied the distributions of $m_J/p_T^J$ in $pp\to {\rm dijet}$ and $Z+1\,{\rm jet}$ processes at NLL,
using the formula in refs.~\cite{Banfi:2010xy,Banfi:2004yd}, and including resummation effects of non-global logrithms (NGLs) in large-$N_c$ approximation.
The jet mass spectrum with the ${\rm Higgs}+1\,{\rm jet}$ process was discussed~\cite{Jouttenus:2013hs} in the $N$-jettiness global event shape \cite{Stewart:2010tn}.
The factorization formula and resummation prediction of the jet mass spectrum for direct photon production in the framework of SCET
was provided in ref.~\cite{Chien:2012ur},
where the soft function was factorized into two pieces with different scales.
Thought the non-global logarithms were not resummed there,
their contribution were estimated and it was found that the NGLs only affect the jet mass spectrum in the peak regions significantly.

Studies of the jet mass can not only help us understand QCD, but also be useful to search for new physics,
especially in the complex QCD environment of the LHC.
In particular, if we want to identify  the mass peak of a highly boosted particle, the jet mass spectrum of QCD background must be calculated precisely.
Actually, the jet invariant mass were explored in both ATALS and CMS collaborations at the 7 TeV LHC \cite{ATLAS:2012am,Chatrchyan:2013vbb}.
From these results, we can see that the jet mass $m_J$ peaks at about 50 GeV, which can be much smaller than the transverse momenta of jet $p_T$.
Therefore there exist large logarithmic terms $\frac{\alpha_s^n}{m_J^2} \ln^{m} (m_J^2/p_T^2)$ with $m\le 2n-1$ in the perturbative calculations near the peak region,
which need to be resummed to all order in order to give reliable predictions.

\begin{figure}
\begin{center}
  \includegraphics[width=0.5\linewidth]{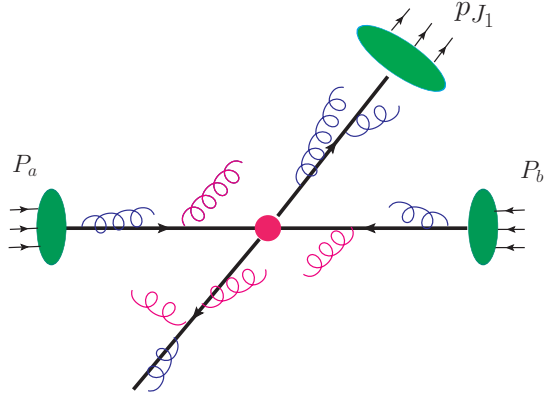}\\
  \caption{The illustrative picture for dijet production at the LHC. The blue and red arcs denote the collinear and soft gluons, respectively. }
  \label{fig:dijetpro}
\end{center}
\end{figure}

In this paper, we study one of the simplest jet substructures, i.e. the invariant mass of a jet, and investigate the factorization and resummation prediction on the jet mass spectrum in SCET for one-jet inclusive production at the LHC. Compared with direct photon process~\cite{Chien:2012ur}, the factorization formula for dijet process is more complicated
due to the nontrivial color structure and associating soft radiation.
The illustrative picture of this process is shown in figure~\ref{fig:dijetpro}.
Since the soft radiation can either be inside or outside the cone of the measured jet, there are two kinematic variables which can lead to large logarithms at threshold limit: one is the invariant mass $m_J$ of the measured jet, and another is the invariant mass $\sqrt{s_4}$ of the partonic system that recoils against the observed jet. In the threshold region $m_J^2\to 0$ and $s_4\to 0$, both of the large logarithms $\ln^n (m_J^2/p_T^2)$ and $\ln^n (s_4/p_T^2)$ need to be resummed to all order. In the threshold limits, the cross section can be factorized as
\begin{equation}\label{eq:genfacform}
\sigma = f_{P_a}\otimes f_{P_b}\otimes {\bm H}\otimes {\bm S}\otimes
J^{\rm obs.}\otimes J^{\rm rec.}\,,
\end{equation}
where ${\bm H}$, ${\bm S}$, $J$, $f_P$ are the hard function, soft function, jet function and parton distribution function (PDF), respectively. Both of the hard and soft function are matrices in color space. The hard function includes the short distance contributions arising from virtual corrections. The jet function presents the collinear radiation in the jet. The indices ``obs.'' and ``rec.'' denote the observed jet and the recoiled one, respectively. The effects from soft gluon emission are incorporated in the soft function and its phase space is constrained by the jet algorithms. It is noteworthy that the large angle soft gluon arising from the initial state radiation (ISR) and recoiled final state radiation are taken into account in this formalism.
In contrast to the cone algorithm adopted in ref.~\cite{Chien:2012ur},
we choose anti-$k_T$ algorithm~\cite{Cacciari:2008gp} to calculate the jet and soft functions,
which is boost-invariant and stable against the change of jet boundary~\cite{Kelley:2012kj}.
Thus, our prediction can be valid for the jet with both small and large rapidity, and is more useful for phenomenological purposes.

This paper is organized as follows. In section~\ref{sec:kinema}, we analyze the kinematics of the one-jet inclusive production at hadron colliders and give the definition of the threshold region. In section\ref{sec:factorize}, we derive the factorization formula. In section~\ref{sec:HardF} and section~\ref{sec:JetF}, we show the results of hard function and jet function at NLO, respectively. We calculate the soft function at NLO and present its refactorization in section~\ref{sec:SoftF}. In section~\ref{sec:RGimp}, we give the final renormalization group (RG) improved cross section analytically. In section~\ref{sec:numres}, we discuss the numerical results of the jet mass distribution for one-jet inclusive production at the LHC, including the leading singular distribution at threshold limit, scale uncertainties, $R$ dependence, distinction between quark jets and gluon jets, and comparison between the RG improved predictions and ATLAS data. We conclude in section~\ref{sec:conc}.
\section{Analysis of kinematics and factorization}\label{sec:kinema}
In this section, we introduce the relevant kinematical variables and the factorization formula needed in our analysis. We consider the process
\begin{align}\label{eq:process}
  N_1(P_a) + N_2(P_a) \to J(p_{J_1}) + X \,,
\end{align}
where $J$ denotes the leading final jet, and $m_{J}$ is its invariant mass. The partonic channels include $qq\to qq$, $gg\to qq$, $gg\to gg$ and their various crossing ones. The Feynman diagrams at leading order (LO) are shown in appendix~\ref{app1}.

It is convenient to introduce two lightlike vectors $n_a^\mu=(1,0,0,1)$ and $n_b^\mu=(1,0,0,-1)$ along the beam directions, and another lightlike vector $n_J=(1,{\hat n}_J)$ along the measured jet direction. In the center-of-mass (CM) frame of the initial partons, for the one-jet inclusive production, the momentum of recoiling parton to the observed jet is along the direction $\bn_J=(1,-{\hat n}_J)$.
In the CM frame of the hadronic collision, the momenta of the incoming hadrons are given by
\begin{equation}
P^\mu_a=E_{\rm CM}\frac{n_a^\mu}{2},\qquad P^\mu_b=E_{\rm CM}\frac{n_b^\mu}{2}.
\end{equation}
Here $E_{\rm CM}$ is the CM energy of the collider and we have neglected the mass of the hadrons. The momenta of the incoming partons, with a light-cone
momentum fraction of the hadronic momenta, are
\begin{equation}
p_a = x_a E_{\rm CM}\frac{n^\mu_a}{2},\qquad
p_b = x_b E_{\rm CM}\frac{n^\mu_b}{2}.
\end{equation}

The hadronic kinematic invariants are defined as
\begin{gather}
  \label{eq:mandelstam}
  s = (P_a+P_b)^2 \, , \quad t_1 = (P_a-p_{J_1})^2-m_{J_1}^2 \, , \quad u_1=(P_b-p_{J_1})^2-m_{J_1}^2 \,,\nn\\
  M_X^2\equiv P_X^2=(P_a+P_b-p_{J_1})^2=s+t_1+u_1+m_{J_1}^2\,,
\end{gather}
and the partonic ones are defined as
\begin{gather}
  \hat{s}=(p_a+p_b)^2= x_a x_b s \, ,
  \quad \hat{t}_1=(p_a-p_{J_1})^2-m_{J_1}^2=x_a t_1 \, ,
  \quad \hat{u}_1=(p_b-p_{J_1})^2-m_{J_1}^2=x_b u_1 \, ,
  \nonumber  \\
  s_4\equiv m_X^2=(p_a+p_b-p_{J_1})^2=\hat{s} + \hat{t}_1 + \hat{u}_1+m_{J_1}^2\, , \label{eq:mandelstampart}
\end{gather}
where $p_{J_1}^2=m_{J_1}^2$. In the threshold limits, we have $p_{J_1}^2\to 0$ and $s_4\to 0$. The kinematic region we are interested in is
\begin{equation}
\hat{s}\,,\hat{t}_1\,,\hat{u}_1\gg m_J^2\,,s_4\gg \Lambda_{\rm QCD}^2
\end{equation}

Any four vector can be decomposed along the light-like reference vector $n_i$
\begin{equation}\label{eq:momdecomp}
\begin{aligned}
p^\mu=(n_i\cdot p)\frac{\bn_i^\mu}{2}+(\bn_i\cdot p)\frac{n_i^\mu}{2}+p_\perp^\mu
=p^+\frac{\bn_i^\mu}{2}+p^-\frac{n_i^\mu}{2}+p_\perp^\mu\,.
\end{aligned}
\end{equation}
Hence the momentum $p^\mu$ can be denoted by $p^\mu=(p^+,p^-,p_\perp)$.
The momentum modes relevant to our discussions are the collinear mode $p_{n_J}^\mu\sim \sqrt{\hat s}(\lambda^2,1,\lambda)$, anti-collinear mode $p_{\bn_J}^\mu\sim\sqrt{\hat s}(1,\lambda^2,\lambda)$ and soft mode $p_s^\mu\sim\sqrt{\hat s}(\lambda^2,\lambda^2,\lambda^2)$, where $\lambda=m_J/\sqrt{\hat s}$ is treated as a small expansion parameter. In the partonic threshold limits $m_J\to 0$ and $s_4\to 0$, the radiation is constrained to be either soft or collinear with the final-state partons.

In order to identify energetic cluster of radiation, the sequential recombination jet algorithms are used. The longitudinal boost invariant distance measures $d_{ij}$ and $d_{iB}$ are defined by
\begin{equation}\label{eq:jetclus}
\begin{aligned}
d_{ij}&={\rm min}(p_{T,i}^\alpha,p_{T,j}^\alpha)\Delta R_{ij}/R\,,\qquad
\Delta R_{ij}=\sqrt{(y_i-y_j)^2+(\phi_i-\phi_j)^2}\,,\\
d_{iB}&=p_{T,i}^\alpha\,,\nn
\end{aligned}
\end{equation}
where $R$ is the jet radius parameter, $y_i$ and $\phi_i$ are rapidity and azimuthal angle of the jet $i$, respectively. $\alpha=-1$, 0 and 1 represent the inclusive anti-${ k_T}$~\cite{Cacciari:2008gp}, Cambridge-Aachen~\cite{Dokshitzer:1997in,Wobisch:1998wt} and ${k_T}$~\cite{Catani:1993hr,Ellis:1993tq} jet algorithms, respectively.
The effects of jet algorithms on the resummation have been studied in refs.~\cite{Banfi:2005gj,Delenda:2006nf,KhelifaKerfa:2011zu,Kelley:2012kj,Kelley:2012zs}, among which
ref.~\cite{Kelley:2012kj} has shown that jet boundary can be changed significantly by boundary clustering for Cambridge-Aachen and $k_T$ algorithms,
while the change of the phase space is power suppressed for anti-$k_T$  algorithm.
In this paper, the anti-$k_T$  algorithm is adopted, and the jet boundary is just a circle of radius $R$ in $\phi-y$ plane around the jet direction.

After clustering jets, the jet invariant mass $m_J$ receives contribution from the radiation inside the jet, whether from collinear and soft gluons.
Thus we split the soft radiation $k^\mu$ to two parts, denoted by $k^\mu=\kin^\mu+\kout^\mu$. Then, the partonic threshold variables take the form
\begin{equation}\label{eq:thresv1}
\begin{aligned}
m_J^2&=(p_{J_1}+\kin)^2=m_{J_1}^2+2\kin\cdot p_{J_1}\,,\\
s_4&=(p_{J_2}+\kout)^2=m_{J_2}^2+2\kout\cdot p_{J_2}\,.
\end{aligned}
\end{equation}
In the kinematic region $m_J^2, s_4\ll {\hat s}$, the momenta of the two jets can be written as $p_{J_1}^\mu=E_{J_1}n_J^\mu$ and $p_{J_2}^\mu=E_{J_2}\bn_J^\mu$ in the partonic CM frame, where $E_{J_1}= E_{J_2}=\sqrt{\hat s}/2$ in the threshold limit. And $m_J$ and $s_4$ can be rewritten as
\begin{equation}\label{eq:thresv2}
\begin{aligned}
m_J^2&=m_{J_1}^2+2E_J(n_J\cdot\kin)\,,\\
s_4&=m_{J_2}^2+2E_J(\bn_J\cdot\kout)\,.
\end{aligned}
\end{equation}
For later convenience, we write $\kin\equiv n_J\cdot\kin$ and $\kout\equiv \bn_J\cdot\kout$.

The hadronic threshold is defined as $M_X^2\to 0$. In this limit ,the final state radiations and beam remnants are highly suppressed, which leads to final states consisting of two narrow jets, as well as the remaining soft radiations. For convenience, we introduce the dimensionless variables
\begin{gather}
v=1+\frac{\hat{t}_1}{\hat{s}}\,,\quad w=-\frac{\hat{u}_1}{\hat{s}+\hat{t}_1}\,,
\quad \overline{v}=1-v\,.
\end{gather}
In terms of $m_X$, $x_1$, $x_2$ and $v$,
\begin{equation}
M_X^2 = \frac{m_X^2}{x_2} + E_{\rm CM}^2 \left[ (1-x_1) v + (1-x_2) \overline{v} \right]+m_J^2\,.
\end{equation}
In the limit $x_1\to 1$, $x_2\to 1$, $m_J^2\to 0$ and $m_X^2\to 0$, we have
\begin{align}\label{eq:MXinthres}
M_X^2  = m_X^2 + m_J^2+\frac{p_T^2}{v \overline{v}}\left[ (1-x_1)v + (1-x_2)\overline{v}  \right] + \dots\,.
\end{align}
This expression is helpful when we derive the RG equation of the soft function by using the RG invariance in section~\ref{sec:SoftF}.
\section{Factorization in SCET}\label{sec:factorize}
To derive a factorization formula for dijet process in SCET, we first have to match the full QCD onto the effective theory~\cite{Bauer:2008jx,Bauer:2010vu}. To illustrate the factorization in detail, we consider the process $qq'\to qq'$. The initial partons are labeled by 1 and 2 and the final partons are labeled by 3 and 4, and the relevant operator in QCD is given by \cite{Kelley:2010fn}
\begin{equation}
 \mathcal{O}^{\rm QCD}_{I\Gamma } =
 ( \bar{\psi}_4^{a_4}  \gamma_{\mu} \Gamma \psi_2^{a_2} )
 ( \bar{\psi}_3^{a_3}  \gamma^{\mu} \Gamma' \psi_1^{a_1} )
 (c_I)_{\{a\}}  \,,
\end{equation}
where $c_I$ denotes a 4 order color tensor with color indices $a_i$, and $\Gamma~(\Gamma')$ denote the chirality ($P_L$ or $P_R$).
In SCET, the $n$-collinear quark field $\psi_n$ can be written as
\begin{eqnarray}\label{eq:colfield}
\chi_n(x) &=&  W_n^\dagger(x) \xi_n(x),\quad\xi_n(x)=\frac{\nslash\bnslash}{4}\psi_n(x),
\end{eqnarray}
where $W_n^\dagger$ is the Wilson line, and $\chi_n$ is the gauge invariant combination of $W_n^\dagger$ and collinear quark field $\xi_n$ in SCET.
At the leading power in $\lambda$, only the $n\mcdot A_s$ component of soft gluons can interact with the $n$-collinear field $\chi_n(x)$, which can be decoupled by a field redefinition~\cite{Bauer:2001yt}:
\begin{eqnarray}\label{eqs:frd}
 \chi_n(x) \to Y_n(x)\chi_n(x),\qquad
\end{eqnarray}
with
\begin{equation}
 Y_n(x) = \mathbf{P} \exp\left( ig_s\int^0_{-\infty}ds\,n\mcdot A^a_s(x+sn)t^a\right)\,,
\end{equation}
Then the effective Lagrangian can be expressed as
\begin{equation} \label{eq:qqops}
 \mathcal{L}_{\rm eff}=\sum_{I,\Gamma}
 \cC_I^\Gamma\,\mathcal{O}^{\rm SCET}_{I\Gamma }\,,
\end{equation}
with
\begin{align}\label{eq:OcOs}
\mathcal{O}^{\rm SCET}_{I\Gamma }&= \sum_{\{a\}}(c_I)_{\{a\}}[O^c(x)]_\Gamma^{b_1b_2b_3b_4} [O^s(x)]^{\{a\},\{b\}}\,,\\
[O^c(x)]_\Gamma^{b_1b_2b_3b_4}&=
\bar{\chi}_{\bn_J}^{b_4}(x) \gamma_{\mu}\Gamma\chi_{\bn}^{b_2}(x)
\bar{\chi}_{n_J}^{b_3}(x) \gamma^{\mu}\Gamma'\chi_n^{b_1}(x)\,,\\
[O^s(x)]^{\{a\},\{b\}}&=
[Y_{\bn_J}^\dagger(x)]^{b_4a_4} [Y_{\bn}(x)]^{a_2b_2}
[Y_{n_J}^\dagger(x)]^{b_3a_3}[Y_n(x)]^{a_1b_1} \,.
\end{align}
Here $\cC_I^\Gamma$ is the hard matching coefficient.
The scattering amplitude for the $qq'\to qq'$ can be written as
\begin{equation}\label{eq:amphad}
|\cM^\Gamma (x)\rangle = \langle X|O_\Gamma^c(x){\bm O}^s(x)|N_1N_2\rangle |C^\Gamma\rangle\,,
\end{equation}
where $|C^\Gamma\rangle$ is the vector of Wilson coefficient combination in color basis $|c_I\rangle$, as following
\begin{equation}\label{eq:amphad}
|C^\Gamma\rangle = \sum_I \cC_I^\Gamma\,|c_I\rangle\,.
\end{equation}
For $qq'\to qq'$, the color basis is chosen as
\begin{gather}\label{eq:color4q}
  | c_1 \rangle  =t_{i_3,i_1}^c t_{i_4,i_2}^c\,,\quad
  | c_2 \rangle  =\delta_{i_3,i_1} \delta_{i_4,i_2}\,.
\end{gather}

The differential cross section can be written as
\begin{equation}\label{eq:xsecdef}
\frac{d\sigma}{dp_T dy dm_J^2} = \frac{1}{2s} \sum_X\sum_\Gamma \int d^4 x
\langle\cM^\Gamma (x)|\widehat{\cM}(m_J^2,p_T,y,R)|\cM^\Gamma(0)\rangle\,,
\end{equation}
where the operator $\widehat{\cM}(p_T,y,R)$ denotes the measurement in the final state, including the jet algorithm. It acts on the
final-state  collinear and soft particles with momenta $\{p_c\},\{k_s\}$ as follows
\begin{equation}
\widehat{\cM}(m_J^2,p_T,y,R) |X_{c+s}\rangle
= \cM(m_J^2,p_T,y,R,{p_c},{k_s}) |X_{c+s}\rangle\,,
\end{equation}
where
\begin{equation}
\begin{aligned}
\cM(m_J^2,p_T,y,R,\{p_c\},\{k_s\})
=&\delta\left((p_c+k_s)^2-m_J^2\right)
\delta\left(|\vec{p}_{cT}|-p_T\right)
\delta\left(y-\frac{1}{2}\ln\frac{p_c^+}{p_c^-}\right)\\
&\times\Theta\left(R^2-(y_s-y_c)^2-(\phi_s-\phi_c)^2\right)\,.
\end{aligned}
\end{equation}
Since the soft and collinear sectors are decoupled due to field redefinition, the matrix element in eq.~(\ref{eq:xsecdef}) can be factorized into a product of several matrices,
\begin{equation}\label{eq:factoredME}
\begin{aligned}
\sum_X \langle\cM(x)|\widehat{\cM}(m_J^2,p_T,y,R)|\cM(0)\rangle=&
\frac{1}{N_{\rm init}} \sum_{\Gamma}\left( \Gamma\gamma_{\nu} \right)_{\alpha1 \gamma1}\left( \gamma_{\mu}\Gamma \right)_{\beta1 \sigma1}
\left( \Gamma'\gamma_{\nu} \right)_{\alpha2 \gamma2}\left( \gamma_{\mu}\Gamma' \right)_{\beta2 \sigma2}  \\
&\times
\langle N_1(P_1)|\bar{\chi}^{\alpha1}_n(x) \chi^{\beta1}_n(0)|N_1(P_1)\rangle
\\
&\times\langle
N_2(P_2)|\bar{\chi}^{\alpha2}_{\bn}(x) \chi^{\beta2}_{\bn}(0)|N_2(P_2)\rangle
\\
&\times \sum_{X_{c1}}\langle 0| \chi_{n_J}^{\gamma1}(x) |X_{c1}\rangle\langle X_{c1}|\bar{\chi}_{n_J}^{\sigma1}(0) |0\rangle\, \\
&\times \sum_{X_{c2}}\langle 0|  \chi_{\bn_J}^{\gamma2}(x) |X_{c2}\rangle\langle X_{c2}|\bar{\chi}_{\bn_J}^{\sigma2}(0) |0\rangle\\
&\times \sum_{X_s}\langle C^{\Gamma}|
\langle 0|{\bm O}^{s\dagger}(x)|X_s\rangle\langle X_s|{\bm O}^s(0)|0\rangle
|C^\Gamma\rangle\\
&\times \cM(m_J^2,p_T,y,R,\{p_c\},\{k_s\})\,,
\end{aligned}
\end{equation}
where $N_{\rm init}=1/(4N^2)$ denotes the average over the colors and spin of the initial-state partons,
and $\alpha1,\beta1$, etc, are Dirac indices.
The initial state collinear sectors match to the conventional PDFs:
\begin{equation}\label{eq:qPDFdef}
\langle  N_i(P_i) | \bar \chi_i^{\alpha1} \left(
n_i \cdot x \frac{ \bar{n}_i^\mu }{2}
\right)   \chi_i^{\beta1}(0) \,|  N_i(P_i) \rangle  = \frac{1}{2}
\bn_i \cdot P_i \,  \left( \frac{\Slash{n}_i}{2}  \right)^{\beta1 \alpha1 } \int_{-1}^1 d\xi\, f_{q/N_i}(\xi)\,e^{i\,\xi\, ( n_i\cdot x)(\bn_i \cdot P_i)/2} \,,
\end{equation}
and the matrix elements of the collinear fields in the final state match to the quark jet function:
\begin{equation}\label{eq:qJetdef}
\sum_{X_{c1}} \langle  0|\, \chi_{n_i}^{\gamma1} \left( x \right) |X_{c1}\rangle\langle X_{c1}|  \bar{\chi}_{n_i}^{\sigma1}(0) \, | 0 \rangle =
 \,\left( \frac{\Slash{n}_i}{2} \right)^{\gamma1 \sigma1}
\int \frac{d^4 p}{(2\pi)^3} \theta(p^0)\,(\bar{n}_J\cdot p)\,J_q(p^2)\, e^{-i\, x\, p} \,.
\end{equation}
The soft function can be defined as the matrix element associated with the soft Wilson line
\begin{equation}\label{eq:S_ope_def}
{\bm S}(x,\mu)=\langle 0|{\bm O}^{s\dagger}(x)|X_s\rangle\langle X_s|{\bm O}^s(0)|0\rangle\,,
\end{equation}
which can be decomposed in the color basis
\begin{equation}\label{eq:SinColorBasis}
S_{IJ}\equiv\langle c_I|{\bm S}|c_J\rangle\,.
\end{equation}
Now the matrix element appearing in eq.~(\ref{eq:factoredME}) can be simplified as
\begin{equation}
\langle C^{\Gamma}|
\langle 0|{\bm O}^{s\dagger}(x)|X_s\rangle\langle X_s|{\bm O}^s(0)|0\rangle
|C^\Gamma\rangle=\sum_{IJ}\cC_I^{\Gamma *}\,S_{IJ}\,\cC_J^{\Gamma}\,.
\end{equation}

All the above components in the factorization form in eq.~(\ref{eq:factoredME}) satisfy certain RG equations, which we will discuss in the following sections. Combining the different parts together, we get the factorized differential cross section in the threshold limits
\begin{align}\label{eq:genxsec}
  \frac{d\sigma}{dp_Tdydm_J^2} &= \frac{p_T}{8\pi s}\sum_{\overset{i,j}{\rm channels}} \int_{x_a^{\text{min}}}^1
  \frac{dx_a}{x_a} \int_{x_b^{\text{min}}}^{1} \frac{dx_b}{x_b} \, f_{i/N_1}(x_a,\mu_f) \,
  f_{j/N_2}(x_b,\mu_f) \,C_{ij}(\hat{s},\hat{t}_1,\hat{u}_1,m_J^2,R,\mu_f) \,,
\end{align}
where $C_{ij}$  is the hard-scattering kernel
\begin{equation}\label{eq:kernelCij}
\begin{aligned}
  C_{ij}(\hat{s},\hat{t}_1,\hat{u}_1,m_J^2,R)
  =&  \sum_{I,J} \int dm_{J_1}^2\,dm_{J_2}^2\,d\kin\,d\kout
  \,H_{IJ}(\hat{s},\hat{t}_1,\hat{u}_1)\,S_{JI}(\kin,\kout) \\  &\times
  \,J_1(m_{J_1}^2)\,J_2(m_{J_2}^2)
   \delta(m_J^2-m_{J_1}^2-2E_J \kin)\,\delta(s_4-m_{J_2}^2-2E_J\kout)
  \,,
\end{aligned}
\end{equation}
with
\begin{equation}\label{eq:Hdef}
H_{IJ}=\sum_{\Gamma}\cC_I^\Gamma\cC_J^{\Gamma *}\,.
\end{equation}
And $H_{IJ}$ is the hard function, the details of which are shown in section~\ref{sec:HardF}.

For other channels, such as $gg\to qq'$ or $gg\to gg$, the formula of factorization is similar to the process $qq'\to qq'$,
except for the different jet functions and PDFs. The definitions of gluon PDF and jet function are given by
\begin{equation}\label{eq:gPDFef}
\langle  N_i(P_i)\, |\,\, (-g_{\mu\nu})\,\, {\cal A}_{i \perp}^\mu\left(
n_i \cdot x \frac{ \bn_i^\mu }{2}\right) {\cal A}_{i \perp}^\nu(0) \,|  N_i(P_i) \rangle  =
\int_{-1}^1 \frac{d\xi}{\xi}\, f_{g/N_i}(\xi)\,e^{i\, \xi (n_i\cdot x)\, (\bn_i \cdot  P_i)/2}\,,
\end{equation}
and
\begin{equation}\label{eq:gJetef}
\langle 0|\, \, {{\cal A}_{J}^a}_\perp^\mu(x){{\cal A}_{J}^b}_\perp^\nu(0)\,  |0\rangle  = \delta^{ab}
(-g_\perp^{\mu\nu})\,g_s^2 \int \frac{d^4 p}{(2\pi)^3}\, \theta(p^0)\,  J_g(p^2) \,e^{-i\, x\, p} \,.
\end{equation}

\section{Hard function}\label{sec:HardF}
The coefficient $\cC_I^\Gamma$ can be obtained by matching the full theory onto SCET. The one loop results for all partonic $2\to 2$ process in QCD have been available in ref.~\cite{Kelley:2010fn}, which are derived in dimensional regularization and the $\overline{\rm MS}$ renormalization scheme. In this section, we show the crossing relations for different channels and the RG evolution briefly. The explicit expressions of hard matching coefficients are shown in appendix~\ref{app2}.
\subsection{Wilsons coefficient at NLO}\label{subsec:HNLO}
First, for the 4-quark processes, there are six channels if two different flavor quarks are involved (e.g. $ud\to ud$)
\begin{gather}
qq'\to qq' \,, \quad q{\bar q}'\to q{\bar q}'\,, \quad q{\bar q}\to q'{\bar q}'\,,
\quad
qq'\to q'q \,, \quad q{\bar q}'\to {\bar q}'q\,, \quad q{\bar q}\to q'{\bar q}'\,.
\end{gather}
The Wilson coefficients for the channel $qq'\to qq'$ are denoted by $\cC_I^\Gamma(s,t,u)$ and the others can be obtained by crossing symmetries, as shown in table~\ref{tab:crosssys}. For example, the Wilson coefficients for the channel $q\bar{q}\to q'\bar{q}'$ are $\cC_I^\Gamma(u,s,t)$.
\begin{table}[t]
\begin{center}
\begin{tabular}{c|c|c|c||c|c|c|c}
\hline
\hline
$12 \to 34$ &\rm{crossing}&
$12 \to 34$ &\rm{crossing}&
$12 \to 34$ &\rm{crossing}&
$12 \to 34$ &\rm{crossing}\\
\hline
$q q' \to q q'$
&$s t u$
&$q q' \to q' q $
&$s u t$
&$g g \to q{\bar q} $
&$s t u$
&$g g \to {\bar q}q $
&$s u t$
\\
\hline
$q \bar{q}' \to q \bar{q}'$
&$u t s$
&$q \bar{q}' \to \bar{q}' q$
&$t u s$
&${\bar q}g \to g{\bar q} $
&$u t s$
&${\bar q}g \to {\bar q}g$
&$t u s$
\\
\hline
$q \bar{q} \to \bar{q}' q'$
& $t s u$
& $q \bar{q} \to q' \bar{q}'$
& $u s t$
&$q g \to q g $
&$t s u$
&$q g \to g q$
&$u s t$
\\
\hline
\hline
\end{tabular}
\end{center}
\caption{Crossing relations for the 4-quark and $gg\to q{\bar q}$ channels.
\label{tab:crosssys}
}
\end{table}
$\Gamma$ in Wilson coefficients denotes the chirality of the incoming and outgoing partons.
In general, there are 16 possible chirality amplitudes. Actually, for the channel $qq'\to qq'$, only 4 chirality amplitudes are non-zero. This is because that chiralities of massless particles 1 and 3 (2 and 4) must be the same. We rewrite the Wilson coefficients as $\cC_I^{\lambda_1,\lambda_2}\equiv\cC_I^{\lambda_1,\lambda_2,\lambda_3,\lambda_4}$ with $\lambda_{1,2}=L \;{\rm or}\; R$. In addition, since the chirality can be changed by charge conjugation, the only two independent chirality amplitudes for $qq'\to qq'$ are $\cC_I^{LL}$ and $\cC_I^{LR}$.

If the 4 quarks are identical, there are two additional non-vanishing chirality amplitudes $\cC_I^{LRRL}$ and $\cC_I^{RLLR}$ because of the contribution of u-channel. The interference between t-channel and u-channel also makes the results different from $qq'\to qq'$ case. The results for $qq\to qq$ can be expressed as
\begin{align}\label{eq:wc4qid}
 \cC_I^{LLLL} &= \cC_{I}^{RRRR}=  \cC_I^{LL}(s,t,u)+ B_{IJ}\cC_J^{LL}(s,u,t)\,,\nn\\
 \cC_I^{LRLR} &=  \cC_{I}^{RLRL} = \cC_I^{LR}(s,t,u)\, ,\\
 \cC_I^{LRRL} &= \cC_I^{RLLR} =   B_{IJ}  \cC_J^{LR} (s,u,t)  \nn\,,
\end{align}
where
\begin{align}\label{eq:matB}
B_{IJ}&=
\begin{pmatrix}
-\frac{1}{C_A}     & 2 \\
-\frac{C_F}{C_A}  & \frac{1}{C_A}
\end{pmatrix}
\,.
\end{align}
The results of other channel associated with $qq\to qq$ can be obtained by crossing symmetry as shown in table~\ref{tab:crosssys}.

Next, we consider the Wilson coefficients for $gg\to q{\bar q}$ channel and its crossing. There are six relevant channels
\begin{equation}
gg \to q\bar{q}, \quad
qg \to qg, \quad
\bar{q}g \to \bar{q}g, \quad
gg \to \bar{q} q, \quad
qg \to gq, \quad
\bar{q} g \to g \bar{q}  \,.
\end{equation}
The Wilson coefficients for the channel $gg\to q{\bar q}$ are denoted by $\cC_I^{\lambda_1,\lambda_2,\lambda_3,\lambda_4}(s,t,u)$ and the others can be obtained by crossing symmetries as shown in table~\ref{tab:crosssys}. According to parity invariance, we have
\begin{equation}
\cC_I^{\lambda_1,\lambda_2,\lambda_3,\lambda_4}
=\cC_I^{-\lambda_1,-\lambda_2,-\lambda_3,-\lambda_4}\,.
\end{equation}
In addition, $\cC_I^{\lambda_1,\lambda_2,\lambda_3,\lambda_4}=0$ when $\lambda_3=\lambda_4$. Thus, the Wilson coefficients for $gg\to q\bar{q}$ can be rewritten as $\cC_I^{\lambda_1,\lambda_2}\equiv\cC_I^{\lambda_1,\lambda_2;+-}$ , and there are only 4 independent chirality amplitudes for each color structure, the explicit expressions of which are shown in appendix \ref{app2}.

Finally, we consider the process $gg\to gg$. In ref.~\cite{Kelley:2010fn}, the Wilson coefficients are obtained by matching to an overcomplete basis of 9 color structures, though there are only 8 independent color structures. Then, 16 possible helicity amplitudes for each color structures give 144 matching coefficients. Basing on the symmetry, the Wilson coefficients can be expressed concisely as follows
\begin{equation}\label{eq:4gWCsys}
\begin{aligned}
\cC_I^\Gamma&= 4g_s^2 \mathcal{M}_I^{\Gamma}\left( 1 + \frac{\alpha_s}{4\pi}\mathcal{Q}_I^{\Gamma} \right)
 &\quad    I = 1 \cdots 6                                                                         &\quad    \Gamma = 1 \cdots 6\,,\\
\cC_I^\Gamma&= 4g_s^2 \frac{\alpha_s}{4\pi} \mathcal{Q}_I^{\Gamma}
 &\quad    I = 7,8,9                                                                         &\quad    \Gamma = 1 \cdots 6\,,\\
\cC_I^\Gamma&= 4g_s^2 \frac{\alpha_s}{4\pi} \mathcal{Q}_I^{\Gamma}\,.
 &\quad    I = 1 \cdots 9                                                                         &\quad    \Gamma = 7 \cdots 16\,.
\end{aligned}
\end{equation}
The explicit expressions of $\mathcal{M}_I^{\Gamma}$ and $\mathcal{Q}_I^{\Gamma}$ are listed in appendix \ref{app2} for the convenience of the reader.

\subsection{RG evolution of the hard function}
The Wilson coefficients $\cC_I^\Gamma$ satisfy the RG equation \cite{Becher:2009qa,Gardi:2009qi,Dixon:2009ur,Catani:1998bh,Sterman:2002qn}
\begin{equation}\label{eq:REGHard}
   \frac{d}{d \ln \mu} \cC_I^\Gamma (\mu)
= \Gamma_{IJ}^H \cC_J^\Gamma (\mu)  \,,
\end{equation}
where $\Gamma_{IJ}^H$ can be expressed as
\begin{align}\label{eq:Gamma_H_def}
\Gamma_{IJ}^H(s,t,u,\mu) &= \left(\gcusp \frac{c_H}{2} \ln\frac{-t}{\mu^2}
+ \gamma_H - \frac{\beta(\alpha_s)}{\alpha_s}\right) \delta_{IJ}
+ \gcusp M_{IJ}(s,t,u)  \,,
\end{align}
with
\begin{align}
c_H&=n_qC_F+n_gC_A\,,
\end{align}
and
\begin{align}
\gamma_H&=n_q\gamma_q+n_g\gamma_g\,,
\end{align}
where $\beta(\alpha_s)$ is the QCD beta function, $\gcusp$ is the cusp anomalous dimension, and $n_q$ and $n_g$ is the number of external quarks and gluons involved in the process, respectively. $M_{IJ}(s,t,u)$ denotes the color mixing terms, and can be written as
\begin{equation}
{\bm M}=-\sum_{i\neq j}\frac{{\bm T}_i\cdot {\bm T}_j}{2}
\left[L(s_{ij})-L(t)\right]\,,
\end{equation}
where $s_{12}=s_{34}=\hat{s}$, $s_{13}=s_{24}=\hat{t}$, $s_{14}=s_{23}=\hat{u}$, and $L(x)$ is defined as
\begin{equation}
L(x) = \ln\frac{|x|}{\mu^2}-i\pi\theta(x)\,.
\end{equation}
The explicit expressions of $M_{IJ}$ for each channel can be found in appendix \ref{app2}. $M_{IJ}$ can be diagonalized with eigenvalues $\lambda_K$. For example, for $qq'\to qq'$ channel, we have
\begin{align}
\left(F\cdot M\cdot F^{-1}\right)_{KK'}
 =\left(
\begin{array}{cc}
 \lambda_1 & 0 \\
 0         & \lambda_2
\end{array}
\right)\,,
\end{align}
where $F(s,t,u)$ denotes the transform matrix, which can be calculated numerically.
The Wilson coefficients in the diagonal basis are denoted by $\hat{\cC}_K^\Gamma\equiv F_{KI}\cC_I^\Gamma$,
which satisfy the RG equation
\begin{align}\label{eq:RGECdiag}
 \frac{d}{d \ln \mu} \hat{\cC}_I^\Gamma (\mu)&= \left[\gcusp \frac{c_H}{2} \ln\frac{-t}{\mu^2}
+ \gamma_H + \gcusp \lambda_K- \frac{\beta(\alpha_s)}{\alpha_s}\right]
 \hcC_I^\Gamma (\mu)\,.
\end{align}
The hard function in the diagonal basis is denoted by
$\hat{H}_{KK'}\equiv(F\cdot H\cdot F^\dagger)_{KK'}$. With eq.~(\ref{eq:Hdef}), the RG equation of the hard function can be obtained,
\begin{equation}\label{eq:diagHREG}
  \frac{ d}{ d \ln \mu} \hH_{KK'}(\mu) =
    \left[\gcusp\left( c_H \ln \left|\frac{{\hat t}_1}{\mu^2}\right|  + \lambda_{K}+ \lambda_{K'}^* \right)
          +2 \gamma_H -  \frac{2\beta(\alpha_s)}{\alpha_s}
    \right]  \hH_{KK'}(\mu)\, ,
\end{equation}
Solving the RG equation, we can get the resummed hard function
\begin{multline}\label{eq:resummedH}
\hat{H}_{KK'}({\hat s},{\hat t},{\hat u},\mu) =
\frac{\alpha_s(\mu_h)^2}{\alpha_s(\mu)^2}
\exp\Big[ 2c_H S (\mu_h, \mu) -2 A_H (\mu_h, \mu) \Big] \\
\times \exp\left[-A_{\Gamma} (\mu_h, \mu)
 \left(\lambda_{K}({\hat s},{\hat t},{\hat u})
 + \lambda_{K'}^*({\hat s},{\hat t},{\hat u})
 +c_H \ln  \left| \frac{t}{\mu_h^2} \right| \right)\right]
  \hat{H}_{KK'}({\hat s},{\hat t},{\hat u},\mu_h)   \,.
\end{multline}
where $S(\nu,\mu)$ and $A_\Gamma(\nu,\mu)$ are defined as
\begin{equation}\label{eq:SudakovS}
\begin{aligned}
 S(\nu,\mu)&=-\int_{\alpha_s(\nu)}^{\alpha_s(\mu)}d\alpha \frac{\gcusp(\alpha)}{\beta(\alpha)}
 \int_{\alpha_s(\nu)}^{\alpha}\frac{d\alpha'}{\beta(\alpha')}\,,\\
 A_\gamma(\nu,\mu)&=-\int_{\alpha_s(\nu)}^{\alpha_s(\mu)}d\alpha \frac{\gamma(\alpha)}{\beta(\alpha)}
\end{aligned}
\end{equation}
Up to $\rm NNLL$ level, we need three loop $\gcusp$ and $\beta$ function and  two loop $\gamma_H$, and their explicit expressions are collected in the appendix A of ref.~\cite{Becher:2009th}.

\section{Jet function}\label{sec:JetF}
\begin{figure}[t]
\begin{center}
\subfigure[]{
	\label{fig:qJetQCD1}\includegraphics[width=0.250\textwidth]{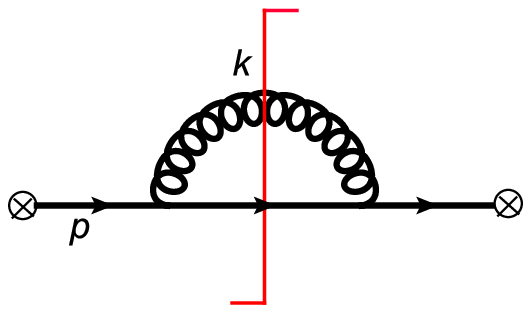}
	}
\subfigure[]{
	\label{fig:qJetSCET1}\includegraphics[width=0.250\textwidth]{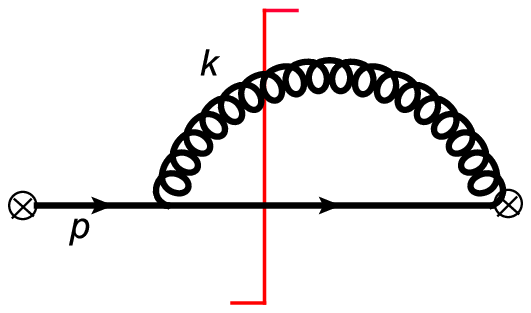}
	}
\subfigure[]{
	\label{fig:qJetSCET2}\includegraphics[width=0.250\textwidth]{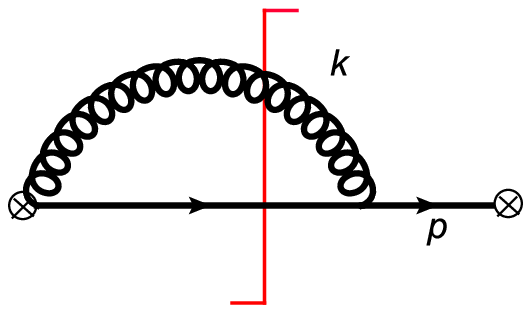}
	}\\
\subfigure[]{
	\label{fig:gJetQCD1}\includegraphics[width=0.250\textwidth]{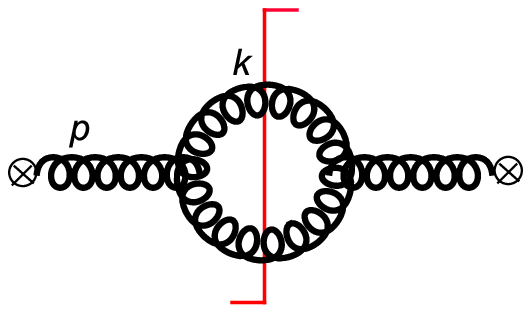}
	}
\subfigure[]{
	\label{fig:gJetQCD2}\includegraphics[width=0.250\textwidth]{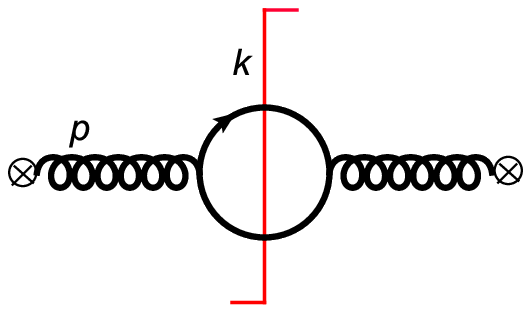}
	}
\subfigure[]{
	\label{fig:gJetQCD3}\includegraphics[width=0.250\textwidth]{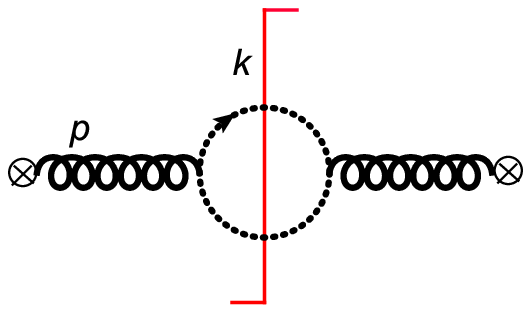}
	}\\
\subfigure[]{
	\label{fig:gJetSCET1}\includegraphics[width=0.250\textwidth]{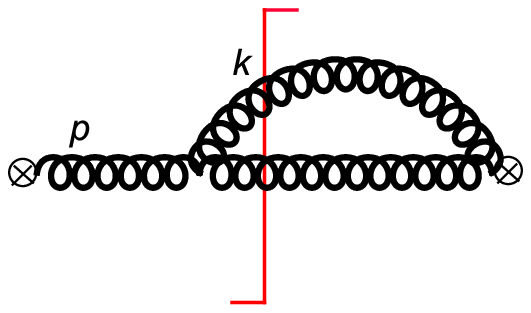}
	}
\subfigure[]{
	\label{fig:gJetSCET2}\includegraphics[width=0.250\textwidth]{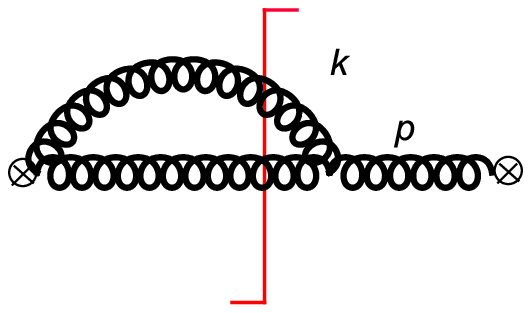}
	}
\end{center}
\vspace{-0.5cm}
\caption{\label{fig:JetFdiag} Nonvanishing diagrams contributing to the quark and gluon jet function at NLO.}
\end{figure}

The jet functions $J(p^2,\mu)$, defined in eqs.~(\ref{eq:qJetdef}) and (\ref{eq:gJetef}), describes a collinear quark or gluon with the invariant mass $p^2$. It is process independent and has been calculated at NLO in ref.~\cite{Manohar:2003vb} and NNLO in refs.~\cite{Becher:2006qw,Becher:2010pd}, respectively.  The nonvanishing diagrams contributing to the NLO jet function in Feynman gauge are given in figure~\ref{fig:JetFdiag}. The relevant diagrams corresponding to quark jet function are shown in the top row, and the ones corresponding to gluon are shown by the others. The RG evolution of the quark jet function is given by
\begin{equation}
 \frac{dJ_q(p^2,\mu)}{d\ln\mu} = \left( -2 C_F\gcusp
\ln\frac{p^2}{\mu^2} - 2 \gamma^{J_q} \right)J_q(p^2,\mu)
+2C_F\gcusp\int^{p^2}_0
dq^2\,\frac{J_q(p^2,\mu)-J_q(q^2,\mu)}{p^2-q^2}.
\end{equation}
The gluon jet function is the same with $C_F$ replaced by $C_A$ and $\gamma^{J_q}$ replaced by $\gamma^{J_g}$, respectively. This evolution equation can be solved by the Laplace transformation~\cite{Becher:2006nr,Becher:2006mr}:
\begin{equation}\label{eq:JetFLapTrans}
 \widetilde{j}(\ln\frac{Q^2}{\mu^2},\mu)=\int^\infty_0
dp^2\,\exp(-\frac{p^2}{Q^2e^{\gamma_E}}) J(p^2,\mu),
\end{equation}
which satisfies the the RG equation
\begin{equation}\label{eq:jetRG}
\begin{aligned}
\frac{d}{d\ln\mu}{\widetilde j}_q(\ln\frac{Q^2}{\mu^2},\mu)&=\left(-2C_F \gcusp
\ln\frac{Q^2}{\mu^2}-2\gamma^{J_q}\right){\widetilde j}_q(\ln\frac{Q^2}{\mu^2},\mu)\,,\\
\frac{d}{d\ln\mu}{\widetilde j}_g(\ln\frac{Q^2}{\mu^2},\mu)&=\left(-2C_A \gcusp
\ln\frac{Q^2}{\mu^2}-2\gamma^{J_g}\right){\widetilde j}_g(\ln\frac{Q^2}{\mu^2},\mu)\,.
\end{aligned}
\end{equation}
Thus the jet function at an arbitrary scale $\mu$ is given by
\begin{equation}\label{eq:evoljet}
\begin{aligned}
 {J}_q(p^2,\mu)&=\exp \bigl[ -4C_FS(\mu_j,\mu)+2A_{J_q}(\mu_j,\mu)
\bigr] {\widetilde j}_q(\partial_{\eta_j}, \mu_j )  \frac{1}{p^2} \left(
\frac{p^2}{\mu^2_j}\right)^{\eta_{j_q}}
\frac{e^{-\gamma_E
\eta_j}}{\Gamma(\eta_{j_q})}\,,\\
 {J}_g(p^2,\mu)&=\exp \bigl[ -4C_A S(\mu_j,\mu)+2A_{J_g}(\mu_j,\mu)
\bigr] {\widetilde j}_g(\partial_{\eta_j}, \mu_j )  \frac{1}{p^2} \left(
\frac{p^2}{\mu^2_j}\right)^{\eta_{j_g}}
\frac{e^{-\gamma_E
\eta_{j_g}}}{\Gamma(\eta_{j_g})},
\end{aligned}
\end{equation}
where $\eta_{j_q}=2 C_F A_\Gamma(\mu_j,\mu)$ and $\eta_{j_q}=2 C_A A_\Gamma(\mu_j,\mu)$.
The $\mu$-dependent part of the Laplace transformed jet function $\widetilde{j}(L,\mu)$
is determined by the anomalous dimensions of the jet function as in eq. (\ref{eq:jetRG}),
while the $\mu$-independent part can be obtained by a fixed-order calculation.
At NLO, it is
\begin{eqnarray}\label{eq:jetFuncNLO}
{\widetilde j}(L,\mu)&=&1+\frac{\alpha_s}{4\pi}\bigg( \frac{\Gamma_0^J}{2}L^2+\gamma^J_0L+c^J_1\bigg)\,.
\end{eqnarray}

When jet algorithm is applied, the phase space for the collinear radiation need to be constrained. For anti-$k_T$  algorithm, the restriction of phase space is given by
\begin{equation}\label{eq:PSrestrict}
\begin{aligned}
\Theta_{\rm anti-k_T}
=\Theta\left(\tan^2\frac{R}{2}>\frac{k^+(p^-)^2}{k^-(p^--k^-)^2}\right)\,.
\end{aligned}
\end{equation}
$p$ and $k$ denote the incoming and loop momenta, respectively. In the threshold limit $p^2\to 0$, the integral region of momentum $k$ can be expressed as
\begin{equation}
\begin{aligned}
\frac{p^2}{p^-\tan^2\frac{R}{2}}<k^-<p^--\frac{p^2}{p^-\tan^2\frac{R}{2}}\,.
\end{aligned}
\end{equation}
To avoid the double counting of the soft region covered by soft function, the zero-bin subtraction \cite{Manohar:2006nz} should be considered. Taking the soft limit of the restriction in eq.~(\ref{eq:PSrestrict}), the phase space of zero-bin region is given by~\cite{Ellis:2010rwa,Cheung:2009sg}
\begin{equation}\label{eq:PSrestrictZB}
\begin{aligned}
\Theta_{\rm anti-k_T}^{(0)}
=\Theta\left(\tan^2\frac{R}{2}>\frac{k^+}{k^-}\right)\,.
\end{aligned}
\end{equation}
After integrating the phase space and taking zero-bin substraction, the jet functions with anti-$k_T$ algorithm are given by
\begin{equation}\label{eq:kTJetFinal}
\begin{aligned}
J_q^{\rm anti-k_T}(p^2,p^-,R,\mu)=&J_q(p^2,\mu)
 +\frac{C_F \alpha_s}{4\pi}\left(\frac{1}{p^2}\right)_\star
 \frac{2p^2}{(p^-)^2\tan^2\frac{R}{2}}\,,\\
J_g^{\rm anti-k_T}(p^2,p^-,R,\mu)=&J_g(p^2,\mu)
  +\frac{\alpha_s}{4 \pi}\left(\frac{1}{p^2}\right)_\star
  \frac{p^2}{(p^-)^2\tan^2\frac{R}{2}}(4C_A-2n_f)\,,
\end{aligned}
\end{equation}
where $J_q(p^2,\mu)$ and $J_g(p^2,\mu)$ are traditional jet functions. $J_q^{\rm anti-k_T}$ and $J_g^{\rm anti-k_T}$ approach to the traditional ones when jet radius $R\to \infty$, which means that there is no restriction to collinear radiation. In addition, it can be seen that the $R$-dependent terms in eq.~(\ref{eq:kTJetFinal}) are suppressed by $m_J^2/p_T^2$ at threshold limit, because $p^2\sim m_J^2$, $p^-\sim 2E_J$ and $m_J\ll E_J$. We have checked numerically that the corrections from $R$-dependent terms to jet mass spectra are below $1\%$ for $m_J<100\,{\rm GeV}$, so we will use the traditional jet functions in the following numerical calculation.
\section{Soft function}\label{sec:SoftF}
The soft function defined in eq.~(\ref{eq:S_ope_def}) describes soft interactions between all colored particles. When calculating the soft function, we need to consider jet algorithm, which imposes a restriction on the phase space of the soft radiation. In ref.~\cite{Chien:2012ur}, the jet was defined as all the radiation in a cone of half-angle $R$ around the jet direction, which is different from the one defined by the standard sequencial recombination jet algorithms at hadron colliders. In this work, we choose anti-$k_T$ algorithm to calculate the boost-invariant soft function. In addition, as discussed in  ref.~\cite{Chien:2012ur}, there are multiple soft scales in soft function, which need to be refactorized. In this section, we first discuss the calculation of the NLO soft function with jet algorithm for all channels, and then show its refactorization. The details of the calculations can be found in appendix~\ref{app3}.
\subsection{NLO calculation}
\begin{figure}[t]
\begin{center}
\subfigure[${\cI}_{12}$]{
	\label{fig:I12}\includegraphics[width=0.250\textwidth]{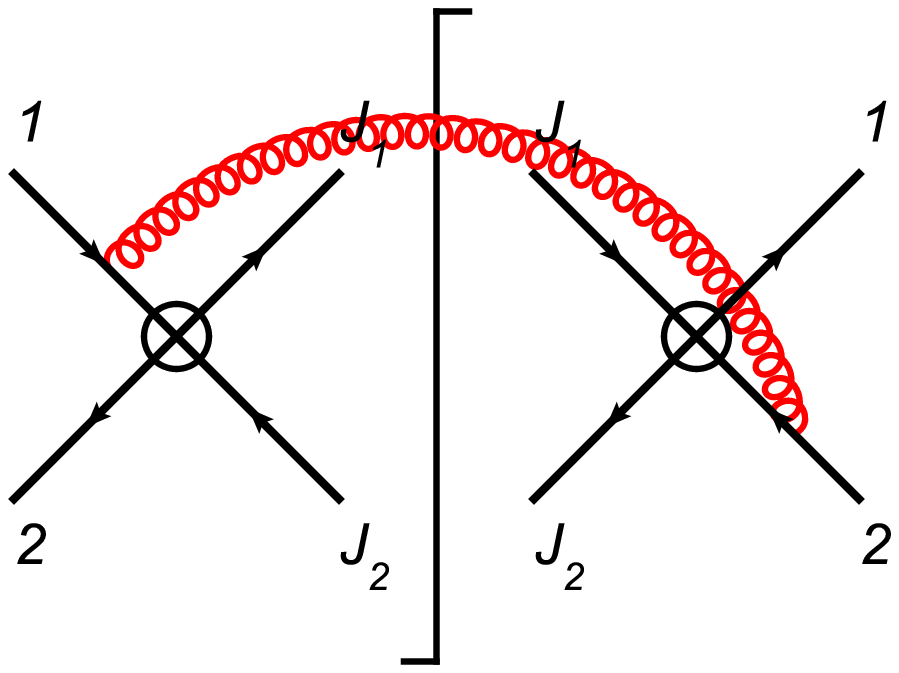}
	}\;\quad
\subfigure[$\cI_{13}$]{
	\label{fig:I13}\includegraphics[width=0.250\textwidth]{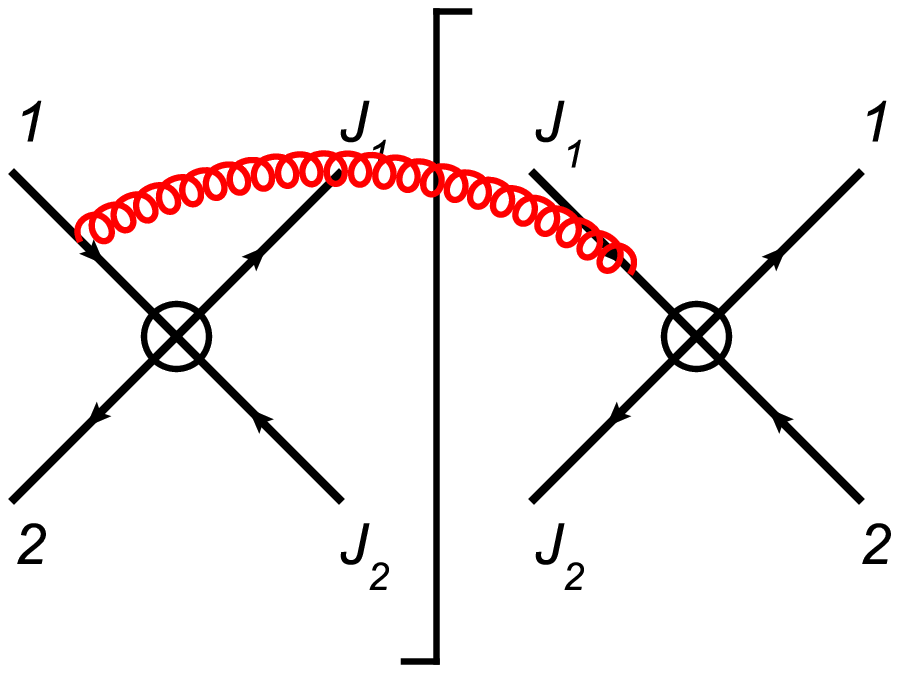}
	}\;\quad
\subfigure[$\cI_{14}$]{
	\label{fig:I14}\includegraphics[width=0.250\textwidth]{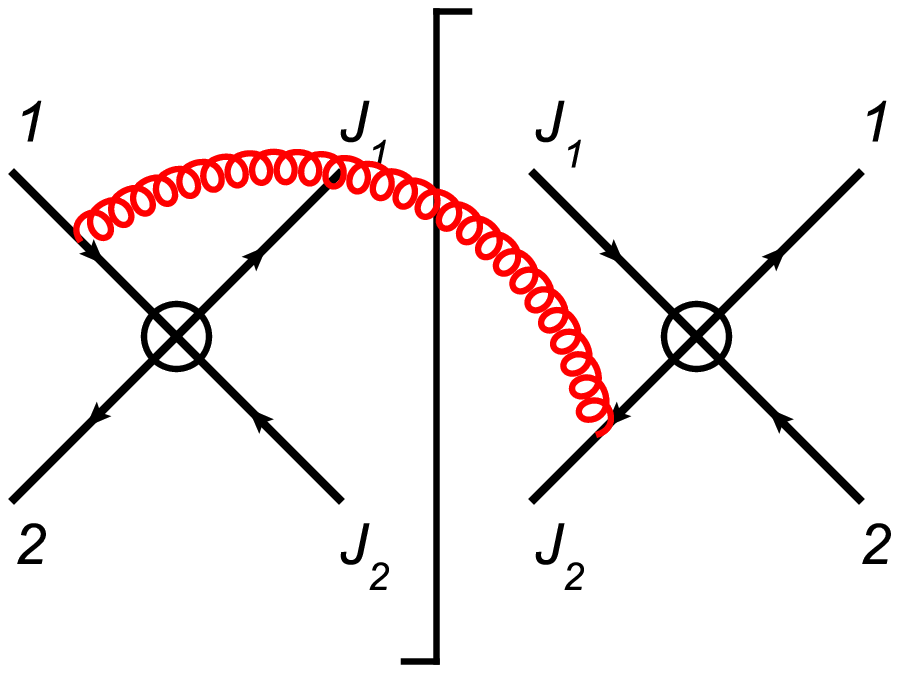}
	}\\
\subfigure[$\cI_{23}$]{
	\label{fig:I23}\includegraphics[width=0.250\textwidth]{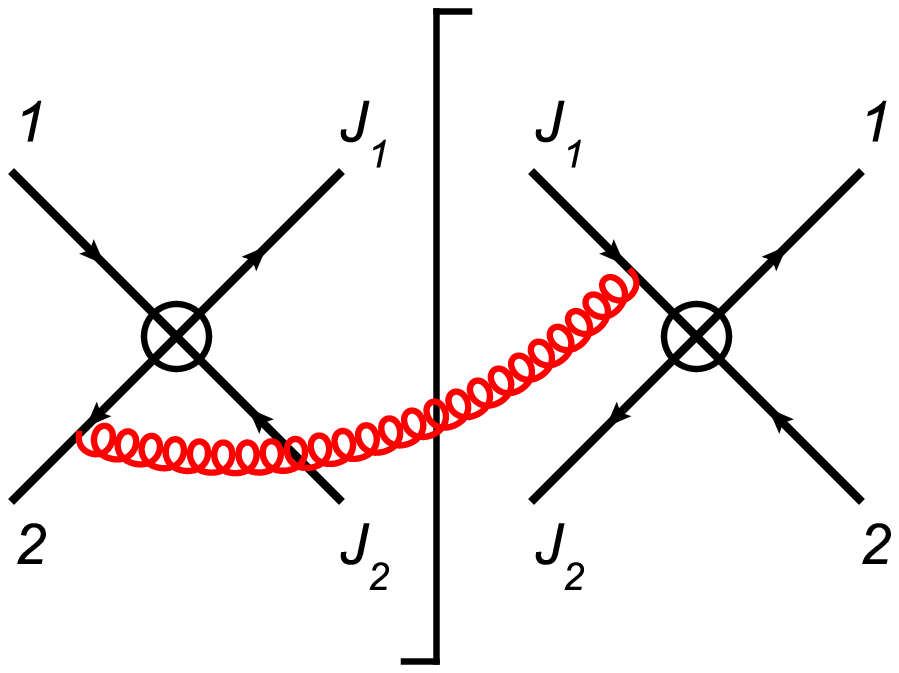}
	}\;\quad
\subfigure[$\cI_{24}$]{
	\label{fig:I24}\includegraphics[width=0.250\textwidth]{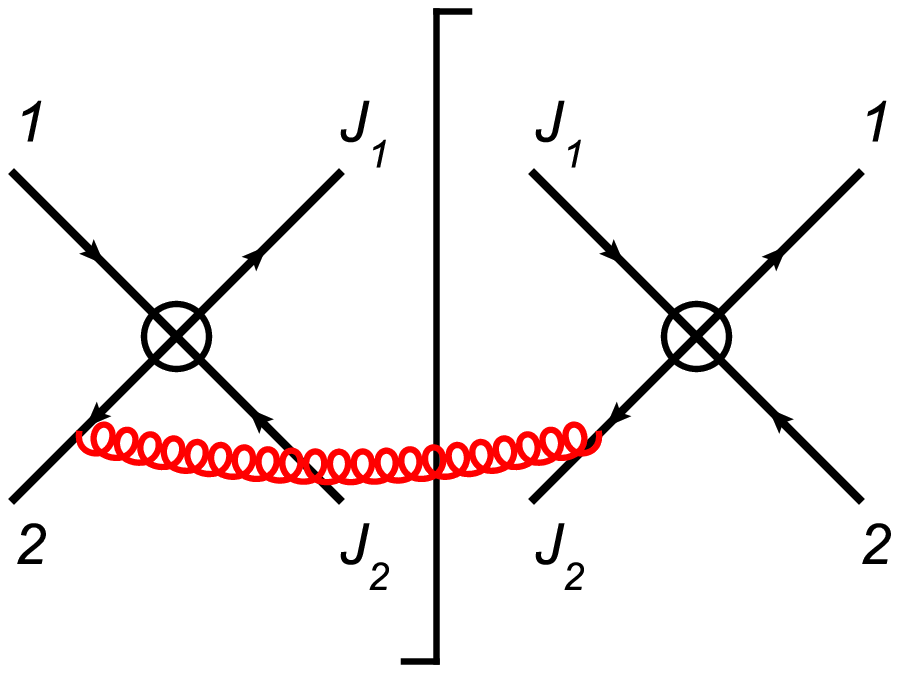}
	}\;\quad
\subfigure[$\cI_{34}$]{
	\label{fig:I34}\includegraphics[width=0.250\textwidth]{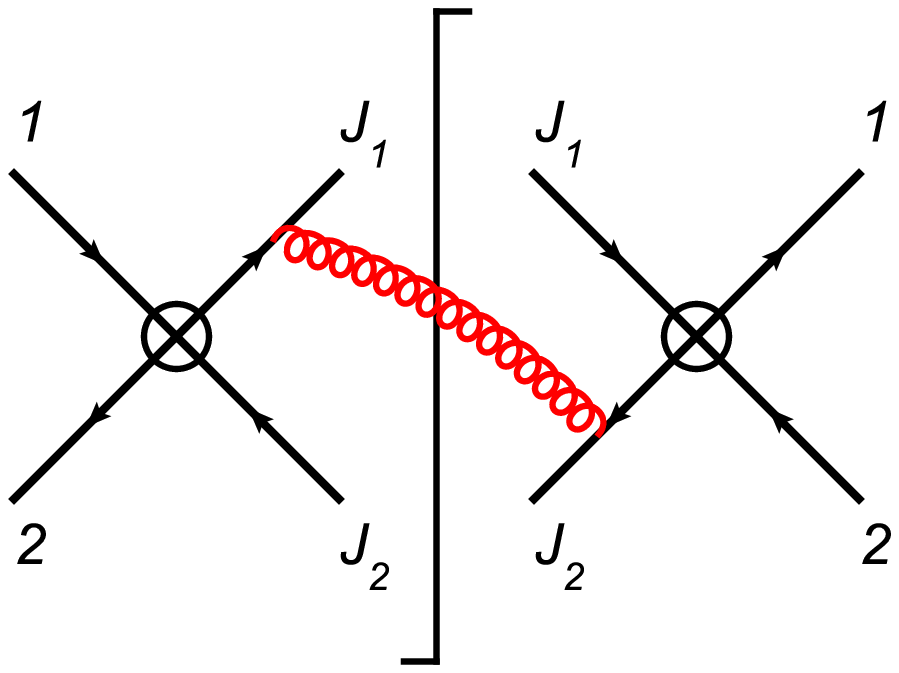}
	}
\end{center}
\vspace{-0.5cm}
\caption{\label{fig:Iijcutdiags} Non-vanishing diagrams contributing to the soft function at NLO. }
\end{figure}

As shown in eq.~(\ref{eq:SinColorBasis}), the soft function ${\bf S}(\kin,\kout,\mu)$ can be decomposed in color space and calculated in the eikonal approximation. Eq.~(\ref{eq:color4q}) has shown the color structures for 4-quark channels. For $gg\to q\bar{q}$ and 4-gluon channels, the color structures are given by
\begin{gather}
  | c_1 \rangle =\left(t^{a_1} t^{a_2}\right)_{i_3,i_4}\,,\quad
  | c_2 \rangle =\left(t^{a_2} t^{a_1}\right)_{i_3,i_4}\,,\quad
  | c_3 \rangle =\delta^{a_1,a_2} \delta_{i_3,i_4}\,,
\end{gather}
and
\begin{gather}
  | c_1\rangle ={\rm Tr}\left(t^{a_1}t^{a_2}t^{a_3}t^{a_4}\right)\,,\quad
  | c_2\rangle ={\rm Tr}\left(t^{a_1}t^{a_2}t^{a_4}t^{a_3}\right)\,,\quad
  | c_3\rangle ={\rm Tr}\left(t^{a_1}t^{a_4}t^{a_3}t^{a_2}\right)\,,\nn\\
  | c_4\rangle ={\rm Tr}\left(t^{a_1}t^{a_4}t^{a_2}t^{a_3}\right)\,,\quad
  | c_5\rangle ={\rm Tr}\left(t^{a_1}t^{a_3}t^{a_4}t^{a_2}\right)\,,\quad
  | c_6\rangle ={\rm Tr}\left(t^{a_1}t^{a_3}t^{a_2}t^{a_4}\right)\,,\nn\\
  | c_7\rangle ={\rm Tr}\left(t^{a_1}t^{a_4}\right)
  {\rm Tr}\left(t^{a_2}t^{a_3}\right)\,,\quad
  | c_8\rangle ={\rm Tr}\left(t^{a_1}t^{a_2}\right)
  {\rm Tr}\left(t^{a_3}t^{a_4}\right)\,,\quad
  | c_9\rangle ={\rm Tr}\left(t^{a_1}t^{a_3}\right)
  {\rm Tr}\left(t^{a_2}t^{a_4}\right)\,,
\end{gather}
respectively. At LO, the soft functions is given by
\begin{equation}\label{eq:LOS}
S_{IJ}^{(0)}=\tilde{s}_{IJ}^{(0)}\delta(\kin)\delta(\kout)
\end{equation}
At NLO, the soft functions can be expressed as follows
\begin{equation}\label{eq:NLOSdef}
S_{IJ}^{(1)}(\kin,\kout,y_J,R,\mu)=
\sum_{i,j}^{i\neq j}\left(w_{ij}\right)_{IJ}{\cI}_{ij}(\kin,\kout,y_J,R,\mu)\,,
\end{equation}
where $i$ and $j$ index the massless external partons, while $I$ and $J$ index the color structures. According to eq.~(\ref{eq:SinColorBasis}), the color matrix $\left(w_{ij}\right)_{IJ}$ can be written as
\begin{equation}\label{eq:ColMSdef}
\left(w_{ij}\right)_{IJ}=\langle c_I|\bm{T}_i\cdot \bm{T}_j|c_J\rangle.
\end{equation}
For ${\cI}_{ij}(\kin,\kout,p_T,y,R,\mu)$, we need to calculate the non-vanishing real emission diagrams in dimension regularization, as shown in figure~\ref{fig:Iijcutdiags}, which is given by
\begin{equation}\label{eq:Iijdef}
{\cI}_{ij}(\kin,\kout,y_J,R,\mu)=
  -\frac{4\pi \alpha_s}{(2\pi)^{d-1}}
  \Big(\frac{\mu^2e^{\gamma_E}}{4\pi}\Big)^{\epsilon}
  \int d^dq \ \delta(q^2)\theta(q_0) \cM_R(\kin,\kout,R,q)
  \frac{n_i\cdot n_j}{(n_i\cdot q)(n_j\cdot q)}\;,
\end{equation}

In the CM of partons, the measurement function $\cM_R(\kin,\kout,R,q)$ for anti-$k_T$ algorithm is
\begin{equation}\label{eq:Smeasfunc}
\begin{aligned}
\cM_R(\kin,\kout,R,q)
=&\Theta\Big(R^2-(y-y_J)^2-(\phi-\phi_J)^2 \Big)\delta(\kin-n_J\cdot q)\\
&+\Theta\Big((y-y_J)^2+(\phi-\phi_J)^2-R^2\Big)\delta(\kout-{\bn_J}\cdot q) \,.
\end{aligned}
\end{equation}
where $y_J$ and $\phi_J$ presents the rapidity and azimuthal angle of the measured jet. And $y$ and $\phi$ are the rapidity and azimuthal angle of the soft gluon with momentum $q^\mu$, respectively. For convenience, we calculate the soft function in the CM frame of initial partons and take $\phi_J$ to be zero. The results of function ${\cI}_{ij}$ are
\begin{align}\label{eq:IijNLOres}
\cI_{12}(\kin,\kout,y_J,R,\mu)=&\left(\frac{\alpha_s}{4\pi}\right)
\Bigg\{\dkin\dkout\Big[-4R^2\ln(2\cosh y_J)-2R^2+4R^2\ln R\nn\\
&-4\log^2(2\cosh y_J)+\frac{\pi^2}{6}\Big]+\dkout\left[-2R^2\kinp\right]\nn\\
&+\dkin\left[-8\logkoutp+2R^2\koutp\right]\Bigg\}\nn\\
\cI_{13}(\kin,\kout,y_J,R,\mu)=&\left(\frac{\alpha_s}{4\pi}\right)
\Bigg\{\dkin\dkout\Big[-R^2\ln(2\cosh y_J)-\frac{R^2}{2}+R^2\ln R\nn\\
&-8\ln R \ln(2\cosh y_J)+4\ln^2 R-4y_J^2+4\ln(2\cosh y_J)\Big]\nn\\
&+\dkout\left[4\logkinp-(\frac{R^2}{2}+4\ln R)\kinp\right]\nn\\
&+\dkin\left[-4\logkoutp+(\frac{R^2}{2}+4\ln R+8y_J)\koutp\right]\Bigg\}\\
\cI_{14}(\kin,\kout,y_J,R,\mu)=&\left(\frac{\alpha_s}{4\pi}\right)
\Bigg\{\dkin\dkout\Big[\frac{1}{32}R^2 e^{2y_J}\sech^2 y_J\nn\\
&\times\left[-4(R^2+8)\ln(2\cosh y_J)-R^2+4(R^2+8)\ln R -16\right]\Big]\nn\\
&+\dkout\left[-\frac{1}{16}e^{2y_J}\sech^2 y_J R^2(R^2+8)\kinp\right]\nn\\
&+\dkin\left[\frac{1}{16}e^{2y_J}\sech^2 y_J R^2(R^2+8)\koutp\right]\Bigg\}\nn\\
\cI_{34}(\kin,\kout,y_J,R,\mu)=&\left(\frac{\alpha_s}{4\pi}\right)
\Bigg\{\dkin\dkout\Big[-8\ln R\,\ln(2\cosh y_J)+4\ln^2 R\nn\\
&+4\ln^2(2\cosh y_J)-\frac{\pi^2}{6}\Big]\nn\\
&+\dkout\left[4\logkinp-4\ln R \kinp\right]\nn\\
&+\dkin\Bigg(4\logkoutp\nn\\
&+2\left[-2\ln R + 4\ln(2\cosh y_J)\right]\koutp\Bigg)\Bigg\}\nn\,.
\end{align}
$\cI_{23}$ and $\cI_{24}$ can be obtained by the relations
\begin{equation}
\begin{aligned}
\cI_{23}(\kin,\kout,y_J,R,\mu)=&\cI_{13}(\kin,\kout,-y_J,R,\mu)\,,\\
\cI_{24}(\kin,\kout,y_J,R,\mu)=&\cI_{14}(\kin,\kout,-y_J,R,\mu)\,.
\end{aligned}
\end{equation}

In the calculation, we take the limit that $R\to 0$,
but we have kept all the terms up to $O(R^2)$ so that our result can can be used for a wider range of $R$.

\subsection{RG equation of the soft function}
Now, we discuss the evolution of the soft function. Its RG equation can be derived by using the RG invariance of the cross section. In the hadronic threshold limit, we have \cite{Becher:2009th}
\begin{equation}\label{eq:xsecinhadthres}
\begin{aligned}
\frac{ d^2 \sigma_{q \bar q}}{ d \MXt  d y} \propto &\int  d x_1 \int  d x_2 \int  d m_{J_1}^2 \int  d m_{J_2}^2 \int  d \kin\int  d \kout\\
&\times \, H_{IJ}(\hat{s},\hat{t},\hat{u},\mu)\,S_{JI} (\kin,\kout, \mu)
f_{q/N_1} (x_1, \mu)  f_{{\bar{q}}/N_2} (x_2, \mu)
J_1(m_{J_1}^2, \mu) J_2(m_{J_2}^2, \mu) \\
&\times
\, \delta\! \left[M_X^2 - \left( m_{J_1}^2+m_{J_2}^2 + 2 E_J(\kin+\kout)  + \frac{p_T^2}{{\bar v}}(1-x_1) + \frac{p_T^2}{v}(1-x_2)\right) \right]\,.
\end{aligned}
\end{equation}
To transform the convolution form to a product form, using the Laplace transformation
\begin{equation}
\frac{ d^2 \widetilde\sigma}{ d Q^2  d y} = \int_0^\infty  d\MXt \,
\exp \left(-\frac{M_X^2}{Q^2 e^{\gamma_E}}\right) \frac{ d^2\sigma}{ d \MXt  d y}\,,
\end{equation}
we can obtain
\begin{equation}
\frac{ d^2 \widetilde\sigma}{ d Q^2  d y} =
H_{IJ} (\hat{s},\hat{t},\hat{u}, \mu)
\widetilde{s}_{JI} \left(\kain,\kaout, \mu \right)
\widetilde{f}_{i_1/N_1} \left( \tau_1, \mu \right)
\widetilde{f}_{i_2/N_2} \left( \tau_2, \mu \right)
\widetilde{j}_1 (Q^2,\mu)\widetilde{j}_2 (Q^2,\mu)\,.
\end{equation}
where $\widetilde{s}$ is the Laplace transformed soft function
\begin{equation}
{\tilde s}_{IJ}(\kain,\kaout,\mu)
 = \int_0^\infty d\kin\int_0^\infty d\kout
    \exp\Big(-\frac{\kin}{\kain e^{\gamma_E}}\Big)\exp\Big(-\frac{\kout}{\kaout e^{\gamma_E}}\Big)
    S_{IJ}(\kin, \kout,\mu) \,.
\end{equation}
The RG invariance requires
\begin{equation}\label{eq:RGinv}
\frac{ d}{ d \ln \mu} \left[
H_{IJ} (\hat{s},\hat{t},\hat{u}, \mu)
\widetilde{s}_{JI} \left(\kain,\kaout, \mu \right)
\widetilde{f}_{i_1/N_1} \left( \tau_1, \mu \right)
\widetilde{f}_{i_2/N_2} \left( \tau_2, \mu \right)
\widetilde{j}_1 (Q^2,\mu)\widetilde{j}_2 (Q^2,\mu)
\right] = 0\, .
\end{equation}
And the RG equation of PDF is
\begin{equation}\label{eq:PDFsRG}
\begin{aligned}
\frac{ d \widetilde{f}_{q/N} (\tau, \mu)}{ d \ln  \mu} &=
\left[2 C_F \gamma_{\rm{cusp}} \ln\left( \tau \right) + 2\gamma^{f_q} \right] \widetilde{f}_{q/N} (\tau, \mu)\,,
\end{aligned}
\end{equation}
with
\begin{gather}
\tau_1=Q^2/(-{\hat u}_1)\,,\quad \tau_2=Q^2/(-{\hat t}_1)\,,
\end{gather}
for beam $N_1$ and $N_2$, respectively. Here
\begin{align}
\hat{t}_1=-\frac{p_T^2}{v}\,,\quad \hat{u}_1=-\frac{p_T^2}{{\bar v}}\,,
\end{align}
in threshold limit $m_{J_1,J_2}^2\to 0$ and $w\to 1$.
The gluon PDF equation is the same with $C_F \to C_A$ and $\gamma^{f_q} \to \gamma^{f_g}$.
With eqs.~(\ref{eq:diagHREG}), (\ref{eq:jetRG}), (\ref{eq:RGinv}) and (\ref{eq:PDFsRG}), the RG equation of the soft function $\hws_{K'K}$ in the diagonal basis is given by
\begin{equation}\label{eq:diagSRGE}
\begin{aligned}
\frac{ d}{ d \ln \mu}\,\hws_{K'K}(\kain,\kaout,\mu)
&=\Big\{\gcusp\left[2C_{i_1}L({\hat u}_1)
+(2C_{i_2}-c_H)L({\hat t}_1)-\lambda_{K}-\lambda_{K'}^*\right]\\
&\qquad -2\,\gcusp \left(C_{i_1}+C_{i_2}-C_{j_1}-C_{j_2}\right)
\ln\frac{Q^2}{\mu^2}-2\gamma^S\Big\}\,\hws_{K'K}\,,
\end{aligned}
\end{equation}
where $\gamma^S=\gamma_H+\gamma^{f_{i_1}}+\gamma^{f_{i_2}}-\gamma^{J_1}-\gamma^{J_2}$,
$C_{i,j}=C_F$ and $C_A$ for quark and gluon, respectively.
The relation between the soft functions $\hws$ and $\widetilde{s}$ is
\begin{align}
  \hws_{K'K} =\left[\left(F^{-1}\right)^\dagger \cdot \widetilde{s} \cdot F^{-1}\right]_{K'K} \,.
\end{align}
We have checked that the NLO soft function in eq.~(\ref{eq:NLOSdef}) satisfies the RG equation (\ref{eq:diagSRGE}), which means our factorization is reasonable.
\subsection{Refactorization of the soft function}\label{sec:refactsoft}
As shown in eq.~(\ref{eq:NLOSdef}), the soft function depends on two variables $\kin$ and $\kout$, which are $\kin\sim m_J^2/p_T$ and $\kout\sim s_4/p_T$, in principle. It means that we should treat the two scales separately to control the convergence of perturbative expansion. However, at two-loop level, a complicated dependence on $\kin / \kout$ will emerge \cite{Kelley:2011aa,Kelley:2011ng}, which represents the nonglobal structure of the soft radiation. Although we could not ideally factorize the soft function into separate two pieces which depend only on $\kin /\mu$ and $\kout / \mu$, respectively, we can at least extract part of the soft function which depend only on a single scale~\cite{Chien:2012ur}. We define an auxiliary soft function ${\bm S}^{\rm in}$ which only depends on $\kin$
\begin{equation}\label{eq:Sin_ope_def}
{\bm S}^{\rm in}(\kin)=\langle 0|{\bm O}^{s\dagger}|X_s^{\rm in}\rangle\langle X_s^{\rm in}|{\bm O}^s(0)|0\rangle
\delta(\kin-n_J\cdot P_{Xs}^{\rm in})\,.
\end{equation}
In color basis, the NLO ${\bm S}^{\rm in}(\kin)$ can be expressed as
\begin{equation}\label{eq:NLOSindef}
S_{IJ}^{\rm in}(\kin)=
\sum_{i,j}^{i\neq j}\left(w_{ij}\right)_{IJ}{\cI}_{ij}^{\rm in}(\kin,y_J,R,\mu)\,,
\end{equation}
where ${\cI}_{ij}^{\rm in}$ can be computed by the similar integration in eq.~(\ref{eq:Iijdef}), except for the measurement function replaced by
\begin{equation}\label{eq:Sinmeasfunc}
\begin{aligned}
\cM_{\rm in}(\kin,R,q)
=&\Theta\Big(R^2-(y-y_J)^2-(\phi-\phi_J)^2 \Big)\delta(\kin-n_J\cdot q)\,.
\end{aligned}
\end{equation}
Besides, it is necessary to introduce the residual soft function to describe the soft radiation excluded by ${\bm S}^{\rm in}(\kin)$
\begin{equation}\label{eq:Sresdef}
\begin{aligned}
{\bm S}^{\rm res}(\kin,\kout,\mu)=\frac{{\bm S}(\kin,\kout,\mu)}{{\bm S}^{\rm in}(\kin,\mu)}\,.
\end{aligned}
\end{equation}
At one-loop level, we consider only one soft gluon emission, which is either inside or outside the jet. It  means that ${\bm S}^{\rm res}$ describe the soft radiation outside the jet, which only depend on $\kout$ at $\cO (\alpha_s)$, and we rewrite it by notation ${\bm S}^{\rm out}(\kout)$
\begin{equation}\label{eq:NLOSindef}
S_{IJ}^{\rm out}(\kout)=
\sum_{i,j}^{i\neq j}\left(w_{ij}\right)_{IJ}{\cI}_{ij}^{\rm out}(\kout,y_J,R,\mu)\,,
\end{equation}
where ${\cI}_{ij}^{\rm out}(\kout,y_J,R,\mu)$ can be calculated according to eq.~(\ref{eq:Iijdef}), through replacing the measurement function by $\cM_{\rm out}(\kout,R,q)$
\begin{equation}\label{eq:Soutmeasfunc}
\begin{aligned}
\cM_{\rm out}(\kout,R,q)
=&\Theta\Big((y-y_J)^2+(\phi-\phi_J)^2 - R^2\Big)\delta(\kout-\bn_J\cdot q)\,.
\end{aligned}
\end{equation}
Now, the soft function at $\cO(\alpha_s)$ in diagonal basis reads
\begin{equation}\label{eq:resummedS}
{\hat S}_{K'K}(\kin,\kout,\muin,\muout,\mu)={\hat S}_{K'L}^{\rm in}(\kin,\muin,\mu)
\left({\hat S}^{(0)}\right)_{LM}^{-1}{\hat S}_{MK}^{\rm out}(\kout,\muout,\mu)\,.
\end{equation}
$\hws^{\rm in}(L_{\rm in},\mu)$ and $\hws^{\rm out}(L_{\rm out},\mu)$ is the Laplace transformation of  ${\hat S}^{\rm in}$ and ${\hat S}^{\rm out}$, respectively, and their RG equations are
\begin{equation}\label{eq:sioRGE}
\begin{aligned}
\frac{d}{d\,\ln\mu}\hws_{K'L}^{\rm in}(L_{\rm in},\mu) =
&\left[-2\widetilde{B}_{K'L}^{\rm in}\gcusp\,L_{\rm in}
-\widetilde{C}_{K'L}^{\rm in}\gcusp
-\widetilde{\gamma}_{K'L}^{{\rm in}}\right]\hws_{K'L}^{\rm in}\,,\\
\frac{d}{d\,\ln\mu}\hws_{MK}^{\rm out}(L_{\rm out},\mu) =
&\left[-2\widetilde{B}_{MK}^{\rm out}\gcusp\,L_{\rm out}
-\widetilde{C}_{MK}^{\rm out}\gcusp
-\widetilde{\gamma}_{MK}^{{\rm out}}\right]\hws_{MK}^{\rm out} \,,
\end{aligned}
\end{equation}
where $\widetilde{\gamma}^{\rm in,out}$ are anomalous dimensions depending on the jet radius $R$, which are given at one-loop level in appendix~\ref{app3}. Solving the RG equation, we get the resummed soft functions ${\bm S}^{\rm in}$ and ${\bm S}^{\rm out}$
\begin{align}\label{eq:resummedSio}
{\hat S}_{K'L}^{\rm in}(\kin,\muin,\mu) =
&\exp\left[-2\widetilde{B}_{K'L}^{\rm in}\,S(\muin,\mu)
+\widetilde{C}_{K'L}^{\rm in}\,A_\Gamma(\muin,\mu)
+A_{\widetilde{\gamma}_{K'L}^{\rm in}}(\muin,\mu)\right]\nn\\
&\times \hws_{K'L}^{\rm in}(\partial_{\etain} ,\muin)
\frac{1}{\kin}\left(\frac{\kin}{\muin}\sqrt{\frac{2n_{12}}{n_{1J}n_{2J}}}\right)^{\etain}
\frac{e^{-\gE\etain}}{\Gamma(\etain)}\,,\\
{\hat S}_{MK}^{\rm out}(\kout,\muout,\mu) =
&\exp\left[-2\widetilde{B}_{MK}^{\rm out}\,S(\muout,\mu)
+\widetilde{C}_{MK}^{\rm out}\,A_\Gamma(\muout,\mu)
+A_{\widetilde{\gamma}_{MK}^{\rm out}}(\muout,\mu)\right]\nn\\
&\times \hws_{MK}^{\rm out}(\partial_{\etaout},\muout)
\frac{1}{\kout}\left(\frac{\kout}{\muout}\sqrt{\frac{2n_{12}}{n_{1J}n_{2J}}}\right)^{\etaout}
\frac{e^{-\gE\etaout}}{\Gamma(\etaout)}\,,
\end{align}
with
\begin{align}
\etain=&2\widetilde{B}_{K'L}\,A_\Gamma(\muin,\mu)\,,\nn\\
\etaout=&2\widetilde{B}_{MK}\,A_\Gamma(\muout,\mu)\,.\nn
\end{align}

As shown in ref.~\cite{Chien:2012ur}, the above procedure, so-called refactorization, is an approximate factorization, because the residual soft function would depend on both $\kin$ and $\kout$ beyond $\cO (\alpha_s)$. At two-loop level, $\ln^n (\kin / \kout)$ would emerge due to the non-global structure, which has been widely studied at the $e^+ e^-$ colliders ~\cite{Banfi:2010pa,Khelifa:2011zu,Kelley:2011aa,Dasgupta:2001sh,Dasgupta:2002bw,Banfi:2002hw,Kelley:2011ng,Hornig:2011iu}, but rarely investigated at hadron colliders with a sequencial recombination jet algorithm~\cite{Dasgupta:2012hg}. A systematical discussion of them requires the complete two-loop results of the soft function with jet algorithms, and is beyond the scope of this paper.
\section{RG improved cross section}\label{sec:RGimp}
From eq.~(\ref{eq:kernelCij}), using eqs.~(\ref{eq:resummedH}), (\ref{eq:evoljet}) and (\ref{eq:resummedSio}), we can obtain
\begin{equation}\label{eq:resummedCij}
\begin{aligned}
C_{ij}(\hat{s},\hat{t},\hat{u},m_{J_1}^2,\mu)
=&\sum_{K,K',L,M}\frac{\alpha_s(\mu_h)^2}{\alpha_s(\mu)^2}
\exp\Big[ 2c_H S (\mu_h, \mu) -2 A_H (\mu_h, \mu) \Big] \\
&\times \exp\left[-A_{\Gamma} (\mu_h, \mu)
\left(\lambda_{K}+\lambda_{K'}^*
+c_H \ln \left | \frac{\hat t}{\mu_h^2} \right | \right)\right]
H_{KK'}({\hat s},{\hat t},{\hat u},\mu_h)\\
&\times\exp\left[-2\widetilde{B}_{K'L}^{\rm in}\,S(\muin,\mu)
+\widetilde{C}_{K'L}^{\rm in}\,A_\Gamma(\muin,\mu)
+A_{\widetilde{\gamma}_{K'L}^{\rm in}}(\muin,\mu)\right]\\
&\times \exp \left[- 4 C_1 S (\mu_{j_1}, \mu) + 2 A_{J_1} (\mu_{j_1}, \mu)\right]
\left(\frac{\mu_{j_1}^2}{\muin p_T}\right)^{\etain}\\
&\quad\times\widetilde{j}_1 (\partial_{\eta_1},\mu_{j_1})
\hws_{K'L}^{\rm in}(\ln\frac{\mu_{j_1}^2}{\muin p_T}+\partial_{\eta_1} ,\muin)
\frac{1}{m_{J_1}^2}\left(\frac{m_{J_1}^2}{\mu_{j_1}^2}\right)^{\eta_1}
\frac{e^{-\gE\eta_1}}{\Gamma(\eta_1)}\\
&\times\left({\hat S}^{(0)}\right)_{LM}^{-1}
\exp\left[- 4 C_2 S (\mu_{j_2}, \mu) + 2 A_{J_2} (\mu_{j_2}, \mu)\right]
\left(\frac{\mu_{j_2}^2}{\muout p_T}\right)^{\etaout}\\
&\times\exp\left[-2\widetilde{B}_{MK}^{\rm out}\,S(\muout,\mu)
+\widetilde{C}_{MK}^{\rm out}\,A_\Gamma(\muout,\mu)
+A_{\widetilde{\gamma}_{MK}^{\rm out}}(\muout,\mu)\right]\\
&\times\widetilde{j}_2 (\partial_{\eta_2},\mu_{j_2})
\hws_{MK}^{\rm out}(\ln\frac{\mu_{j_2}^2}{\muout p_T}+\partial_{\eta_2} ,\muout)
\frac{1}{s_4}\left(\frac{s_4}{\mu_{j_2}^2}\right)^{\eta_2}
\frac{e^{-\gE\eta_2}}{\Gamma(\eta_2)}
\,,
\end{aligned}
\end{equation}
with
\begin{align}
\eta_1=\etain+\eta_{j_1}\,,\quad
\eta_2=\etaout+\eta_{j_2}\,.
\end{align}
And the resummed cross section (\ref{eq:genxsec}) can be written as
\begin{align}\label{eq:intx1s4}
  \frac{d\sigma^{{\rm NNLL}_{p}}}{dp_Tdydm_{J_1}^2} = &\frac{p_T}{8\pi s} \sum_{i,j} \int_{\frac{-u_1-m_{J_1}^2}{s+t_1}}^1 \frac{dx_1}{x_1}
  \int_0^{x_1 s+x_1 t_1+u_1+m_{J_1}^2} \frac{ds_4}{s_4-x_1t_1-m_{J_1}^2} \nn\\
  &\qquad \times f_{i/N_1}(x_1,\mu_f) \,f_{j/N_2}(x_2,\mu_f) \,C_{ij}(\hat{s},p_T,y,m_{J_1}^2,\mu_f) \, ,
\end{align}
where ${\rm NNLL}_p$ denotes the approximate $\rm NNLL$ resummation, which means that the NGLs are ignored in this paper. Here, we have changed the integration variables from $x_2$ to $s_4$, which have relation
\begin{align}
x_2(s_4)=\frac{s_4-x_1t_1-m_{J_1}^2}{x_1s+u_1}\,.
\end{align}

To give precise predictions, we resum the singular terms $\ln^n(m_J^2/p_T^2)$ and $\ln^n(s_4/p_T^2)$ in threshold limits to all orders and include the nonsingular terms up to NLO. And the RG improved differential cross section is given by
\begin{align}\label{eq:matching}
\frac{d\sigma^{{\rm NNLL}_{p}+{\rm NLO}}}{dp_Tdydm_{J_1}^2} = \frac{d\sigma^{\rm NLO}}{dp_Tdydm_{J_1}^2}
+ f(m_J)\left(\frac{d\sigma^{{\rm NNLL}_{p}}}{dp_Tdydm_{J_1}^2}
-\frac{d\sigma^{{\rm NNLL}_{p}}}{dp_Tdydm_{J_1}^2}\bigg|_{\text{expanded to NLO}}\right)\,,
\end{align}
where
\begin{align}\label{eq:matchfunc}
f(m_J)=\frac{1}{1+\left(m_J/m_J^{\rm match}\right)^i}
\end{align}
is the weight function, as defined in refs.~\cite{Plehn:2000be,Han:2009ya}. $m_J^{\rm match}$ denotes the scale above which the fixed order calculation is reliable. For small $m_J$, $f(m_J)$ approximates to one, and $\sigma^{\rm NLO}$ and $\sigma^{{\rm NNLL}_p}|_{\rm expanded}$ will cancel each other, and the resummation result dominates the cross section. With increasing $m_J$ above $m_J^{\rm match}$, $f(m_J)$ goes to zero quickly, and the main contributions are from the fixed-order results.
When the power index $i$ becomes larger, the translation from the resummation results  to the fixed-order ones is faster.
In this work, $m_J^{\rm match}$ is chosen at 100 GeV and $i$ is taken as 4. But the numerical results are not sensitive to the choices of these parameters.

\section{Numerical results}\label{sec:numres}
In this section, we discuss the numerical results for the jet mass distribution in dijet process at the LHC. Throughout the numerical calculations, we use the MSTW2008 PDF sets~\cite{Martin:2009bu} and associated strong coupling $\alpha_s$. In order to compare with Monte Carlo tools, we use \texttt{PYTHIA8} \cite{Sjostrand:2007gs} with its default "Tune 4C" input. \texttt{FASTJET} \cite{Cacciari:2011ma} is used to perform jet clustering, and the anti-$k_T$ algorithm is chosen unless specified otherwise.
\subsection{Leading singular spectrum of jet mass}\label{sec:leadsingcheck}
\begin{figure}[h]
\begin{center}
\includegraphics[width=0.45\textwidth]{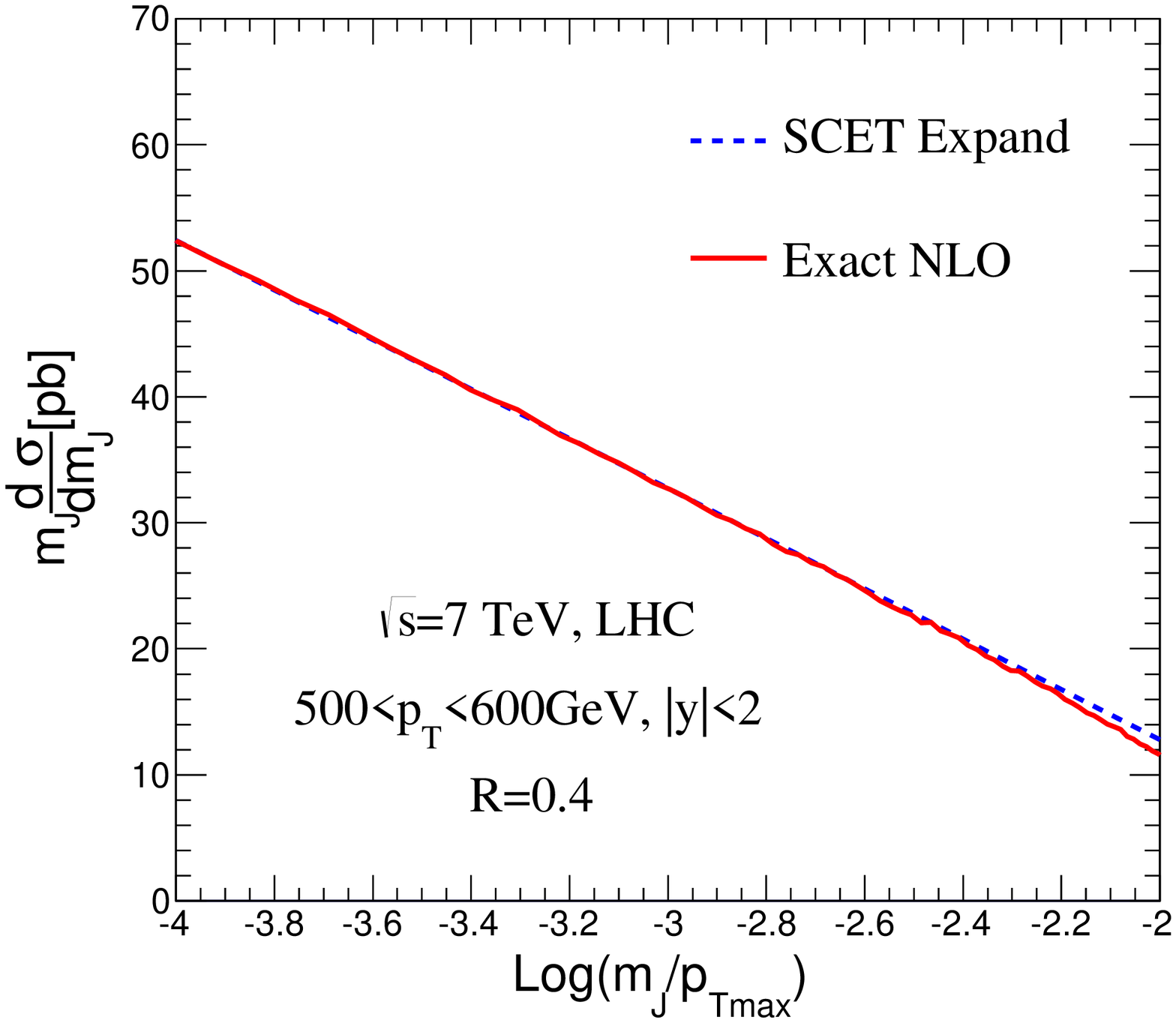}\quad
\includegraphics[width=0.45\textwidth]{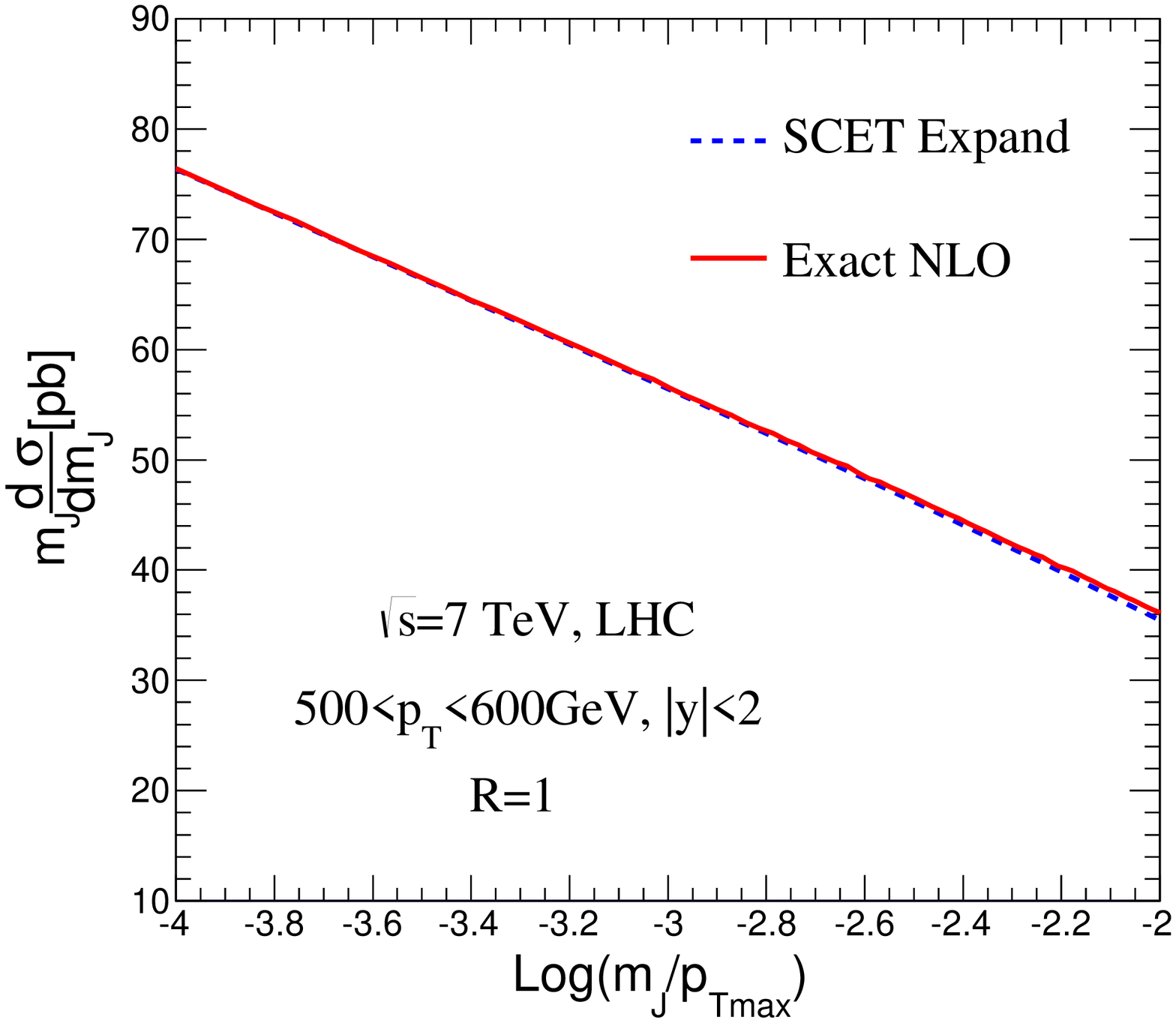}
\end{center}
\vspace{-0.5cm}
\caption{\label{fig:FOexpand}
The jet mass distributions from the exact NLO calculation and the resummed results expanded to leading order (SCET Expand). Here, $p_{ T \rm max}=$ 600 GeV.}
\end{figure}
To verify the correctness of the factorizatoin formula, we expand the eq.~(\ref{eq:kernelCij}) to the leading singular terms (blue dashed line), and compare with the exact NLO results (red solid line), which are obtained from ref.~\cite{Ellis:1985er}. From figure~\ref{fig:FOexpand}, we can see that the leading singular terms of the jet mass distribution can reproduce the exact NLO jet mass spectrum in small $m_J$ region. As $m_J$ increases, the difference between the leading singular terms and the exact NLO results increase.
We find that in both cases of $R=0.4$ and 1, the expanded results agree with the fixed-order ones.
This means that our soft function is applicable for not only small $R$.

\subsection{Scale choices and uncertainties}
\begin{figure}
\begin{center}
\subfigure[${\bm H}$]{
\label{fig:muhch}
\includegraphics[width=0.30\textwidth]{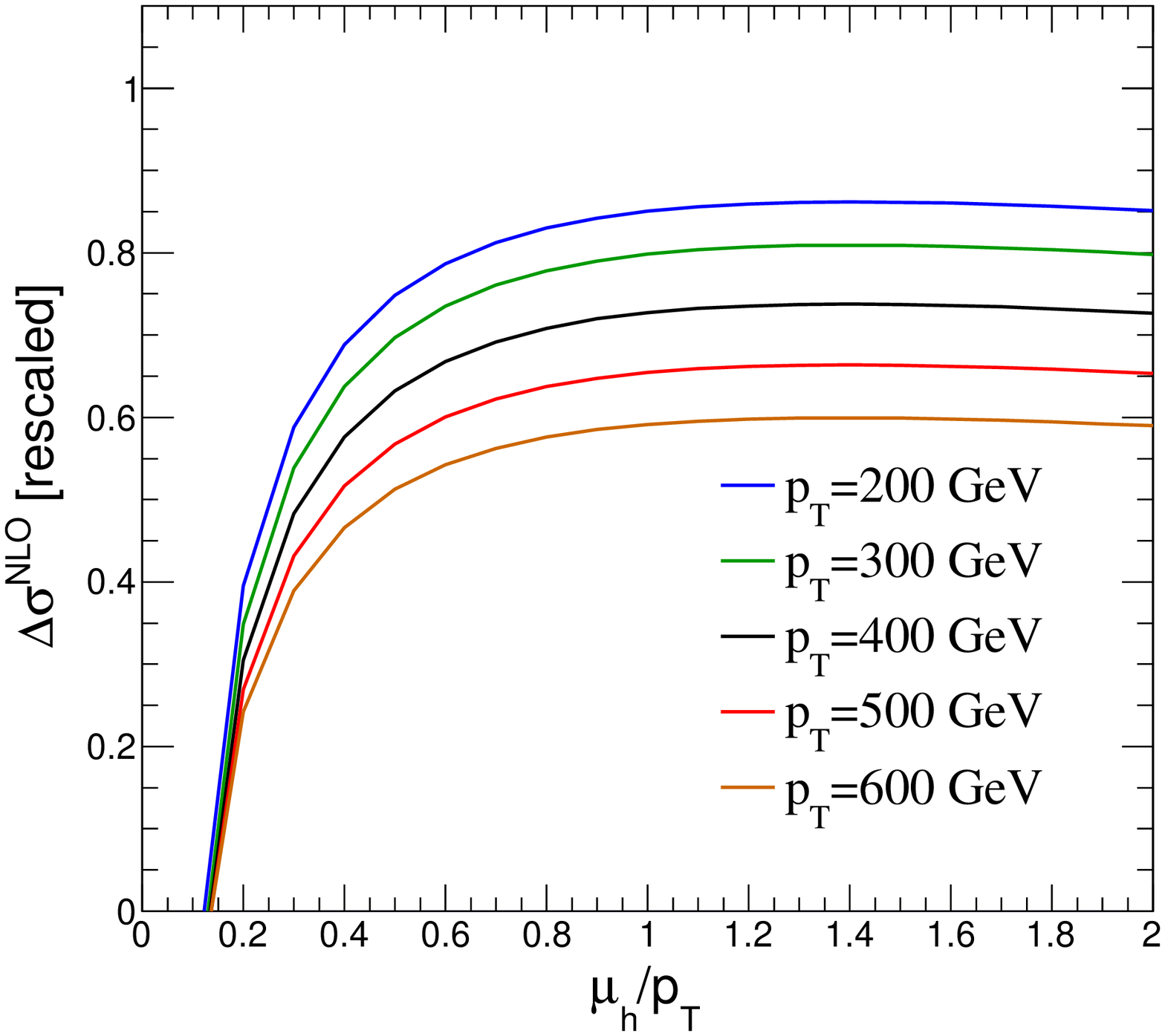}
}
\subfigure[${\bm S}^{\rm out}$ and $J^{\rm rec.}$]{
\label{fig:muoutj2ch}
\includegraphics[width=0.30\textwidth]{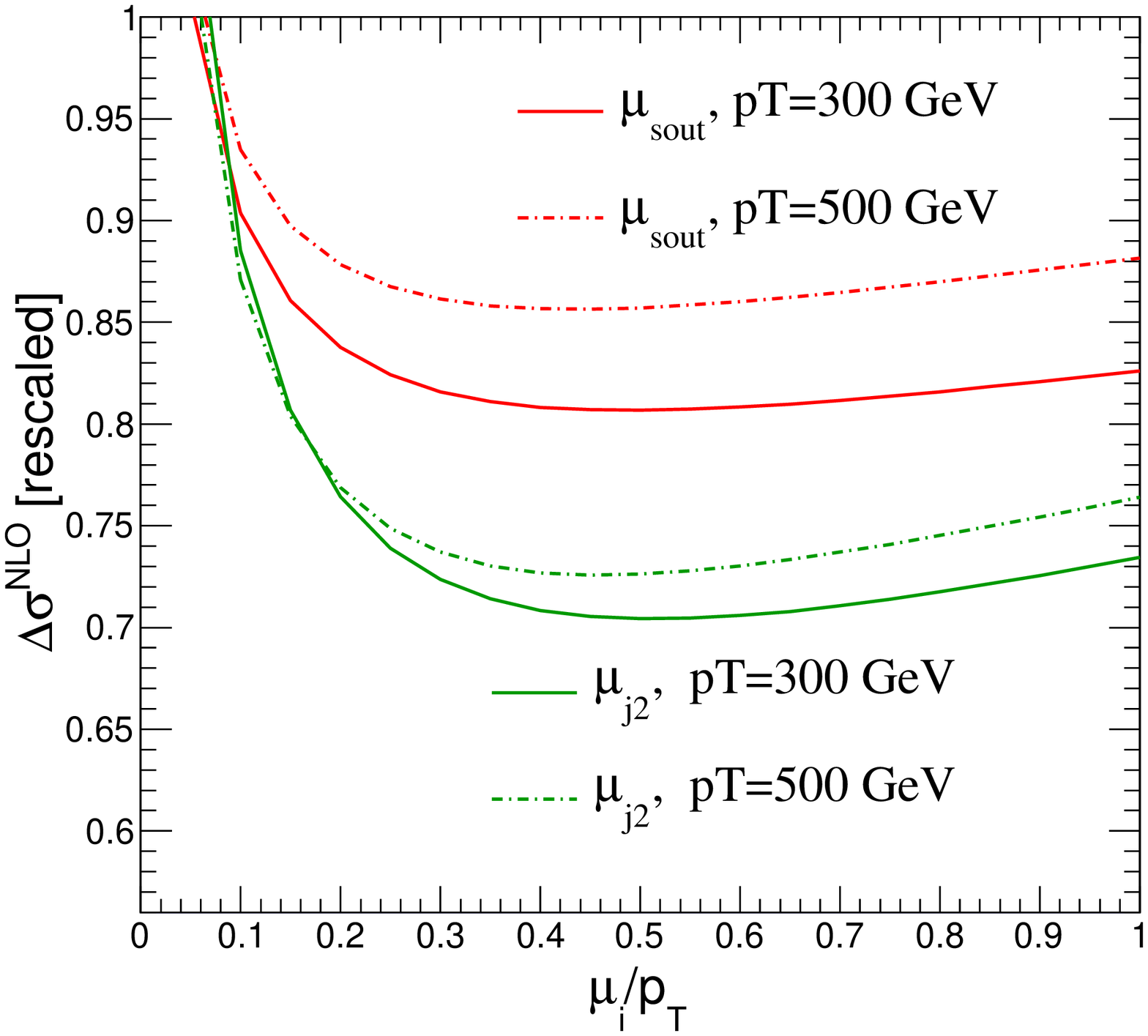}
}
\subfigure[${\bm S}^{\rm in}$ and $J^{\rm obs.}$]{
\label{fig:muinj1ch}
\includegraphics[width=0.30\textwidth]{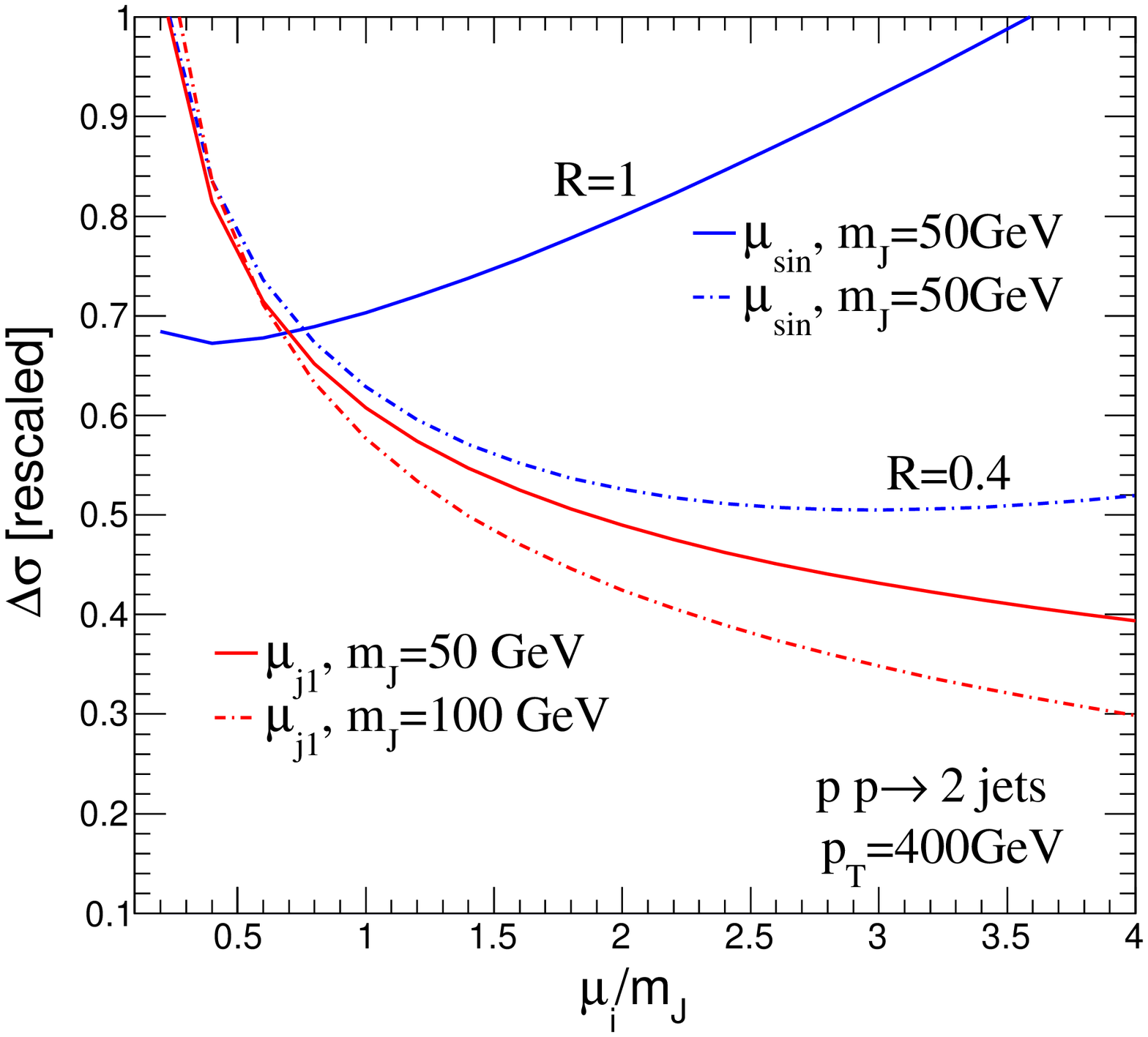}
}
\end{center}
\vspace{-0.5cm}
\caption{\label{fig:much}
The rescaled contribution from hard, soft and jet functions, as a function of their corresponding scales.}
\end{figure}

The factorization scales are set at $p_T$ unless specified otherwise. Besides, there are five other matching scales, $\mu_{h}$, $\muin$, $\mu_{j_1}$, $\muout$ and $\mu_{j_2}$, which need to be chosen properly so that the corresponding hard, soft and jet functions have stable perturbative expansions. The matching scales can be determined by examining the contribution of the NLO matching coefficients as a function of their corresponding scales \cite{Becher:2009th,Becher:2007ty,Becher:2011fc,Becher:2012xr}. As shown in Figs.~\ref{fig:muhch} and \ref{fig:muoutj2ch}, the values of the scale $\mu_{h}$, $\muout$ and $\mu_{j_2}$ are chosen as
\begin{gather}\label{eq:muchoice}
\mu_h=1.4\,p_T\,,\quad \muout=0.2\,p_T+80\, {\rm GeV}\,,\quad \mu_{j_2}=0.5\,p_T\,,
\end{gather}
where the relevant one-loop contributions get the extreme values.

\begin{figure}
\begin{center}
\subfigure[$\mu_f$]{
\label{fig:mufdep}
\includegraphics[width=0.46\textwidth]{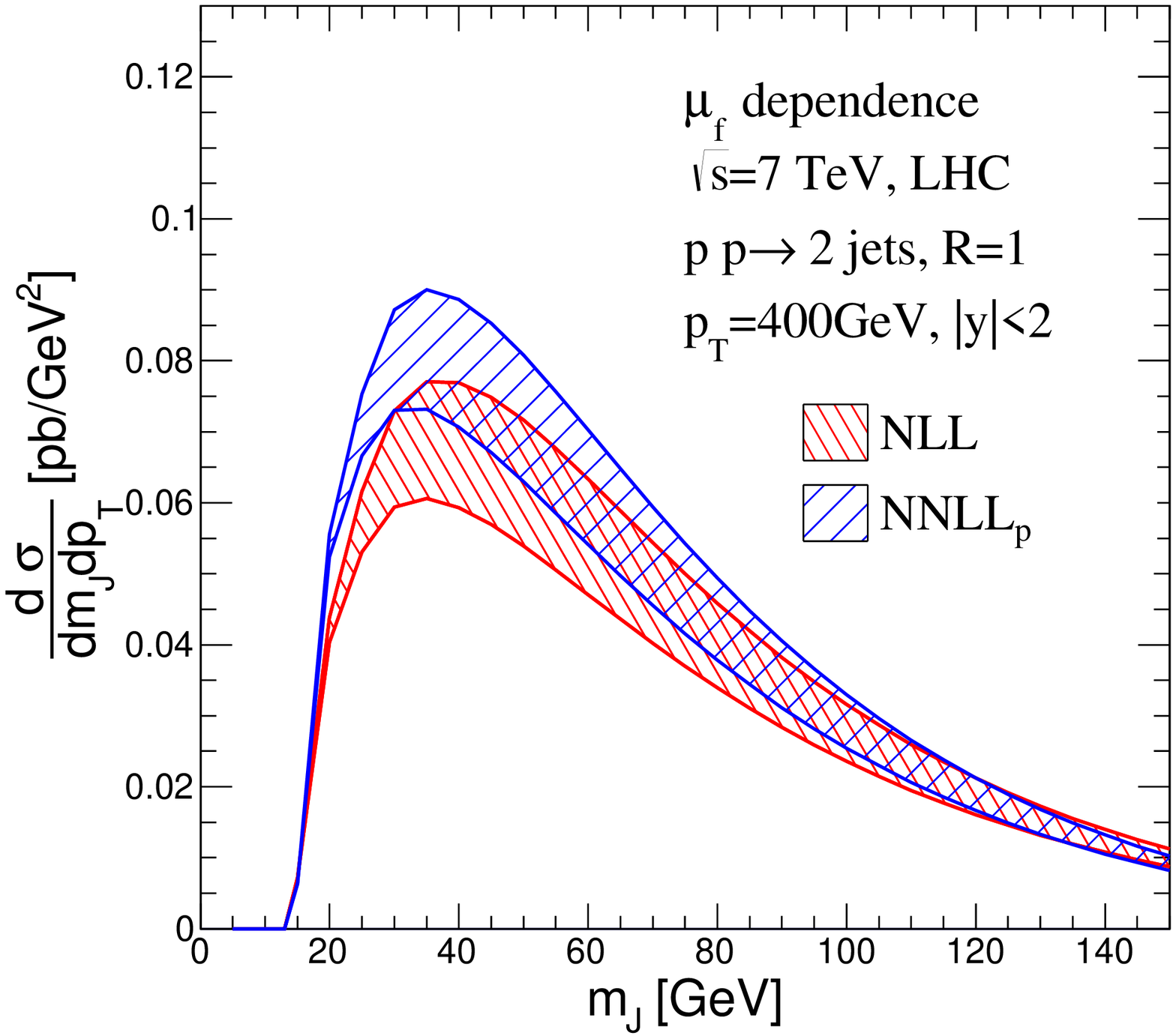}}\quad
\subfigure[$\mu_h$]{
\label{fig:muhdep}
\includegraphics[width=0.46\textwidth]{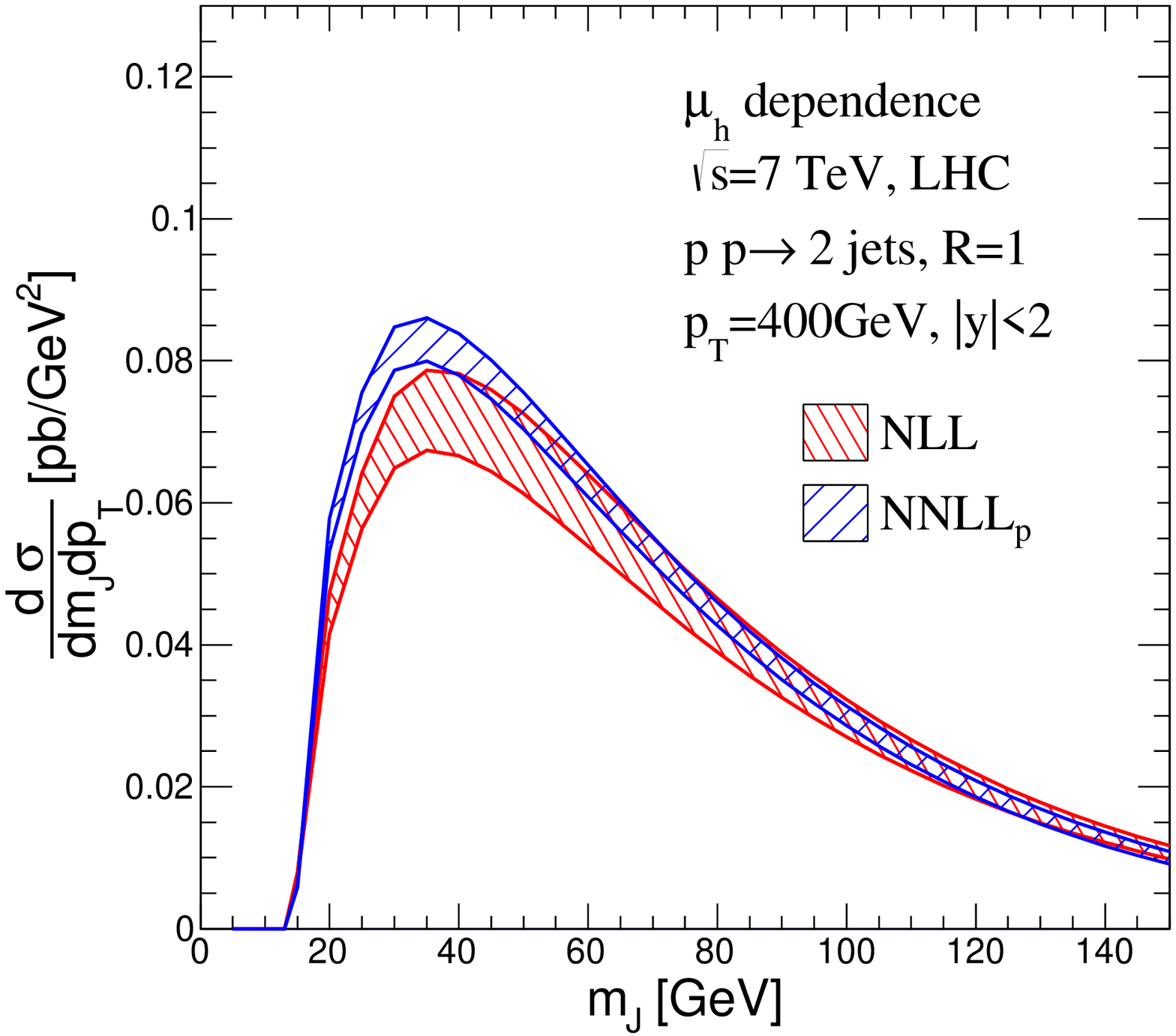}}\\
\subfigure[$\mu_{j_2}$]{
\label{fig:muj2dep}
\includegraphics[width=0.46\textwidth]{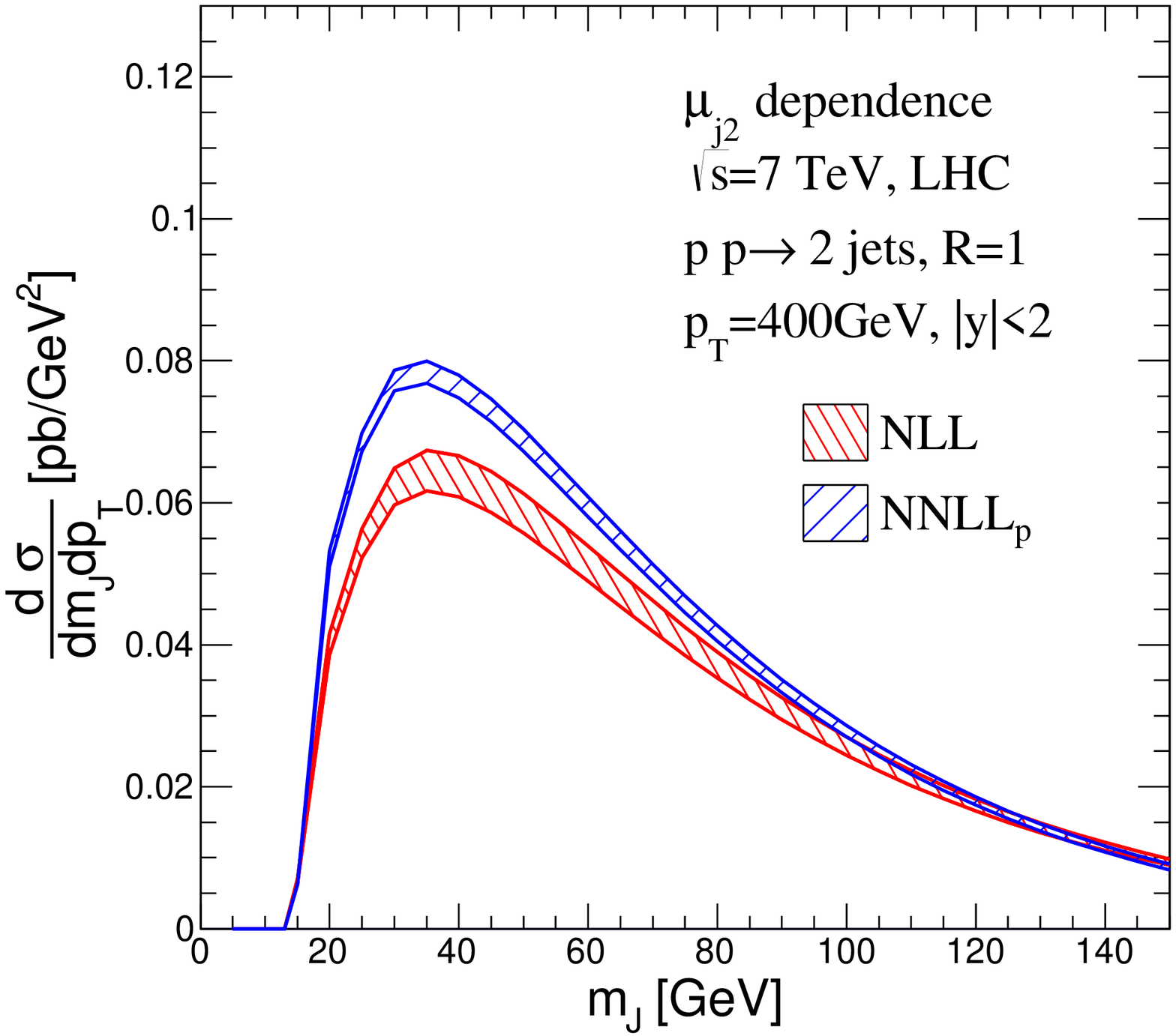}}\quad
\subfigure[$\muout$]{
\label{fig:muoutdep}
\includegraphics[width=0.46\textwidth]{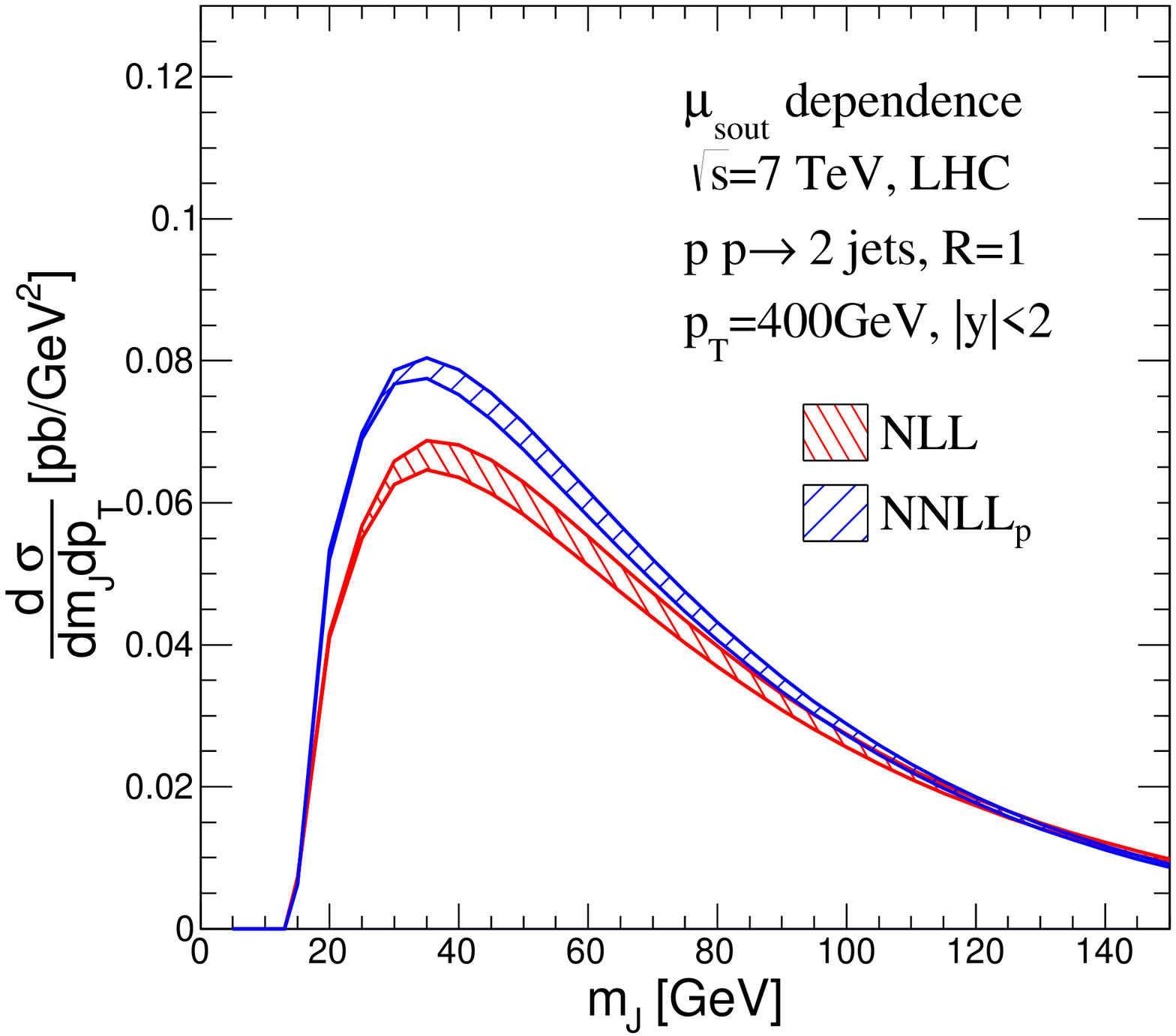}}\\
\subfigure[$\mu_{j_1}$]{
\label{fig:muj1dep}
\includegraphics[width=0.46\textwidth]{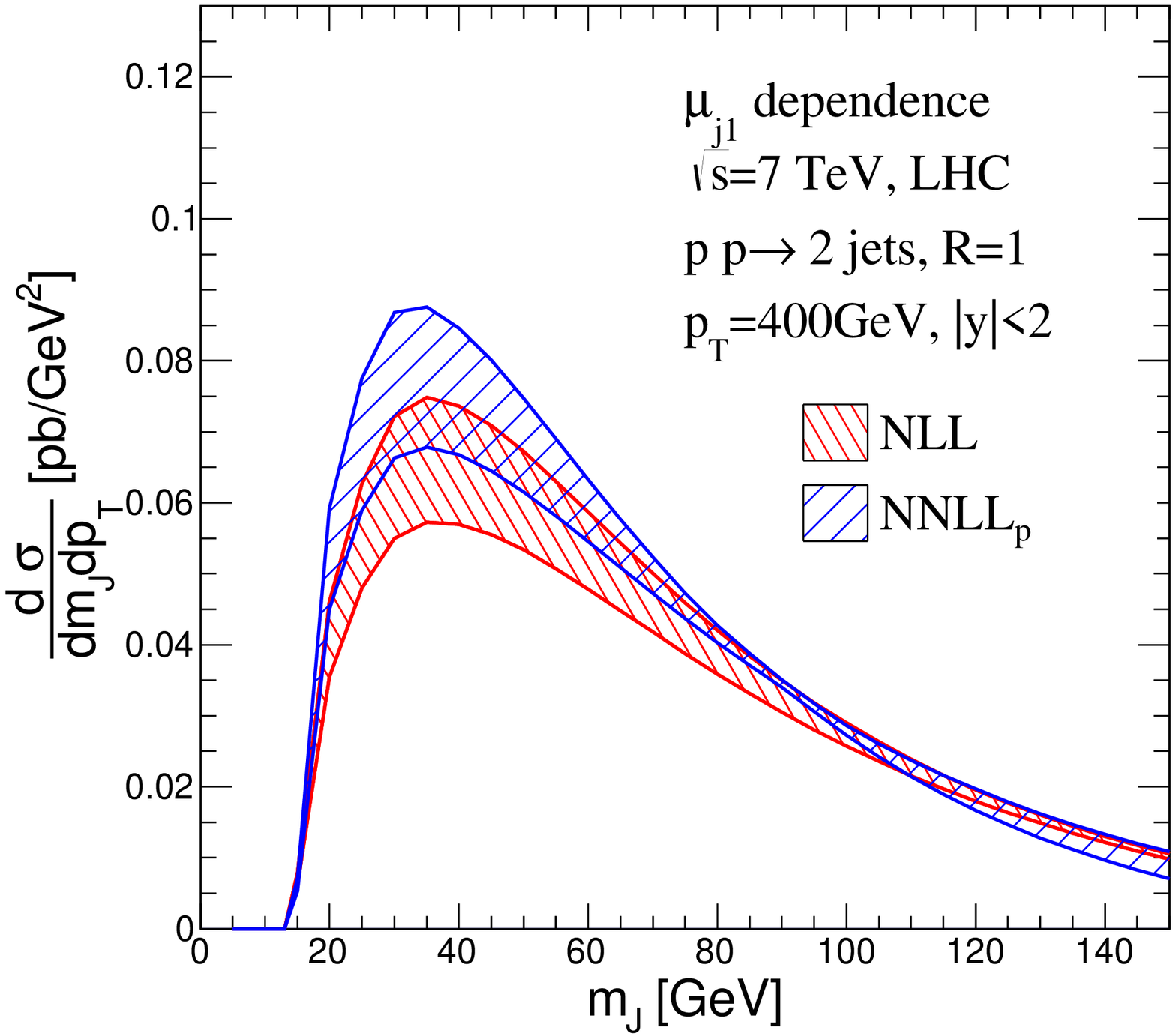}}\quad
\subfigure[$\muin$]{
\label{fig:muindep}
\includegraphics[width=0.46\textwidth]{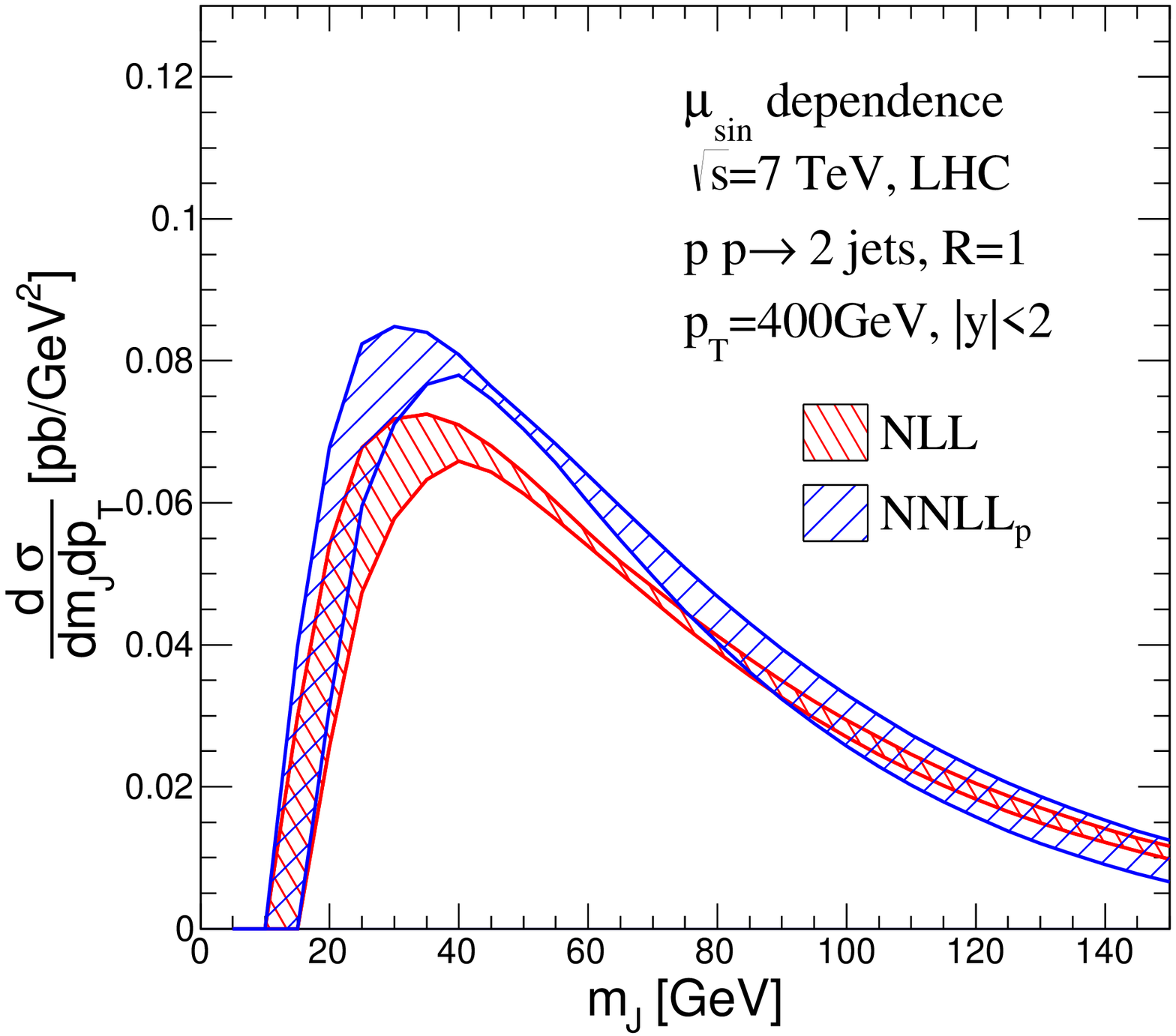}}
\end{center}
\vspace{-0.5cm}
\caption{\label{fig:mudep}
The scale uncertainties of the resummation results for $p_T=400$GeV and $R=1$.}
\end{figure}

However, the extreme points of the one-loop contributions of the observed jet function $J_1(\mu_{j_1})$ and soft function ${\bm S}_{\rm in}(\muin)$ do not exist. It can be seen from their NLO corrections
\begin{gather}
\Delta\sigma^{\rm NLO} \sim \frac{1}{m_J^2}\left(A\ln\frac{m_J^2}{\mu_{J_1}^2}+B\ln\frac{m_J^2}{p_T\muin}+ C\right)\,,
\end{gather}
where $A$, $B$, $C$ are scale independent coefficients. If we measure the jet mass $m_J$, it should not be integrated so that there is no quadratic logarithm term of $\muin$ and $\mu_{j_1}$ in the one-loop corrections, which is different from the cases of ${\bm S}_{\rm out}(\muout)$ and $J_2(\mu_{j_2})$. As shown from the red line in figure~\ref{fig:muinj1ch}, the NLO corrections to $J_1$ always decrease with increasing $\mu_{j_1}$. For $\mu_{j_1}=3m_J$, we can see that the corrections decrease slowly. The blue lines in figure~\ref{fig:muinj1ch} show the variations of resummed results as a function of $\muin$ for different jet radius $R$. The extreme points emerge because higher order contribution of ${\bm S}^{\rm in}$ are included. The variations of $\muin$ are minimized at about $20$~GeV and $120$~GeV for $R=1$ and $0.4$, respectively. Using the method in ref.~\cite{Chien:2012ur}, we can determine $\muin$ numerically by the power function of $m_J$
\begin{equation}
\muin = \frac{{\mu_*}^2}{c_R}\frac{p_T^*}{p_T}\,,
\end{equation}
where $c_R$ is an $R$-dependent parameter, $p_T^*=400{\rm GeV}$ and $\mu_*=1.67 m_J^{1.47}$ ($m_J$ in GeV) ~\cite{Chien:2012ur}. According to the extreme points of the variations of ${\bm S}^{\rm in}$, $c_R$ is numerically determined as $14000$ and $2400$ for $R=1$ and $R=0.4$, respectively.

After all of the natural scales involved in this process have been chosen, we discuss the scale dependence of the resummation results of jet mass spectra in figure~\ref{fig:mudep}. At ${\rm NNLL}_{p}$ level, three loop cusp anomalous dimension and two loop normal anomalous dimension are used. For the $R$-dependent pieces, the one-loop soft anomalous dimensions are used. At ${\rm NLL}$ level, two loop cusp anomalous dimension and one loop normal anomalous dimension are used. Figure~\ref{fig:mudep} shows the scale uncertainties for variation of each scales by a factor of 2 about its default value. It can be seen that the scale uncertainties for $\mu_h$, $\mu_{j_1}$, $\mu_{j_2}$ and $\muout$ reduce significantly from NLL to ${\rm NNLL}_{p}$. But for scale $\muin$, the ${\rm NNLL}_{p}$ bands are broader than the NLL ones at large $m_J$ region. The reason may be that in large $m_J$ region non-singular terms become important and the resummation results are unreliable. In addition, we can also see that the distribution is enhanced by about 23\% from NLL to ${\rm NNLL}_{p}$ at the peak region.
We confirm numerically that this enhancement mainly comes from the one-loop corrections of the hard function, which are included at ${\rm NNLL}_{p}$ order, but not at NLL. This means that if we want to obtain accurate theoretical predictions, the high order corrections of the hard function must be included.

\subsection{$R$ dependence}
\begin{figure}[t]
\begin{center}
\subfigure[$R$ dependence]{
	\label{fig:Rdep}\includegraphics[width=0.46\textwidth]{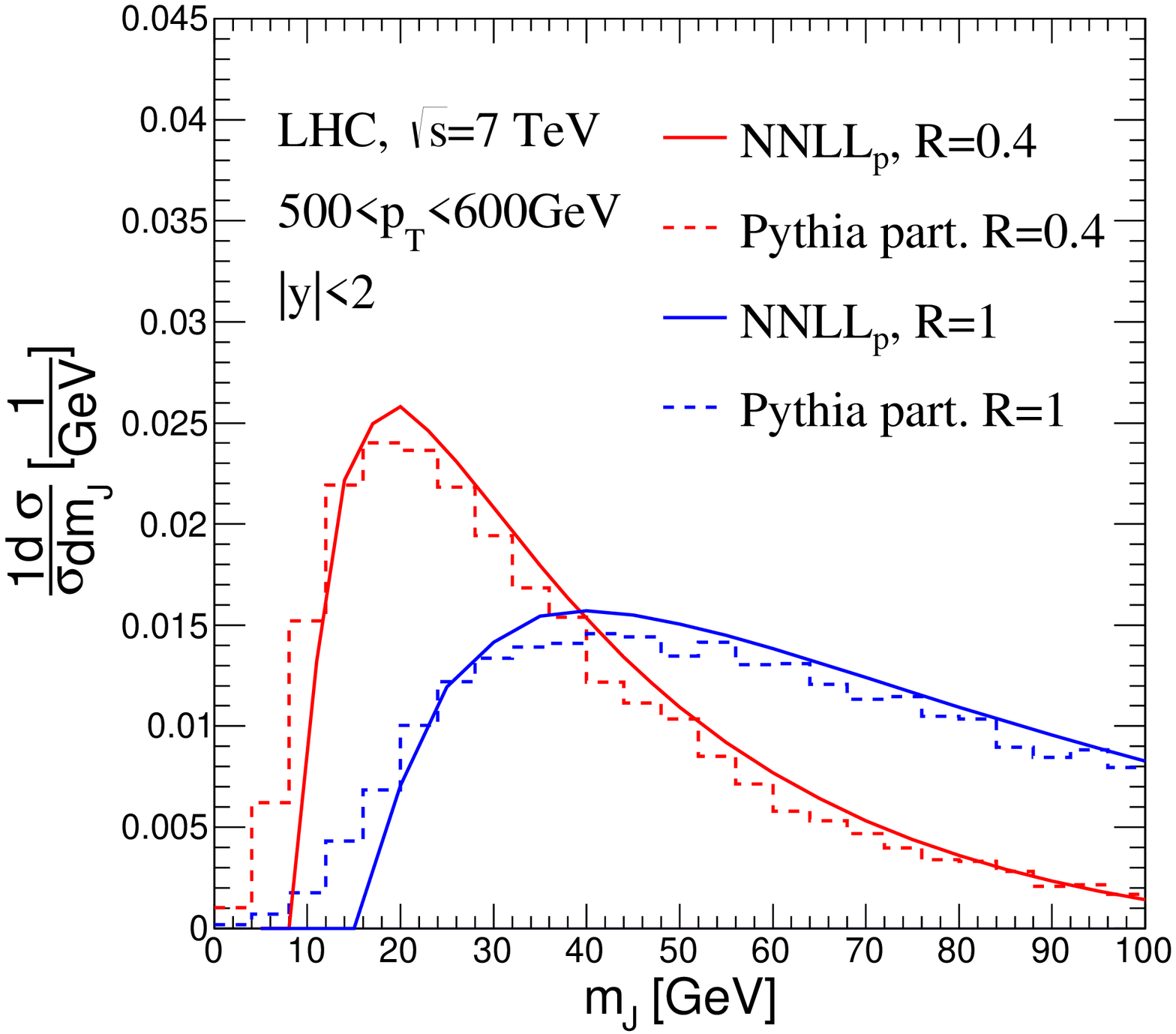}
	}\quad
\subfigure[Jet mass spectra of quark jet and gluon jet]{
	\label{fig:qvsg}\includegraphics[width=0.46\textwidth]{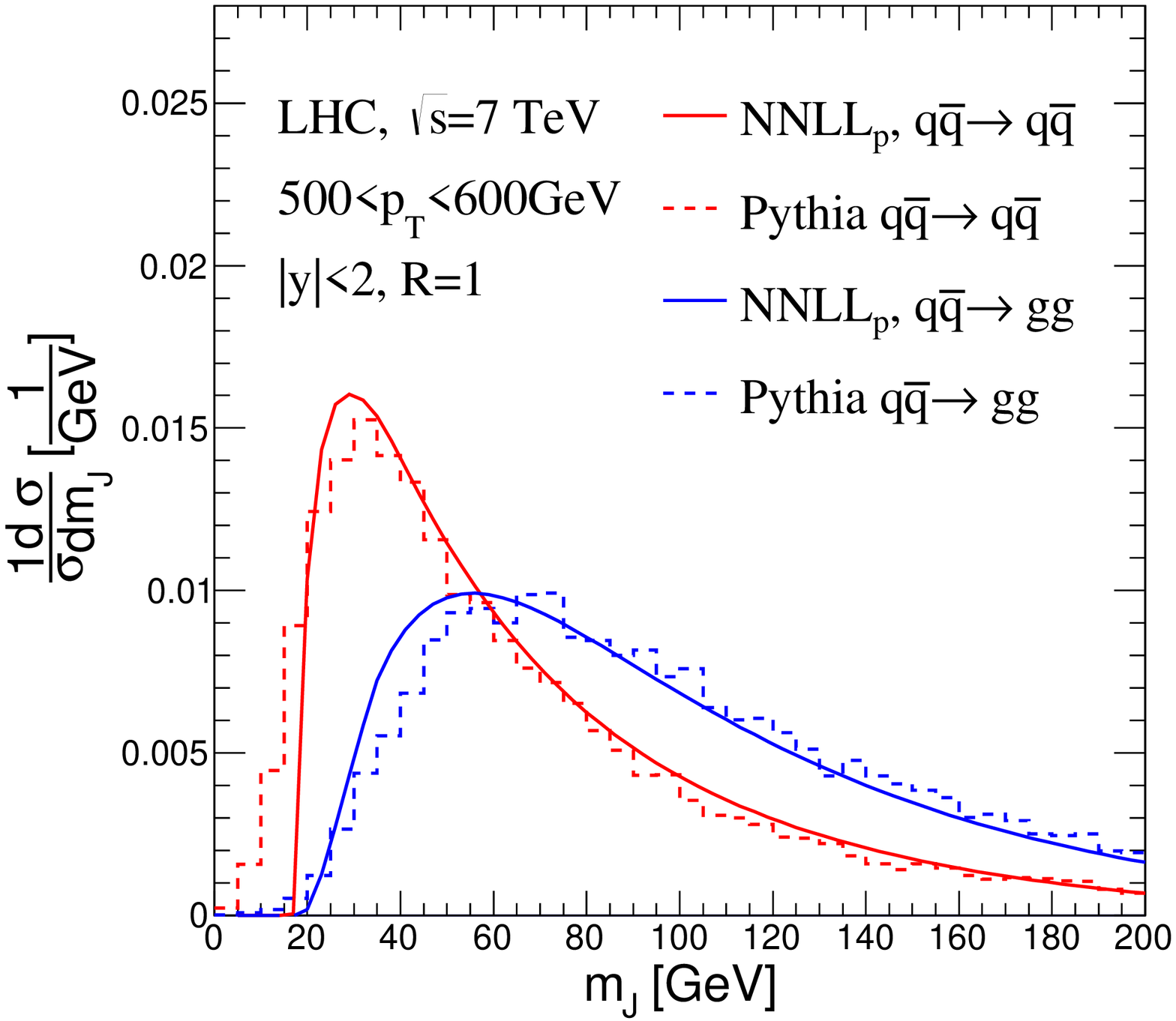}
	}
\end{center}
\vspace{-0.5cm}
\caption{\label{fig:Rdepandqvsg}(a) $R$ dependence of the jet mass distribution. (b) Comparison of jet mass spectrum between quark jet and gluon jet.}
\end{figure}
In figure~\ref{fig:Rdep}, the blue and red solid lines show the results of ${\rm NNLL}_p$ resummation for $R=1$ and 0.4 , respectively. We can see that the jet mass spectra shift to right with increasing $R$, and peak at about 20 GeV for $R=0.4$ and 40~GeV for $R=1$, respectively. This is due to the fact that when $R$ increases, more large angle soft radiation can be combined into the jet, so that the invariant mass of jet $m_J=\sqrt{(p_c+k_s)^2}$ become larger. The results from \texttt{PYTHIA} are shown as dashed histograms. Figure~\ref{fig:Rdep} shows that the peak positions and shapes of our resummation results agree with the ones of \texttt{PYTHIA} at parton level.

\subsection{The difference of jet mass spectra between quark and gluon}
In order to study the difference between quark jet and gluon jet, we show the jet mass distributions for processes with quark and gluon final state separately.
In figure~\ref{fig:qvsg}, the blue and red solid lines correspond to $q{\bar q}\to q{\bar q}$ and $q{\bar q}\to gg$, respectively. The jet mass spectra for quark and gluon jet peak at about $30$ GeV and $55$ GeV , respectively, which is helpful to distinguish between the quark and gluon jet. The peak positions and shapes of our resummation results agree with the ones of \texttt{PYTHIA}.

\subsection{Phenomenological studies of jet mass spectrum at the LHC}\label{subsec:phdis}
\begin{figure}[t]
\begin{center}
\includegraphics[width=0.46\textwidth]{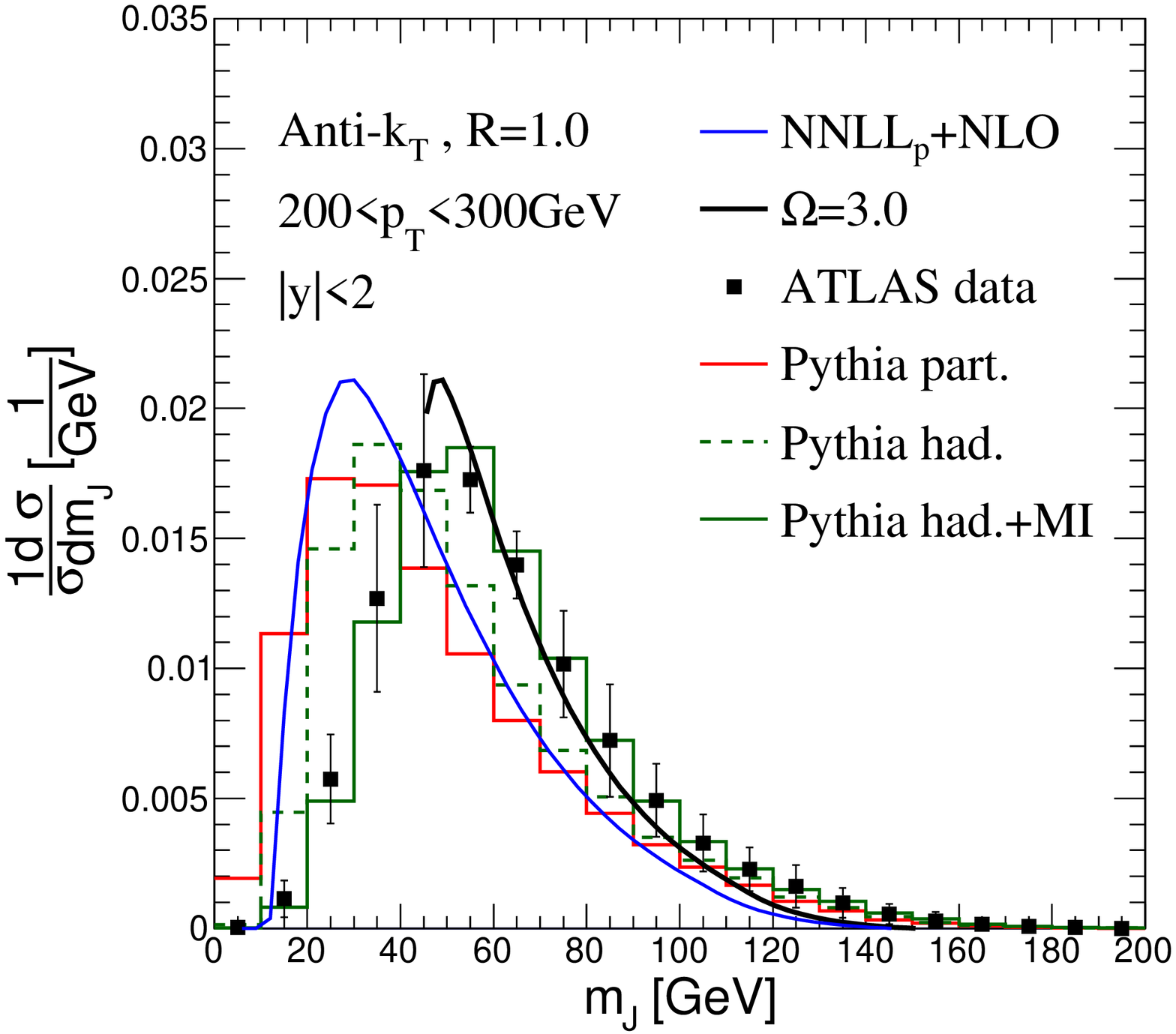}\quad
\includegraphics[width=0.46\textwidth]{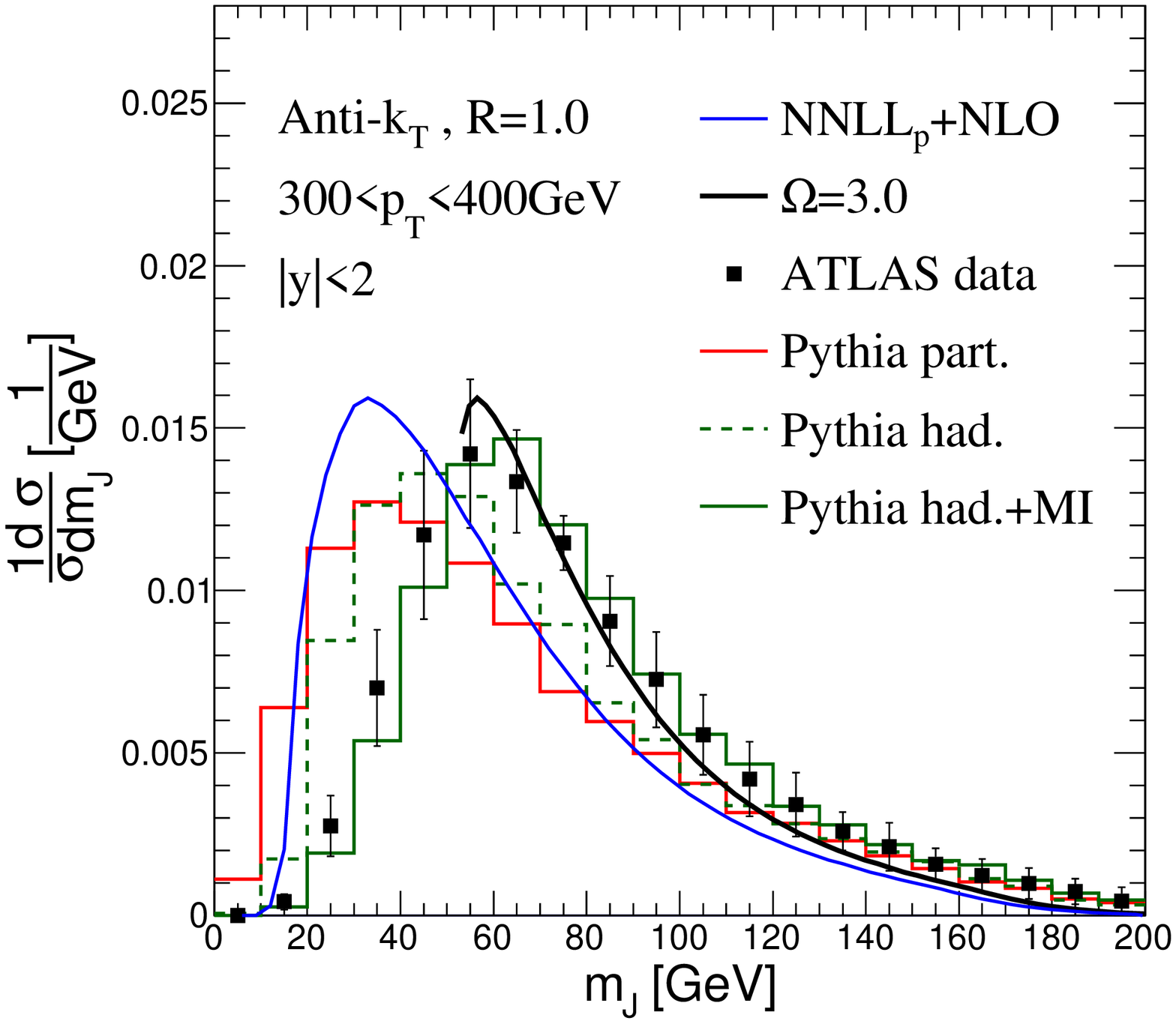}\\
\includegraphics[width=0.46\textwidth]{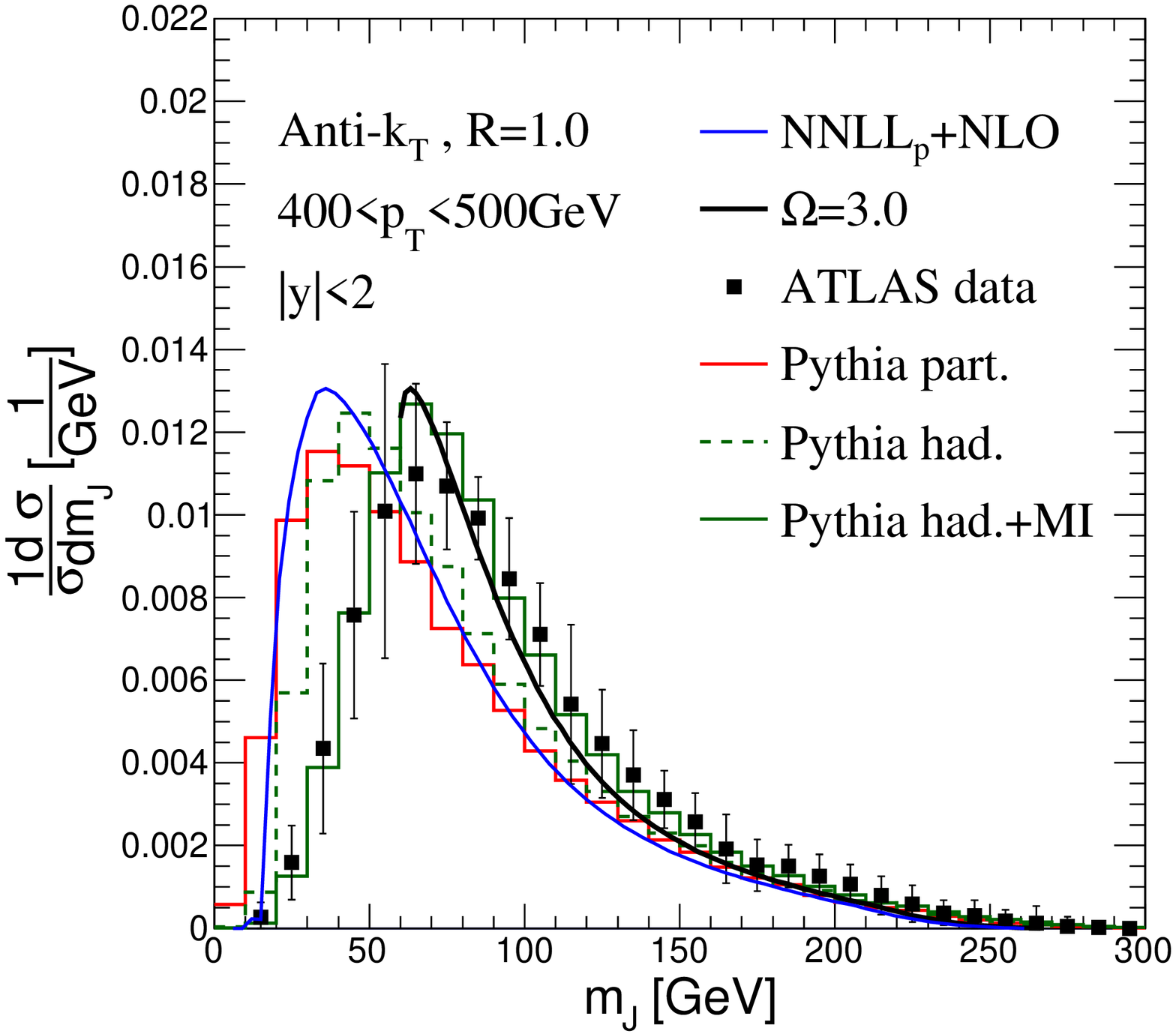}\quad
\includegraphics[width=0.46\textwidth]{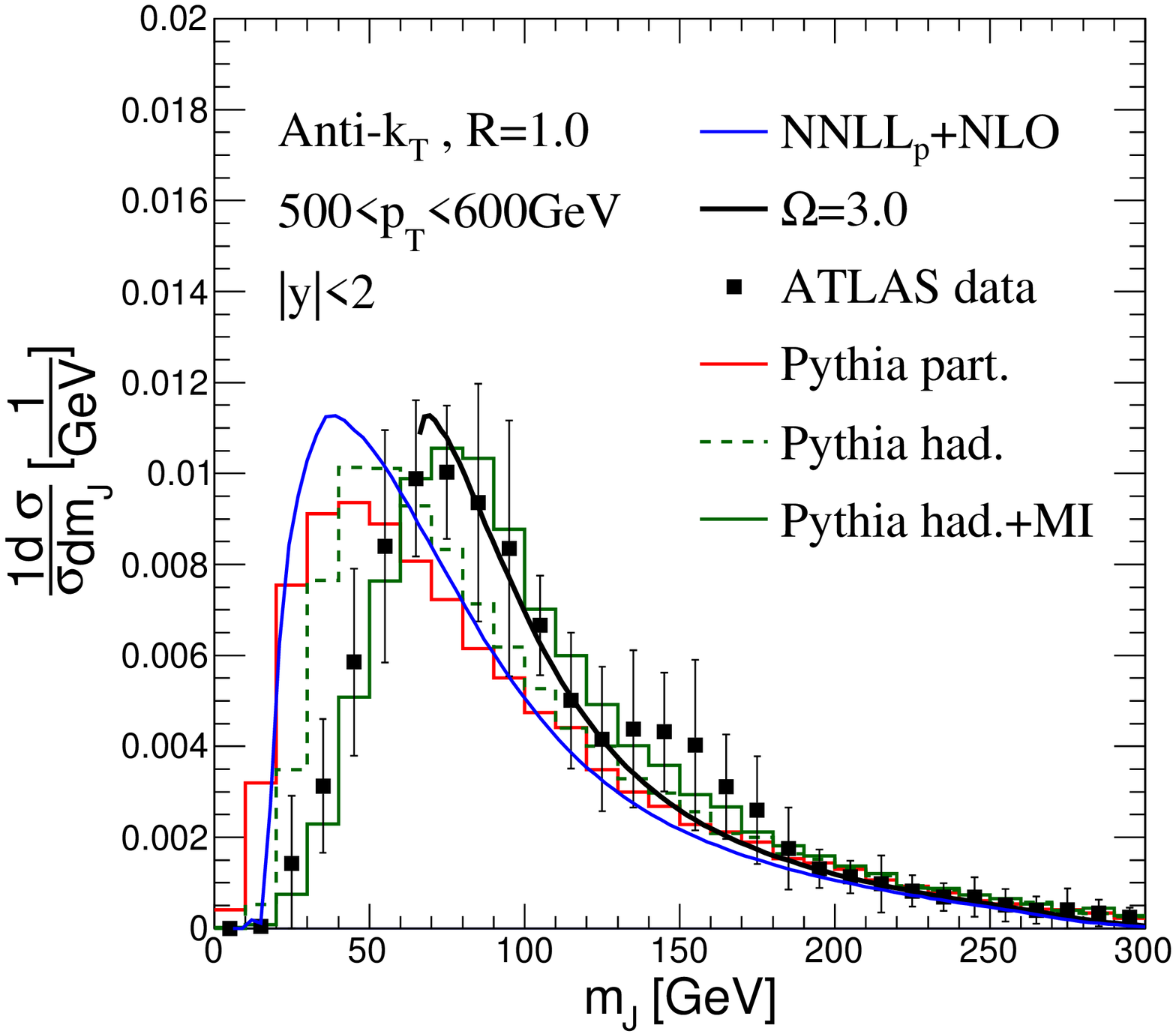}
\end{center}
\vspace{-0.5cm}
\caption{\label{fig:comdata}
Comparison between theoretical predictions and ATLAS data at the 7 TeV LHC. The label MI in the plots denotes the multiparton interactions. The blue lines represent our ${\rm NNLL}_p+{\rm NLO}$ predictions, and the black solid lines represent the results with non-pertubative effects. The red solid, green dashed and green solid histograms represent the results from \texttt{PYTHIA}.
}
\end{figure}

In this section, we give the RG improved predictions of jet mass spectra at the LHC, and compare them with the results of \texttt{PYTHIA} and the ATLAS data~\cite{ATLAS:2012am}. Figure~\ref{fig:comdata} shows the normalized jet mass distributions with $R=1$ in four different $p_T$ bins. At ${\rm NNLL}_p+{\rm NLO}$ level, the jet mass spectra  peak around 25-40 GeV, and shift to right with increasing jet $p_T$. The peak positions agree with the ones of \texttt{PYTHIA} at parton level. In addition, we can see from the results of \texttt{PYTHIA} that the additional hadronization and multiparton interaction shift the spectra to right by about 10 GeV and 20 GeV, respectively.
This means that if we want to obtain predictions which are comparable to data\footnote{Here we have included the multiparton interaction to the non-perturbative effects for simplicity thought it is not necessarily true.}, the non-perturbative effects must be considered.
Ref.~\cite{Dasgupta:2007wa} has computed the non-perturbative corrections to jet mass and their results have been used for $Z$+1 jet process in ref.~\cite{Dasgupta:2012hg},
where a shift $m_J^2\to m_J^2+2\Omega R\,p_T$ for jet mass has been used to account for the non-perturbative effects.
However, as discussed in ref.~\cite{Dasgupta:2012hg}, this shift in small jet mass region is not meaningful,
so we truncate the spectrum in the left side of the peak.
Figure~\ref{fig:comdata} also shows that the ${\rm NNLL}_p+{\rm NLO}$ results with a shift of $\Omega=3.0{\rm ~GeV}$ (the black solid lines)
are consistent with the ATLAS data~\cite{ATLAS:2012am} in all of four $p_T$ bins. Here the shift accounts for the total effects of hadronization and multiparton interaction, so the value of $\Omega$ in this work is larger than the one in ref.~\cite{Dasgupta:2012hg}, where only hadronization is concerned.
Notice that our treatment of the non-perturbative interaction effects here is just an approximation.
A precise estimate of these effects require some modification of the resummation scheme and global fitting with certain precise data.
The further discussion of the non-perturbative effects is beyond the scope of this work, and left in future study.

\begin{figure}[h]
\begin{center}
\includegraphics[width=0.46\textwidth]{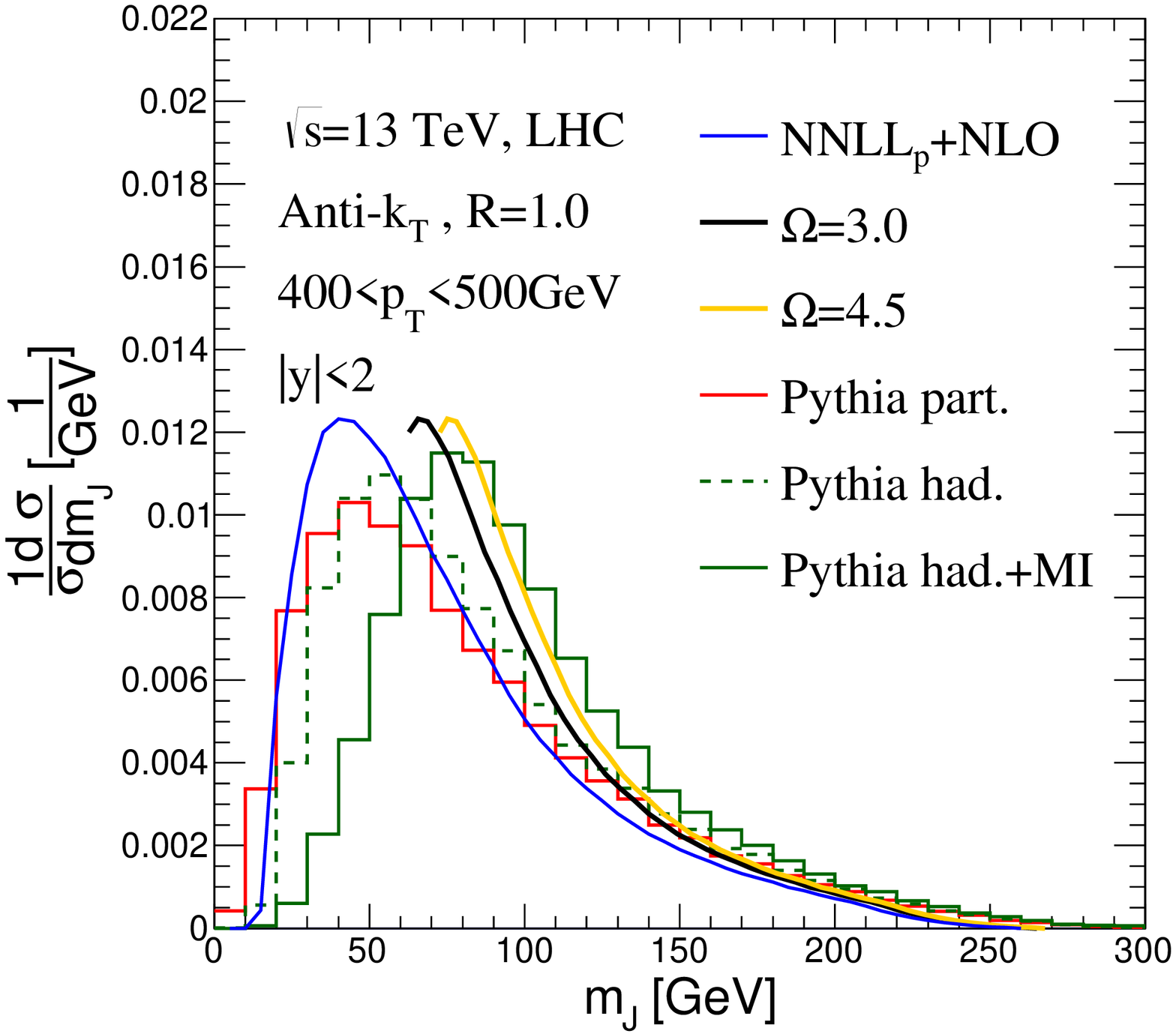}\quad
\includegraphics[width=0.46\textwidth]{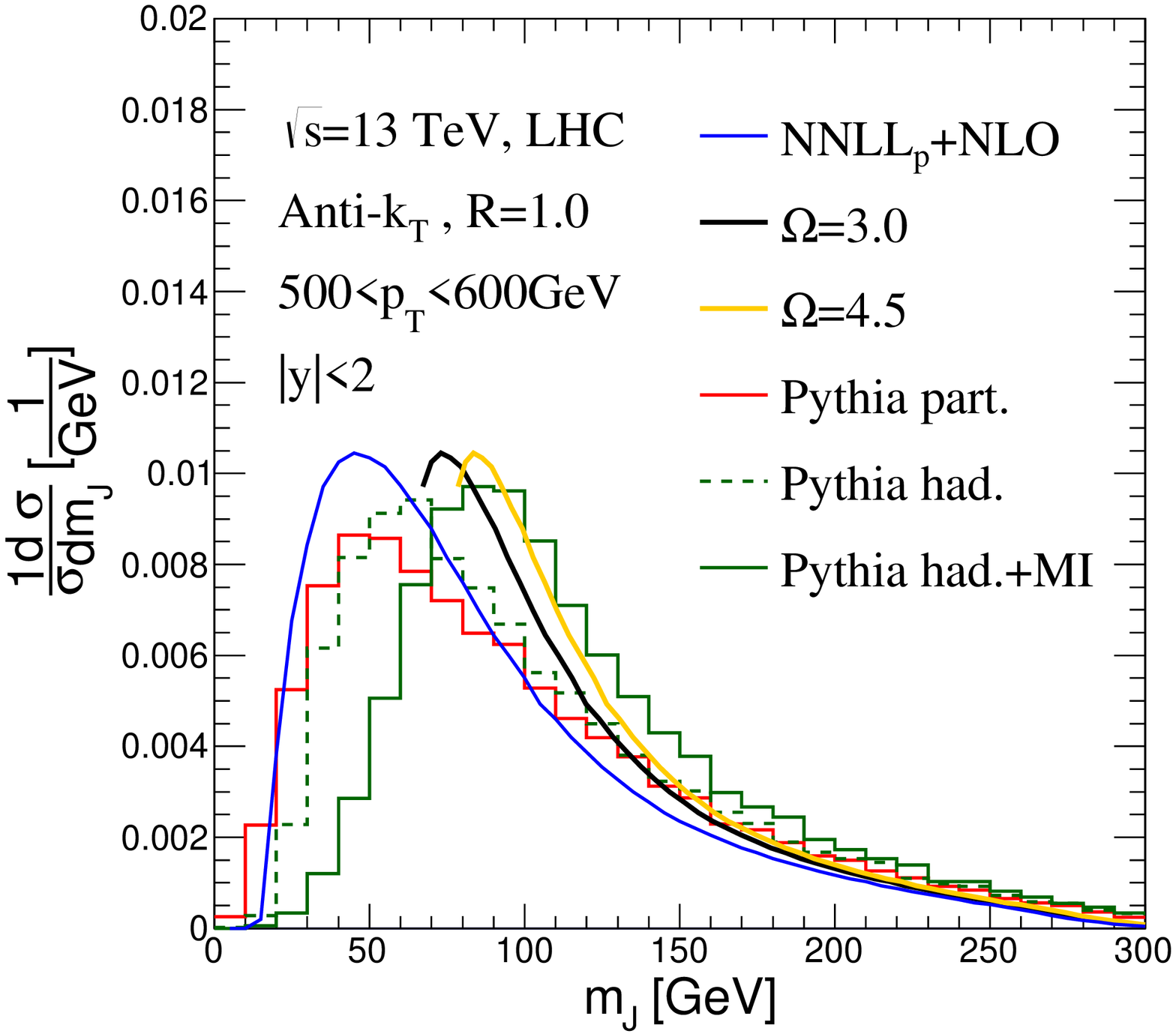}
\end{center}
\vspace{-0.5cm}
\caption{\label{fig:LHC13TeV}
Predictions of jet mass spectrum at 13 TeV run of the LHC.}
\end{figure}
In figure~\ref{fig:LHC13TeV}, we give our RG improved predictions at the 13 TeV LHC. Comparing with the results at the 7 TeV LHC, the jet mass spectra at parton level in the same kinematic region shift to right by about 5 GeV. The reason is that the dominated contributions is from $qg\to qg$ and $gg\to gg$ channel for 7 TeV and 13 TeV LHC, respectively, and the latter channel gives more gluon final states, the average jet mass of which is larger than the one of quark final states.
After including the non-perturbative effects (hadronization and multiparton interaction), the \texttt{PYTHIA} results are closer to the SCET predictions with $\Omega=4.5\,{\rm GeV}$ than  $\Omega=3\,{\rm GeV}$, which implies that the  non-perturbative effects become more significant at hadron colliders with higher CM energy.

Moreover, we can see this more clearly with the mean values of the jet mass squared, defined as
\begin{equation}
    \langle M^2\rangle\equiv \int m_J^2 \frac{1}{\sigma}\frac{d\sigma}{dm_J} dm_J\,,
\end{equation}
which can be changed by non-perturbative  effects in collisions.
In table~\ref{tab:disOmega}, we list the mean jet mass squared at parton level, including hadronization, and both hadronization and underlying event (described by multiparton interactions in \texttt{PYTHIA}), which are denoted by $\langle M_{\rm part.}^2\rangle$, $\langle M_{\rm had.}^2 \rangle$ and $\langle M_{\rm had.+MI}^2\rangle$, respectively. We can see that $\delta M_{\rm had.}^2$ and $\delta M_{\rm UE}^2$ increase by about $12\%$ and $26\%$, respectively, with CM energy from 7 TeV to 13 TeV. Besides, $\delta M_{\rm had.}^2$ in gluon final states is much larger than in quark final states (because of the color factor difference between quark final state and gluon final state~\cite{Dasgupta:2007wa}), and nearly insensitive to the CM energy. Because the $g\,g\to g\,g$ channel is more dominant at higher CM energy collision, $\delta M_{\rm had.}^2$ increases apparently in the $p\,p\to {\rm dijet}$ production. In contrast, $\delta M_{\rm UE}^2$ is almost the same for the quark and gluon final states and sensitive to the CM energy, which increases by about $40\%-50\%$ from 7\,TeV to\,13 TeV with \texttt{PYTHIA}. This is just the improvement from $\Omega=3\,{\rm GeV}$ to $\Omega=4.5\,{\rm GeV}$ in our resummation predictions, as shown in figure~\ref{fig:LHC13TeV}.


\begin{table}
\begin{center}
\begin{tabular}{c|c|c|c|c|c|c}
\hline
\hline
   & $\sqrt{s} $&  ~~$\langle M_{\rm part.}^2\rangle $ ~~  &
~~$\langle M_{\rm had.}^2 \rangle$ ~~& $\langle M_{\rm had.+MI}^2\rangle$ &
~$\delta M_{\rm had.}^2$ ~ & ~~$\delta M_{\rm UE}^2$~~
\\
\hline
$p\,p\to {\rm dijet}$ & 7 TeV  & 7893 & 8689 & 10460 &  796  &  1771 \\
                 & 13 TeV & 9295 & 10190 & 12420 &  895  & 2230 \\
\hline
$q\,{\bar q}\to q'\,{\bar q}'$ & 7 TeV  & 4777 & 5295 & 6989 &  518  & 1694 \\
                              & 13 TeV & 5183 & 5731 & 8101 &  548  & 2370 \\
\hline
$g\,g\to g\,g$ & 7 TeV  & 11370 & 12490 & 14060 & 1120 & 1570 \\
               & 13 TeV & 12020 & 13120 & 15430 & 1100 & 2310 \\
\hline
\hline
\end{tabular}
\end{center}
\caption{\label{tab:disOmega}
The comparison of mean values of jet mass squared from \texttt{PYTHIA} at the 7 TeV and 13 TeV LHC. $\delta M_{\rm had.}^2=\langle M_{\rm had.}^2\rangle - \langle M_{\rm part.}^2\rangle$ and $\delta M_{\rm UE}^2=\langle M_{\rm had.+MI}^2\rangle - \langle M_{\rm had.}^2\rangle$. The observed jets are selected with $400<p_T<500\,{\rm GeV}$ and $|y|<2$. Unit is ${\rm GeV}^2$.}
\end{table}

\section{Conclusion}\label{sec:conc}
We have studied the factorization and resummation of  jet mass for the one-jet inclusive production at the LHC with SCET. The factorization formula is derived systematically. The NLO soft function with anti-$k_T$ algorithm is calculated and its validity is demonstrated by checking the agreement between the expanded leading singular terms with the fixed order results. The soft function is refactorized into two pieces corresponding two different scales. The RG invariance of the cross section is checked at NLO for all channels, which demonstrates the correctness of the factorization. By ignoring the NGLs, we first carry out the resummation at approximate NNLL level. From the numerical results, we find that the jet mass spectrum is enhanced by about 23\% from NLL to ${\rm NNLL}_p$ at the peak region. The enhancement mainly comes from one-loop correction of the hard function. The jet mass spectra shift to right with increasing jet radius $R$ and transverse momentum $p_T$. In addition, we show that there is a significant difference in jet mass spectra between quark and gluon jets. Finally, the normalized jet mass distributions with $R=1$ are given in four different transverse momentum regions. We show that the ${\rm NNLL}_p+{\rm NLO}$ spectra peak at 25-40\,GeV and shift to right with jet $p_T$ increasing. The peak positions agree with the ones of \texttt{PYTHIA} at parton level. Including the non-perturbative effects, our results are consistent with the ATLAS data. We also give the RG improved predictions at the 13 TeV LHC and find that the peak shift to right by about 5 GeV comparing with the results at the 7 TeV LHC. Our results are helpful to precisely study jet mass spectrum at hadron colliders and test the validity of the Monte Carlo tools.

\begin{acknowledgments}
We would like to thank Hua Xing Zhu, Ding Yu Shao, Zhao Li and Hsiang-nan Li for helpful discussions. This work was supported in part by the National Natural Science Foundation of China under Grants No.~11375013 and No.~11135003.
The research of J.W. has been supported by
the Cluster of Excellence {\it Precision Physics, Fundamental Interactions and Structure of Matter} (PRISMA-EXC 1098).
\end{acknowledgments}

\appendix
\section{LO Feynman diagrams}\label{app1}
\begin{figure}[h]
\begin{center}
\subfigure[$q_i+q_j\to q_i+q_j (i\neq j)$]
{\label{fig:cha}\includegraphics[width=0.28\textwidth]{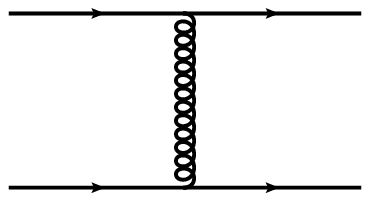}}%
\\
\subfigure[$q_i+q_i\to q_i+q_i$]
{\label{fig:chb}\includegraphics[width=0.56\textwidth]{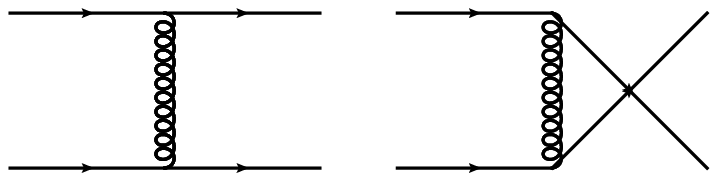}}%
\\
\subfigure[$g+g\to q+\bar q$]
{\label{fig:chc}\includegraphics[width=0.84\textwidth]{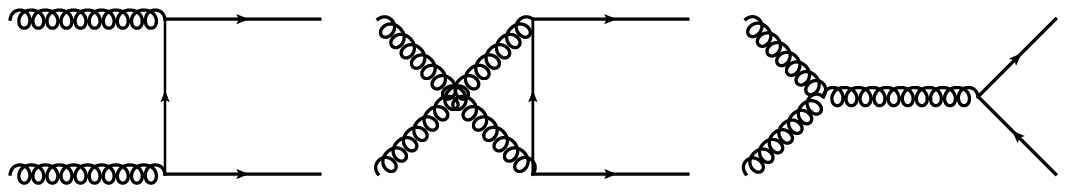}}%
\\
\subfigure[$g+g\to g+g$]
{\label{fig:chd}\includegraphics[width=0.96\textwidth]{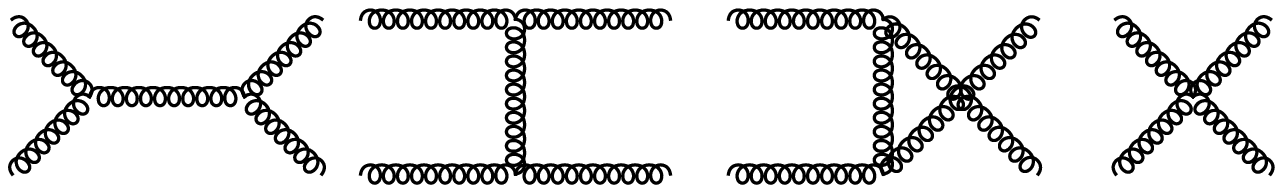}}%
\vspace{-0.5cm}
\end{center}
\caption{\label{fig:LOFeynDiag}Feynman diagrams contributing to the $2\to 2$ process at leading order.}
\end{figure}

The Feynman diagrams for dijet process at LO are shown in figure~\ref{fig:LOFeynDiag}.

\section{Explicit expressions of hard Wilson coefficients}\label{app2}
All the expression of Wilson coefficients can be found in ref.~\cite{Kelley:2010fn}. We list them below the convenience of the reader.

For $qq'\to qq'$ channels, the expressions of the Wilson coefficients in eq.~(\ref{eq:wc4qid}) are given by
\begin{equation}\label{eq:WC4qexp}
\begin{aligned}
\cC_{1}^{LL}(s,t,u) &= 2g_s^2\frac{s}{t}
\left\{1 + \frac{\alpha_s}{4 \pi}
\left[-2C_FL(t)^2 + X_1(s,t,u)L(t) +Y+\left(\frac{1}{2}C_A-2C_F\right) Z(s,t,u)\right]
\right\}  \,, \\
\cC_1^{LR}(s,t,u) &=2g_s^2\frac{u}{t}
\left\{1 + \frac{\alpha_s}{4 \pi}\left[-2C_F L(t)^2 + X_1(s,t,u)L(t)+Y+\left(2C_F-C_A\right) Z(u,t,s)\right]
\right\}  \,, \\
\cC_2^{LL}(s,t,u) &= 2g_s^2\frac{s}{t}
 \left\{\frac{\alpha_s}{4 \pi}
\left[ X_2(s,t,u)L(t)-\frac{C_F}{2C_A} Z(s,t,u)\right]
\right \} \,,\\
\cC_2^{LR}(s,t,u) &=  2g_s^2\frac{u}{t}
 \left\{ \frac{\alpha_s}{4 \pi}
\left[ X_2(s,t,u)L(t) + \frac{C_F}{2C_A} Z(u,t,s)\right]
\right\} \,,
\end{aligned}
\end{equation}
with
\begin{equation}
\begin{aligned}
X_1(s,t,u) &= 6 C_F - \beta_0 + 8 C_F [L(s) - L(u)] - 2C_A[2L(s)-L(t)-L(u)]\\
X_2(s,t,u) &= \frac{2C_F}{C_A} [L(s)-L(u)]
\\
Y &=C_A\left(\frac{10}{3} + \pi^2\right) + C_F\left(\frac{\pi^2}{3} - 16\right) + \frac{5}{3}\beta_0\\
Z(s,t,u) &= \frac{t}{s}\left(\frac{t+2 u}{s}[L(u) - L(t)]^2 +2[ L(u)-L(t)] + \pi^2\frac{t + 2 u }{s}\right)   \,.
\end{aligned}
\end{equation}
If the 4 quarks are identical, the corresponding Wilson coefficients can be obtained by using eq.~(\ref{eq:wc4qid}). The other crossed channels, the Wilson coefficients can be obtained by using crossing relations shown in table~\ref{tab:crosssys}.

For $gg\to q {\bar q}$ channel, the Wilson coefficients are given by
\begin{equation}\label{eq:WCggqqqexp}
\begin{aligned}
\cC_1^{-+}(s,t,u) &=  2g_s^2\frac{\sqrt{t u}}{s}
\left\{ 1 + \frac{\alpha_s}{4 \pi} \left[
-(C_A+C_F)L(s)^2 + V_1(s,t,u)L(s) + W_1 (s,t,u)\right]
\right\}  \\
\cC_1^{+-}(s,t,u) &=  2g_s^2\frac{u}{s}\sqrt{\frac{u}{t}}
\left\{ 1 + \frac{\alpha_s}{4 \pi} \left[
-(C_A+C_F)L(s)^2 + V_1(s,t,u)L(s) + W_2 (s,t,u)\right]
\right\} \\
\cC_1^{++}(s,t,u) &= \cC_1^{--}(s,t,u) = 2g_s^2\sqrt{\frac{u}{t}}
\frac{\alpha_s}{4 \pi} W_3(s,t,u)\\
\cC_2^{-+}(s,t,u) &=\cC_1^{+-}(s,u,t)  \\
\cC_2^{+-}(s,t,u) &=\cC_1^{-+}(s,u,t)\\
\cC_2^{++}(s,t,u) &= \cC_2^{--}(s,t,u)=\cC_1^{++}(s,u,t) \\
\cC_3^{-+}(s,t,u) &= 2g_s^2\sqrt{\frac{t}{u}}  \frac{\alpha_s}{4 \pi}
\left\{ V_2(s,t,u) L(s) +W_4(s,t,u)\right\} \\
\cC_3^{+-}(s,t,u) &=\cC_3^{-+}(s,u,t) \\
\cC_3^{++}(s,t,u) &=\cC_3^{--}(s,t,u)=0  \,,
\end{aligned}
\end{equation}
where
\begin{equation}
\begin{aligned}
W_1 (s,t,u) =& (C_A-C_F)\frac{s}{u} \Big( [L(s) - L(t)]^2 + \pi^2\Big) + C_A - 8 C_F + \left( \frac{7C_A+C_F}{6}\right)\pi^2\\
W_2(s,t,u) =& \left(-C_F \frac{s^3}{u^3} - C_A\frac{t^3+u^3-s^3}{2u^3} \right)
\Big( [ L(s) - L(t)]^2 + \pi^2 \Big)  \\
&+\left(2C_A \frac{t s}{u^2} + C_F \frac{s(2s-u)}{u^2}\right)[ L(t) - L(s)]
+C_F \frac{t-7u}{u} - C_A\frac{t}{u} + \left(\frac{7 C_A + C_F}{6}\right)\pi^2 \\
W_3(s,t,u) =& 2C_F-2C_A - \frac{2t}{3s} (C_A- n_f) \\
W_4(s,t,u) =& -\frac{3u}{4t} [ L(s)-L(u)]^2 - [L(s) - L(t)][L(s)-L(u)] + \frac{3\pi^2}{2} \frac{u^2}{t s}\\
V_1(s,t,u) =& 3 C_F - 2 C_A[ L(t)-L(s)]\\
V_2(s,t,u) =& [L(s)-L(u)]+\frac{t}{s}[L(t)-L(u)]  \,.
\end{aligned}
\end{equation}
For the other crossed channels, the Wilson coefficients can be obtained by using crossing relations shown in table~\ref{tab:crosssys}.

For 4-gluon channel, the Wilson coefficients can be obtained by eq.~(\ref{eq:4gWCsys}).
The LO matching coefficients $\cM_I^\Gamma$ can be obtained in table \ref{tab:4gMexp}.
\begin{table}\label{tab:4gMexp}
\begin{center}
\begin{tabular}{c|c|c|c||c|c|c|c}
\hline
\hline
$~~~\cM_I^\Gamma$~~~ &   $\Gamma = 1,2$  &  $~~~~ 3,4 ~~~~$ & $~~~~5,6~~~~$ &
~~~~~& $~~~~ 1,2 ~~~~$ & $~~~~3,4~~~~$ & $~~~~5,6~~~~$
\\
\hline
$I = 1$ & $\dfrac{s}{u}$    & $\dfrac{u}{s}    $ & $\dfrac{t^2}{su} $ &
$4$  & $\dfrac{s^2}{tu}$ & $\dfrac{u}{t}    $ & $\dfrac{t}{u}    $
\\
\hline
$2$  & $\dfrac{s}{t}$    & $\dfrac{u^2}{st} $ & $\dfrac{t}{s}    $ &
$5$  & $\dfrac{s}{t}$    & $\dfrac{u^2}{st} $ & $\dfrac{t}{s}    $
\\
\hline
$3$  & $\dfrac{s}{u}$    & $\dfrac{u}{s}    $ & $\dfrac{t^2}{su} $ &
$6$  & $\dfrac{s^2}{tu}$ & $\dfrac{u}{t}    $ & $\dfrac{t}{u}    $
\\
\hline
\hline
\end{tabular}
\end{center}
\caption{LO matching coefficients $\cM_1^\Gamma$ for the 4-gluon channel. }
\end{table}
At NLO, we also need $\cQ$.
They can be expressed in terms of $\cA$, $\cB$ and $\cF$, the expressions of which are
\begin{equation}\label{eq:4gWCexp}
\begin{aligned}
\cA(s,t,u)=&-2 C_A L(u)^2
+\Big( - 2C_A [L(s)-L(u)] + \beta_0 \Big) L(u)
+\left( \frac{4\pi^2}{3} - \frac{67}{9} \right) C_A
+\frac{10}{9}n_f\\
\cB(s,t,u)=&\cA(s,t,u)+ \beta_0 \frac{u}{t}[L(u) - L(s)]
- \frac{3n_f}{2} \frac{su}{t^2} \Big( [L(u) - L(s)]^2 + \pi^2 \Big)\\
&+ (C_A-n_f)\frac{su}{t^2}\left[ \frac{s-u}{t} [L(u) - L(s)] + \left(\frac{su}{t^2} - 2 \right)\Big( [L(u) - L(s)]^2 + \pi^2 \Big) - 1 \right]\\
\cF(s,t,u)=&\frac{1}{C_A} \left(\frac{s^2}{t u}\cB(t,s,u)
+\frac{s^2}{t u}\cB(u,s,t)
+\frac{2s}{u} \cA(s,t,u)
+\frac{2s}{t} \cA(s,u,t)\right)\,.
\end{aligned}
\end{equation}
\section{Calculation of the soft functions }\label{app3}
\subsection{Color Matrix}
The color matrix of NLO soft function has been defined in eq.~(\ref{eq:ColMSdef}). At tree level, the color matrix is
\begin{equation}
\begin{aligned}
&{\tilde{\bm s}}_{q q'\to q q'}^{(0)}= \left(
\begin{array}{cc}
  \frac{1}{2}C_A C_F & 0 \\
                   0 & C_A^2
\end{array}
\right)\,.
\end{aligned}
\end{equation}
The NLO color matrix is
\begin{gather}
\bm{w}_{12}=\left(
\begin{array}{cc}
 -\frac{C_F}{2} & \frac{C_A C_F}{2} \\
 \frac{C_A C_F}{2} & 0
\end{array}
\right)\,,\quad
\bm{w}_{13}=\left(
\begin{array}{cc}
 \frac{C_F}{4} & 0 \\
 0 & -C_A^2 C_F
\end{array}
\right)\,,\nn\\
\bm{w}_{14}=\left(
\begin{array}{cc}
 \frac{C_F}{2}-\frac{1}{4} C_A^2 C_F\, & -\frac{1}{2} C_A C_F \\
 -\frac{1}{2} C_A C_F & 0
\end{array}
\right)\,,\nn\\
\bm{w}_{23}=\bm{w}_{14}\,,\qquad
\bm{w}_{24}=\bm{w}_{13}\,,\qquad
\bm{w}_{34}=\bm{w}_{12}\,.
\end{gather}

For $gg\to q{\bar q}$ channel, the color matrix at tree level is
\begin{equation}
\begin{aligned}
&{\tilde{\bm s}}_{gg\to q \bar q}^{(0)}=\left(
\begin{array}{ccc}
 C_A C_F^2\, & -\frac{C_F}{2}\, & C_A C_F \\
 -\frac{C_F}{2}\, & C_A C_F^2\, & C_A C_F \\
 C_A C_F\, & C_A C_F\, & 2 C_A^2 C_F
\end{array}
\right)\,.
\end{aligned}
\end{equation}
The NLO color matrix is
\begin{gather}\label{eq:ggqqcolm}
\bm{w}_{12}=\left(
\begin{array}{ccc}
 -\frac{1}{4} C_A^3 C_F & 0 & -C_A^2 C_F \\
 0 & -\frac{1}{4} C_A^3 C_F & -C_A^2 C_F \\
 -C_A^2 C_F & -C_A^2 C_F & -2 C_A^2 C_F C_A
\end{array}
\right)\,,\nn\\
\bm{w}_{13}=\left(
\begin{array}{ccc}
 \frac{1}{12} C_A^2 C_F (1-3 C_A)\, & \frac{1}{12} C_A^2 C_F &
   -\frac{1}{2} C_A^2 C_F \\
 \frac{1}{12} C_A^2 C_F & \frac{1}{12} C_A^2 C_F & \frac{1}{2} C_A^2 C_F \\
 -\frac{1}{2} C_A^2 C_F & \frac{1}{2} C_A^2 C_F & 0
\end{array}
\right)\,,\nn\\
\bm{w}_{14}=\left(
\begin{array}{ccc}
 \frac{1}{12} C_A^2 C_F & \frac{1}{12} C_A^2 C_F & \frac{1}{2} C_A^2 C_F \\
 \frac{1}{12} C_A^2 C_F\, & \frac{1}{12} C_A^2 C_F (1-3 C_A)\, &
   -\frac{1}{2} C_A^2 C_F \\
 \frac{1}{2} C_A^2 C_F & -\frac{1}{2} C_A^2 C_F & 0
\end{array}
\right)\,,\nn\\
\bm{w}_{34}=\left(
\begin{array}{ccc}
 -\frac{C_F}{12} & -\frac{1}{12} \left(3 C_A+1\right) C_F & -C_A C_F^2 \\
 -\frac{1}{12} \left(3 C_A+1\right) C_F & -\frac{C_F}{12} & -C_A C_F^2 \\
 -C_A C_F^2 & -C_A C_F^2 & -2 C_A^2 C_F^2
\end{array}
\right)\,,\nn\\
\bm{w}_{23}=\bm{w}_{14}\,,\qquad
\bm{w}_{24}=\bm{w}_{14}\,.
\end{gather}

For $gg\to gg$ channel the color matrix at tree level is
\begin{equation}
\begin{aligned}
&{\tilde {\bm s}}_{gg\to gg}^{(0)}=\frac{C_F}{8C_A}\left(
\begin{array}{ccccccccc}
 a_0 & b_0 & c_0 & b_0 & b_0 & b_0 & d_0 & d_0 & -e_0 \\
 b_0 & a_0 & b_0 & b_0 & c_0 & b_0 & -e_0 & d_0 & d_0 \\
 c_0 & b_0 & a_0 & b_0 & b_0 & b_0 & d_0 & d_0 & -e_0 \\
 b_0 & b_0 & b_0 & a_0 & b_0 & c_0 & d_0 & -e_0 & d_0 \\
 b_0 & c_0 & b_0 & b_0 & a_0 & b_0 & -e_0 & d_0 & d_0 \\
 b_0 & b_0 & b_0 & c_0 & b_0 & a_0 & d_0 & -e_0 & d_0 \\
 d_0 & -e_0 & d_0 & d_0 & -e_0 & d_0 & d_0e_0 & e_0^2 & e_0^2 \\
 d_0 & d_0 & d_0 & -e_0 & d_0 & -e_0 & e_0^2 & d_0e_0 & e_0^2 \\
 -e_0 & d_0 & -e_0 & d_0 & d_0 & d_0 & e_0^2 & e_0^2 & d_0e_0
\end{array}
\right)\,,
\end{aligned}
\end{equation}
with
\begin{gather}
a_0=C_A^4-3C_A^2+3\,,\quad b_0=3-C_A^2\,,\quad c_0=3+C_A^2\,,\quad d_0=2C_A^2C_F\,,\quad e_0=C_A\,.
\end{gather}
The NLO color matrix is
\begin{gather}\label{eq:ggggcolm}
\bm{w}_{12}=\left(
\begin{array}{ccccccccc}
 a & h & c & b & h & b & -f & d & 0 \\
 h & a & h & b & c & b & 0 & d & -f \\
 c & h & a & b & h & b & -f & d & 0 \\
 b & b & b & g & b & g & f & k & f \\
 h & c & h & b & a & b & 0 & d & -f \\
 b & b & b & g & b & g & f & k & f \\
 -f & 0 & -f & f & 0 & f & 0 & -e & e \\
 d & d & d & k & d & k & -e & m & -e \\
 0 & -f & 0 & f & -f & f & e & -e & 0
\end{array}
\right)\,, \quad
\bm{w}_{13}=\left(
\begin{array}{ccccccccc}
 g & b & g & b & b & b & f & f & k \\
 b & a & b & h & c & h & 0 & -f & d \\
 g & b & g & b & b & b & f & f & k \\
 b & h & b & a & h & c & -f & 0 & d \\
 b & c & b & h & a & h & 0 & -f & d \\
 b & h & b & c & h & a & -f & 0 & d \\
 f & 0 & f & -f & 0 & -f & 0 & e & -e \\
 f & -f & f & 0 & -f & 0 & e & 0 & -e \\
 k & d & k & d & d & d & -e & -e & m
\end{array}
\right)\,,\nn\\
\bm{w}_{14}=\left(
\begin{array}{ccccccccc}
 a & b & c & h & b & h & d & -f & 0 \\
 b & g & b & b & g & b & k & f & f \\
 c & b & a & h & b & h & d & -f & 0 \\
 h & b & h & a & b & c & d & 0 & -f \\
 b & g & b & b & g & b & k & f & f \\
 h & b & h & c & b & a & d & 0 & -f \\
 d & k & d & d & k & d & m & -e & -e \\
 -f & f & -f & 0 & f & 0 & -e & 0 & e \\
 0 & f & 0 & -f & f & -f & -e & e & 0
\end{array}
\right)\,, \nn\\
\bm{w}_{23}=\bm{w}_{14}\,,\quad
\bm{w}_{24}=\bm{w}_{13}\,,\quad
\bm{w}_{34}=\bm{w}_{12}\,,
\end{gather}
with
\begin{gather}
  a=-\frac{1}{16} \left(C_A^4-2 C_A^2+2\right) C_F\,,\quad
  b=-\frac{1}{16} \left(2-C_A^2\right) C_F\,,\quad
  c=-\frac{1}{8} \left(C_A^2+1\right) C_F\,,\nn\\
  d=-\frac{1}{4} C_A^2 C_F^2\,,\quad
  e=\frac{1}{8} C_A^2 C_F\,,\quad
  f=\frac{1}{16} C_A^3 C_F\,,\quad
  g=\frac{1}{4} C_A C_F^2\,,\\
  h=-\frac{C_F}{8}\,,\quad
  k=\frac{C_A C_F}{8}\,,\quad
  m=-\frac{1}{4} C_A^3 C_F^2\,.\nn
\end{gather}

\subsection{Calculation of $\cI_{ij}$}
Here, we show the detail of the calculation of the $\cI_{ij}$ function.
First, in order to compute ${\cI}_{ij}^{\rm out}$ conveniently, we define an auxiliary function ${\cI}_{ij}^{\rm aux}(\kout)$ with the measurement function $\cM_{\rm aux}(\kout,R,q)$,
\begin{equation}\label{eq:Iijaux}
\begin{aligned}
\cM_{\rm aux}(\kout,R,q)
=&\Theta\Big(R^2 - (y-y_J)^2 -(\phi-\phi_J)^2 \Big)\delta(\kout-\bn_J\cdot q)\,,
\end{aligned}
\end{equation}
which is the same as $\cM_{\rm in}$ in eq.~(\ref{eq:Sinmeasfunc}) except for the delta function. Then ${\cI}_{ij}^{\rm out}$ can be obtained by
\begin{equation}\label{eq:Iijaux}
\begin{aligned}
{\cI}_{ij}^{\rm out}(\kout)
={\cI}_{ij}^{\rm full}(\kout)-{\cI}_{ij}^{\rm aux}(\kout)\,,
\end{aligned}
\end{equation}
where ${\cI}_{ij}^{\rm full}(\kout)$ denote the soft radiation without constraints from jet algorithm, the results of which are
\begin{equation}\label{eq:Iijaux}
\begin{aligned}
{\cI}_{12}^{\rm full}(\kout,\mu)&=-\Big(\frac{\alpha_s}{4\pi}\Big)
\Big\{\delta(k)\Big[\ln^2\frac{2n_{12}}{n_{14}\,n_{24}}-\frac{\pi^2}{6}\Big]
+8\left[\frac{1}{k}\ln\left(\frac{k}{\mu}\sqrt{\frac{2n_{12}}{n_{14}\,n_{24}}}
\right)\right]_\star\Big\}\,,\\
{\cI}_{13}^{\rm full}(\kout,\mu)&=-\Big(\frac{\alpha_s}{4\pi}\Big)
\Big\{\delta(k)\Big[\ln^2\frac{2n_{13}}{n_{14}\,n_{34}}-\frac{\pi^2}{6}\Big]
+8\left[\frac{1}{k}\ln\left(\frac{k}{\mu}\sqrt{\frac{2n_{13}}{n_{14}\,n_{34}}}
\right)\right]_\star\Big\}\,,\\
{\cI}_{23}^{\rm full}(\kout,\mu)&=-\Big(\frac{\alpha_s}{4\pi}\Big)
\Big\{\delta(k)\Big[\ln^2\frac{2n_{23}}{n_{24}\,n_{34}}-\frac{\pi^2}{6}\Big]
+8\left[\frac{1}{k}\ln\left(\frac{k}{\mu}\sqrt{\frac{2n_{23}}{n_{24}\,n_{34}}}
\right)\right]_\star\Big\}\,,\\
{\cI}_{14}^{\rm full}(\kout,\mu)&=
{\cI}_{24}^{\rm full}(\kout,\mu)=
{\cI}_{34}^{\rm full}(\kout,\mu)=0\,,
\end{aligned}
\end{equation}
where $n_{ij}=n_i\cdot n_j$.

In partonic CM frame, the four vectors of initial and final partons can be written as
\begin{equation}\label{eq:4vrap}
\begin{aligned}
n_1^\mu &=(1,0,0,1)\,,\\
n_2^\mu &=(1,0,0,-1)\,,\\
p_{J_1}^\mu &=p_T(\cosh y_{J}, 0, 1, \sinh y_{J})\,,\\
p_{J_2}^\mu &=p_T(\cosh y_{J}, 0, -1, -\sinh y_J)\,,\\
q^\mu &=q_T(\cosh y, \sin\phi, \cos\phi, \sinh y)\,.
\end{aligned}
\end{equation}
This choice of frame makes the measurement functions simple but leaves the complexity in delta function. The phase space integration can be written as
\begin{equation}\label{eq:transintv}
\begin{aligned}
\int d^d q \delta(q^2)\Theta(q^0)&=
\frac{\pi^{\frac{1}{2}-\epsilon}}{\Gamma(\frac{1}{2}-\epsilon)}
\int_0^\pi d\phi\sin^{-2\epsilon}\phi\int dy\int dq_T q_T^{1-2\epsilon}
\,.\\
\end{aligned}
\end{equation}
Integrating over the delta function, we can get
\begin{equation}
q_T = \frac{\kin\cosh y_J}{\cosh(y-y_J)-\cos \phi}\,,
\end{equation}
for soft emission inside jet, and
\begin{equation}
q_T = \frac{\kout\cosh y_J}{\cosh(y+y_J)+\cos \phi}\,,
\end{equation}
for the one outside the jet.
For ${\cI}_{ij}^{\rm in}$ and ${\cI}_{ij}^{\rm aux}$, the integral region of rapidity $y$ and azimuthal angle $\phi$ are constrained by measurement function is a circle with radius $R$. We redefine the integration variables
\begin{gather}
y=y'+y_J\,,\quad
y' = r\,\cos\varphi\,,\quad
\phi = r\,\sin\varphi
\end{gather}
and then
\begin{gather}
\int dy \int_0^\pi d\phi \Theta (R^2-(y-y_J)^2-\phi^2) = \int_0^R dr r \int _0^\pi d\varphi\,.
\end{gather}
For $\cI_{12}^{\rm in}$, we can get
\begin{equation}\label{eq:I12incalc}
\begin{aligned}
\cI_{12}^{\rm in}(\kin,y_J,R,\mu)=&-\frac{4\pi\alpha_s}{(2\pi)^{3-2\epsilon}}
\left(\frac{e^{\gamma_E}}{4\pi}\right)^\epsilon
\frac{2\pi^{\frac{1}{2}-\epsilon}}{\Gamma(\frac{1}{2}-\epsilon)}
\frac{1}{\kin}\left(\frac{\kin}{\mu}\right)^{-2\epsilon}\\
&\times\int_0^R dr r \int _0^\pi d\phi \sin^{-2\epsilon}\phi\, (\cosh y'- \cos\phi)^{2\epsilon}\cosh^{-2\epsilon}y_J\,.
\end{aligned}
\end{equation}
This integration can be computed analytically by approximation at small $R$
\begin{equation}
\begin{aligned}
\sin \phi &\approx \phi = r\sin \varphi\,,\\
\cosh y' - \cos\phi &\approx \frac{1}{2}y'^2+\frac{1}{2}\phi^2 = \frac{1}{2} r^2\,.
\end{aligned}
\end{equation}
From figure~\ref{fig:FOexpand}, we can see that the approximation is validity at even larger $R$, i.e. $R=1$. The other $I_{ij}^{\rm in}$ and $I_{ij}^{\rm aux}$ functions can be calculated by similar method.

The results of the refactorized soft function in Laplace space are
\begin{equation}\label{eq:LapIio}
\begin{aligned}
{\tilde I}_{12}^{in}(\kain,y_J,R,\mu)
=&\left(\frac{\alpha_s}{4\pi}\right)\left(A_{I_{12}}^{\rm in }
+\gamma_{I_{12}}^{(0)}\,L_{\rm in}\right)\,,\nn\\
{\tilde I}_{12}^{out}(\kaout,y_J,R,\mu)
=&\left(\frac{\alpha_s}{4\pi}\right)\left(
A_{I_{12}}^{\rm out}-\gcusp^{(0)}\,L_{\rm out}^2-\gamma_{I_{12}}^{(0)}\,L_{\rm out}
\right)\,,\nn\\
{\tilde I}_{13}^{in}(\kain,y_J,R,\mu)=&
\left(\frac{\alpha_s}{4\pi}\right)
\left(A_{I_{13}}^{\rm in}+\frac{1}{2}\gcusp^{(0)} L_{\rm in}^2+\gamma_{I_{13}}^{(0)}\,L_{\rm in}\right)\,,\nn\\
{\tilde I}_{13}^{out}(\kaout,y_J,R,\mu)=&
\left(\frac{\alpha_s}{4\pi}\right)\left(A_{I_{13}}^{\rm out}
-\frac{1}{2}\gcusp^{(0)} L_{\rm out}^2 + 2 \gcusp^{(0)}y_J L_{\rm out}
-\gamma_{I_{13}}^{(0)}\,L_{\rm out}\right)\,,\nn\\
{\tilde I}_{14}^{in}(\kain,y_J,R,\mu)=&
\,\left(\frac{\alpha_s}{4\pi}\right)
\left(A_{I_{14}}^{\rm in}+\gamma_{I_{14}}^{(0)}\,L_{\rm in}\right)\,,\nn\\
{\tilde I}_{14}^{out}(\kaout,y_J,R,\mu)=&
\left(\frac{\alpha_s}{4\pi}\right)
\left(A_{I_{14}}^{\rm out}-\gamma_{I_{14}}^{(0)}\,L_{\rm out}\right)\,,\nn\\
{\tilde I}_{34}^{in}(\kain,y_J,R,\mu)=&
\left(\frac{\alpha_s}{4\pi}\right)
\left(A_{I_{34}}^{\rm in}+\frac{1}{2}\gcusp^{(0)}\,L_{\rm in}^2
+\gamma_{I_{34}}^{(0)}\,L_{\rm in}\right)\,,\nn\\
{\tilde I}_{34}^{out}(\kaout,y_J,R,\mu)=&
\left(\frac{\alpha_s}{4\pi}\right)
\Big[A_{I_{34}}^{\rm out}+\frac{1}{2}\gcusp^{(0)}\,L_{\rm out}^2
-2\gcusp^{(0)}\ln\left(2\,\cosh y_J\right)\,L_{\rm out}
-\gamma_{I_{34}}^{(0)}\,L_{\rm out}\Big]\,,\nn\\
{\tilde I}_{23}^{in}(\kain,y_J,R,\mu)
=& {\tilde I}_{13}^{in}(\kain,-y_J,R,\mu)\,,\qquad
{\tilde I}_{23}^{out}(\kaout,y_J,R,\mu)
={\tilde I}_{13}^{out}(\kaout,-y_J,R,\mu)\,,\nn\\
{\tilde I}_{24}^{in}(\kain,y_J,R,\mu)
=&{\tilde I}_{14}^{in}(\kain,-y_J,R,\mu)\,,\qquad
{\tilde I}_{24}^{out}(\kaout,y_J,R,\mu)
={\tilde I}_{14}^{out}(\kaout,-y_J,R,\mu)\,,
\end{aligned}
\end{equation}
with $L_{\rm in}=\ln\left(2\kain\,\cosh y_J/\mu\right)$ and $L_{\rm out}=\ln\left(2\kaout\,\cosh y_J/\mu\right)$.
The one-loop $R$-dependent anomalous dimensions $\gamma_{I_{ij}}$ are
\begin{gather}
\gamma_{I_{12}}^{(0)}(y_J,R)=-2\,R^2\,,\quad
\gamma_{I_{13}}^{(0)}(y_J,R)=-\frac{R^2}{2}-4\ln R\,,\nn\\
\gamma_{I_{14}}^{(0)}(y_J,R)= -\frac{1}{16} R^2 \left(R^2+8\right)
   e^{2 y_J} \sech^2y_J\,,\quad
\gamma_{I_{34}}^{(0)}(y_J,R)= -4\ln R\,,\nn\\
\gamma_{I_{23}}^{(0)}(y_J,R)=\gamma_{I_{13}}^{(0)}(-y_J,R)\,,\quad
\gamma_{I_{24}}^{(0)}(y_J,R)=\gamma_{I_{14}}^{(0)}(-y_J,R)\,,
\end{gather}
and the constant terms $A_{I_{ij}}^{\rm in,out}$ are
\begin{eqnarray}
\begin{aligned}
A_{I_{12}}^{\rm in}(y_J,R)=& R^2\left(-1+2\ln R \right)\,,\nn\\
A_{I_{12}}^{\rm out}(y_J,R)=& -4R^2\ln\left(2\cosh y_J\right)+R^2(-1+2\ln R)-\frac{\pi^2}{2}\,,\nn\\
A_{I_{13}}^{\rm in}(y_J,R)=&\frac{1}{2}R^2\ln R + 2\ln^2 R +\frac{\pi^2}{4}\,,\nn\\
A_{I_{13}}^{\rm out}(y_J,R)=&\frac{1}{2} \left[R^2-16 \ln (2 \cosh y_J)\right]
 \ln R -\left(R^2+8y_J\right) \ln (2 \cosh y_J)\nn\\
&-\frac{R^2}{2}+2 \ln ^2 R -4 y_J^2
+4 \ln ^2(2 \cosh y_J)-\frac{\pi ^2}{4} \,,\nn\\
A_{I_{14}}^{\rm in}(y_J,R)=& \frac{1}{64} R^2
\left[R^2+4\left(R^2+8\right) \ln R-16\right]e^{2 y_J} \sech^2y_J\,,\nn\\
A_{I_{14}}^{\rm out}(y_J,R)=& \frac{1}{64} R^2
\left[-8\left(R^2+8\right) \ln (2 \cosh y_J)-3 R^2+4 \left(R^2+8\right)
\ln R-16\right]e^{2y_J}\sech^2y_J\,,\nn\\
A_{I_{34}}^{\rm in}(y_J,R)=& 2\ln^2 R +\frac{\pi^2}{4}\,,\nn\\
A_{I_{34}}^{\rm out}(y_J,R)=& -8 \ln (2 \cosh y_J) \ln R +2 \ln^2 R
+8 \ln ^2(2 \cosh y_J)+\frac{\pi ^2}{4}\,,\nn\\
A_{I_{23}}^{\rm in}(y_J,R)=&A_{I_{13}}^{\rm in}(-y_J,R)\,,\qquad
A_{I_{23}}^{\rm out}(y_J,R)=A_{I_{13}}^{\rm out}(-y_J,R)\,,\nn\\
A_{I_{24}}^{\rm in}(y_J,R)=&A_{I_{14}}^{\rm in}(-y_J,R)\,,\qquad
A_{I_{24}}^{\rm out}(y_J,R)=A_{I_{14}}^{\rm out}(-y_J,R)\,.
\end{aligned}
\end{eqnarray}
\bibliographystyle{JHEP}
\bibliography{dijet}

\begin{thebibliography}{10}

\bibitem{Butterworth:2002tt}
J.M. Butterworth, B.E. Cox, and Jeffrey~R. Forshaw.
\newblock {$W W$ scattering at the CERN LHC}.
\newblock {\em Phys.Rev.}, D65:096014, 2002.

\bibitem{Butterworth:2008iy}
Jonathan~M. Butterworth, Adam~R. Davison, Mathieu Rubin, and Gavin~P. Salam.
\newblock {Jet substructure as a new Higgs search channel at the LHC}.
\newblock {\em Phys.Rev.Lett.}, 100:242001, 2008.

\bibitem{Kaplan:2008ie}
David~E. Kaplan, Keith Rehermann, Matthew~D. Schwartz, and Brock Tweedie.
\newblock {Top Tagging: A Method for Identifying Boosted Hadronically Decaying
  Top Quarks}.
\newblock {\em Phys.Rev.Lett.}, 101:142001, 2008.

\bibitem{Ellis:2009me}
Stephen~D. Ellis, Christopher~K. Vermilion, and Jonathan~R. Walsh.
\newblock {Recombination Algorithms and Jet Substructure: Pruning as a Tool for
  Heavy Particle Searches}.
\newblock {\em Phys.Rev.}, D81:094023, 2010.

\bibitem{Thaler:2008ju}
Jesse Thaler and Lian-Tao Wang.
\newblock {Strategies to Identify Boosted Tops}.
\newblock {\em JHEP}, 0807:092, 2008.

\bibitem{Krohn:2009th}
David Krohn, Jesse Thaler, and Lian-Tao Wang.
\newblock {Jet Trimming}.
\newblock {\em JHEP}, 1002:084, 2010.

\bibitem{Gallicchio:2010dq}
Jason Gallicchio, John Huth, Michael Kagan, Matthew~D. Schwartz, Kevin Black,
  et~al.
\newblock {Multivariate discrimination and the Higgs + W/Z search}.
\newblock {\em JHEP}, 1104:069, 2011.

\bibitem{Thaler:2010tr}
Jesse Thaler and Ken Van~Tilburg.
\newblock {Identifying Boosted Objects with N-subjettiness}.
\newblock {\em JHEP}, 1103:015, 2011.

\bibitem{Gallicchio:2010sw}
Jason Gallicchio and Matthew~D. Schwartz.
\newblock {Seeing in Color: Jet Superstructure}.
\newblock {\em Phys.Rev.Lett.}, 105:022001, 2010.

\bibitem{Cui:2010km}
Yanou Cui, Zhenyu Han, and Matthew~D. Schwartz.
\newblock {W-jet Tagging: Optimizing the Identification of Boosted
  Hadronically-Decaying W Bosons}.
\newblock {\em Phys.Rev.}, D83:074023, 2011.

\bibitem{Gallicchio:2011xq}
Jason Gallicchio and Matthew~D. Schwartz.
\newblock {Quark and Gluon Tagging at the LHC}.
\newblock {\em Phys.Rev.Lett.}, 107:172001, 2011.

\bibitem{Altheimer:2012mn}
A.~Altheimer, S.~Arora, L.~Asquith, G.~Brooijmans, J.~Butterworth, et~al.
\newblock {Jet Substructure at the Tevatron and LHC: New results, new tools,
  new benchmarks}.
\newblock {\em J.Phys.}, G39:063001, 2012.

\bibitem{Ellis:2012sn}
Stephen~D. Ellis, Andrew Hornig, Tuhin~S. Roy, David Krohn, and Matthew~D.
  Schwartz.
\newblock {Qjets: A Non-Deterministic Approach to Tree-Based Jet Substructure}.
\newblock {\em Phys.Rev.Lett.}, 108:182003, 2012.

\bibitem{Gleisberg:2003xi}
Tanju Gleisberg, Stefan Hoeche, Frank Krauss, Andreas Schalicke, Steffen
  Schumann, et~al.
\newblock {SHERPA 1. alpha: A Proof of concept version}.
\newblock {\em JHEP}, 0402:056, 2004.

\bibitem{Gleisberg:2008ta}
T.~Gleisberg, Stefan. Hoeche, F.~Krauss, M.~Schonherr, S.~Schumann, et~al.
\newblock {Event generation with SHERPA 1.1}.
\newblock {\em JHEP}, 0902:007, 2009.

\bibitem{Sjostrand:2006za}
Torbjorn Sjostrand, Stephen Mrenna, and Peter~Z. Skands.
\newblock {PYTHIA 6.4 Physics and Manual}.
\newblock {\em JHEP}, 0605:026, 2006.

\bibitem{Sjostrand:2007gs}
Torbjorn Sjostrand, Stephen Mrenna, and Peter~Z. Skands.
\newblock {A Brief Introduction to PYTHIA 8.1}.
\newblock {\em Comput.Phys.Commun.}, 178:852--867, 2008.

\bibitem{Bahr:2008pv}
M.~Bahr, S.~Gieseke, M.A. Gigg, D.~Grellscheid, K.~Hamilton, et~al.
\newblock {Herwig++ Physics and Manual}.
\newblock {\em Eur.Phys.J.}, C58:639--707, 2008.

\bibitem{Gieseke:2011na}
S.~Gieseke, D.~Grellscheid, K.~Hamilton, A.~Papaefstathiou, S.~Platzer, et~al.
\newblock {Herwig++ 2.5 Release Note}.
\newblock 2011.

\bibitem{ATLAS:2012am}
Georges Aad et~al.
\newblock {Jet mass and substructure of inclusive jets in $\sqrt{s}=7$ TeV $pp$
  collisions with the ATLAS experiment}.
\newblock {\em JHEP}, 1205:128, 2012.

\bibitem{Becher:2008cf}
Thomas Becher and Matthew~D. Schwartz.
\newblock {A precise determination of $\alpha_s$ from LEP thrust data using
  effective field theory}.
\newblock {\em JHEP}, 0807:034, 2008.

\bibitem{Cheung:2009sg}
William Man-Yin Cheung, Michael Luke, and Saba Zuberi.
\newblock {Phase Space and Jet Definitions in SCET}.
\newblock {\em Phys.Rev.}, D80:114021, 2009.

\bibitem{Ellis:2009wj}
Stephen~D. Ellis, Andrew Hornig, Christopher Lee, Christopher~K. Vermilion, and
  Jonathan~R. Walsh.
\newblock {Consistent Factorization of Jet Observables in Exclusive Multijet
  Cross-Sections}.
\newblock {\em Phys.Lett.}, B689:82--89, 2010.

\bibitem{Ellis:2010rwa}
Stephen~D. Ellis, Christopher~K. Vermilion, Jonathan~R. Walsh, Andrew Hornig,
  and Christopher Lee.
\newblock {Jet Shapes and Jet Algorithms in SCET}.
\newblock {\em JHEP}, 1011:101, 2010.

\bibitem{Jouttenus:2009ns}
Teppo~T. Jouttenus.
\newblock {Jet Function with a Jet Algorithm in SCET}.
\newblock {\em Phys.Rev.}, D81:094017, 2010.

\bibitem{Kelley:2011tj}
Randall Kelley, Matthew~D. Schwartz, and Hua~Xing Zhu.
\newblock {Resummation of jet mass with and without a jet veto}.
\newblock 2011.

\bibitem{Kelley:2011aa}
Randall Kelley, Matthew~D. Schwartz, Robert~M. Schabinger, and Hua~Xing Zhu.
\newblock {Jet Mass with a Jet Veto at Two Loops and the Universality of
  Non-Global Structure}.
\newblock {\em Phys.Rev.}, D86:054017, 2012.

\bibitem{Chien:2012ur}
Yang-Ting Chien, Randall Kelley, Matthew~D. Schwartz, and Hua~Xing Zhu.
\newblock {Resummation of Jet Mass at Hadron Colliders}.
\newblock {\em Phys.Rev.}, D87:014010, 2013.

\bibitem{Chien:2014nsa}
Yang-Ting Chien and Ivan Vitev.
\newblock {Jet Shape Resummation Using Soft-Collinear Effective Theory}.
\newblock 2014.

\bibitem{Banfi:2010pa}
Andrea Banfi, Mrinal Dasgupta, Kamel Khelifa-Kerfa, and Simone Marzani.
\newblock {Non-global logarithms and jet algorithms in high-$p_T$ jet shapes}.
\newblock {\em JHEP}, 1008:064, 2010.

\bibitem{Li:2011hy}
Hsiang-nan Li, Zhao Li, and C.-P. Yuan.
\newblock {QCD resummation for jet substructures}.
\newblock {\em Phys.Rev.Lett.}, 107:152001, 2011.

\bibitem{Li:2012bw}
Hsiang-nan Li, Zhao Li, and C.-P. Yuan.
\newblock {QCD resummation for light-particle jets}.
\newblock {\em Phys.Rev.}, D87:074025, 2013.

\bibitem{Dasgupta:2012hg}
Mrinal Dasgupta, Kamel Khelifa-Kerfa, Simone Marzani, and Michael Spannowsky.
\newblock {On jet mass distributions in Z+jet and dijet processes at the LHC}.
\newblock {\em JHEP}, 1210:126, 2012.

\bibitem{Dasgupta:2013ihk}
Mrinal Dasgupta, Alessandro Fregoso, Simone Marzani, and Gavin~P. Salam.
\newblock {Towards an understanding of jet substructure}.
\newblock {\em JHEP}, 1309:029, 2013.

\bibitem{Jouttenus:2013hs}
Teppo~T. Jouttenus, Iain~W. Stewart, Frank~J. Tackmann, and Wouter~J.
  Waalewijn.
\newblock {Jet mass spectra in Higgs boson plus one jet at
  next-to-next-to-leading logarithmic order}.
\newblock {\em Phys.Rev.}, D88:054031, 2013.

\bibitem{Banfi:2010xy}
Andrea Banfi, Gavin~P. Salam, and Giulia Zanderighi.
\newblock {Phenomenology of event shapes at hadron colliders}.
\newblock {\em JHEP}, 1006:038, 2010.

\bibitem{Banfi:2004yd}
Andrea Banfi, Gavin~P. Salam, and Giulia Zanderighi.
\newblock {Principles of general final-state resummation and automated
  implementation}.
\newblock {\em JHEP}, 0503:073, 2005.

\bibitem{Stewart:2010tn}
Iain~W. Stewart, Frank~J. Tackmann, and Wouter~J. Waalewijn.
\newblock {N-Jettiness: An Inclusive Event Shape to Veto Jets}.
\newblock {\em Phys.Rev.Lett.}, 105:092002, 2010.

\bibitem{Chatrchyan:2013vbb}
Serguei Chatrchyan et~al.
\newblock {Studies of jet mass in dijet and W/Z + jet events}.
\newblock {\em JHEP}, 1305:090, 2013.

\bibitem{Cacciari:2008gp}
Matteo Cacciari, Gavin~P. Salam, and Gregory Soyez.
\newblock {The Anti-$k_t$ jet clustering algorithm}.
\newblock {\em JHEP}, 0804:063, 2008.

\bibitem{Kelley:2012kj}
Randall Kelley, Jonathan~R. Walsh, and Saba Zuberi.
\newblock {Abelian Non-Global Logarithms from Soft Gluon Clustering}.
\newblock {\em JHEP}, 1209:117, 2012.

\bibitem{Dokshitzer:1997in}
Yuri~L. Dokshitzer, G.D. Leder, S.~Moretti, and B.R. Webber.
\newblock {Better jet clustering algorithms}.
\newblock {\em JHEP}, 9708:001, 1997.

\bibitem{Wobisch:1998wt}
M.~Wobisch and T.~Wengler.
\newblock {Hadronization corrections to jet cross-sections in deep inelastic
  scattering}.
\newblock 1998.

\bibitem{Catani:1993hr}
S.~Catani, Yuri~L. Dokshitzer, M.H. Seymour, and B.R. Webber.
\newblock {Longitudinally invariant $K_t$ clustering algorithms for hadron
  hadron collisions}.
\newblock {\em Nucl.Phys.}, B406:187--224, 1993.

\bibitem{Ellis:1993tq}
Stephen~D. Ellis and Davison~E. Soper.
\newblock {Successive combination jet algorithm for hadron collisions}.
\newblock {\em Phys.Rev.}, D48:3160--3166, 1993.

\bibitem{Banfi:2005gj}
A.~Banfi and M.~Dasgupta.
\newblock {Problems in resumming interjet energy flows with $k_t$ clustering}.
\newblock {\em Phys.Lett.}, B628:49--56, 2005.

\bibitem{Delenda:2006nf}
Yazid Delenda, Robert Appleby, Mrinal Dasgupta, and Andrea Banfi.
\newblock {On QCD resummation with $k_t$ clustering}.
\newblock {\em JHEP}, 0612:044, 2006.

\bibitem{KhelifaKerfa:2011zu}
Kamel Khelifa-Kerfa.
\newblock {Non-global logs and clustering impact on jet mass with a jet veto
  distribution}.
\newblock {\em JHEP}, 1202:072, 2012.

\bibitem{Kelley:2012zs}
Randall Kelley, Jonathan~R. Walsh, and Saba Zuberi.
\newblock {Disentangling Clustering Effects in Jet Algorithms}.
\newblock 2012.

\bibitem{Bauer:2008jx}
Christian~W. Bauer, Andrew Hornig, and Frank~J. Tackmann.
\newblock {Factorization for generic jet production}.
\newblock {\em Phys.Rev.}, D79:114013, 2009.

\bibitem{Bauer:2010vu}
Christian~W. Bauer, Nicholas~Daniel Dunn, and Andrew Hornig.
\newblock {Factorization of Boosted Multijet Processes for Threshold
  Resummation}.
\newblock {\em Phys.Rev.}, D82:054012, 2010.

\bibitem{Kelley:2010fn}
Randall Kelley and Matthew~D. Schwartz.
\newblock {1-loop matching and NNLL resummation for all partonic 2 to 2
  processes in QCD}.
\newblock {\em Phys.Rev.}, D83:045022, 2011.

\bibitem{Bauer:2001yt}
Christian~W. Bauer, Dan Pirjol, and Iain~W. Stewart.
\newblock {Soft collinear factorization in effective field theory}.
\newblock {\em Phys.Rev.}, D65:054022, 2002.

\bibitem{Becher:2009qa}
Thomas Becher and Matthias Neubert.
\newblock {On the Structure of Infrared Singularities of Gauge-Theory
  Amplitudes}.
\newblock {\em JHEP}, 0906:081, 2009.

\bibitem{Gardi:2009qi}
Einan Gardi and Lorenzo Magnea.
\newblock {Factorization constraints for soft anomalous dimensions in QCD
  scattering amplitudes}.
\newblock {\em JHEP}, 0903:079, 2009.

\bibitem{Dixon:2009ur}
Lance~J. Dixon, Einan Gardi, and Lorenzo Magnea.
\newblock {On soft singularities at three loops and beyond}.
\newblock {\em JHEP}, 1002:081, 2010.

\bibitem{Catani:1998bh}
Stefano Catani.
\newblock {The Singular behavior of QCD amplitudes at two loop order}.
\newblock {\em Phys.Lett.}, B427:161--171, 1998.

\bibitem{Sterman:2002qn}
George~F. Sterman and Maria~E. Tejeda-Yeomans.
\newblock {Multiloop amplitudes and resummation}.
\newblock {\em Phys.Lett.}, B552:48--56, 2003.

\bibitem{Becher:2009th}
Thomas Becher and Matthew~D. Schwartz.
\newblock {Direct photon production with effective field theory}.
\newblock {\em JHEP}, 1002:040, 2010.

\bibitem{Manohar:2003vb}
Aneesh~V. Manohar.
\newblock {Deep inelastic scattering as $x \to 1$ using soft collinear
  effective theory}.
\newblock {\em Phys.Rev.}, D68:114019, 2003.

\bibitem{Becher:2006qw}
Thomas Becher and Matthias Neubert.
\newblock {Toward a NNLO calculation of the ${\bar B} \to X_s \gamma$ decay
  rate with a cut on photon energy. II. Two-loop result for the jet function}.
\newblock {\em Phys.Lett.}, B637:251--259, 2006.

\bibitem{Becher:2010pd}
Thomas Becher and Guido Bell.
\newblock {The gluon jet function at two-loop order}.
\newblock {\em Phys.Lett.}, B695:252--258, 2011.

\bibitem{Becher:2006nr}
Thomas Becher and Matthias Neubert.
\newblock {Threshold resummation in momentum space from effective field
  theory}.
\newblock {\em Phys.Rev.Lett.}, 97:082001, 2006.

\bibitem{Becher:2006mr}
Thomas Becher, Matthias Neubert, and Ben~D. Pecjak.
\newblock {Factorization and Momentum-Space Resummation in Deep-Inelastic
  Scattering}.
\newblock {\em JHEP}, 0701:076, 2007.

\bibitem{Manohar:2006nz}
Aneesh~V. Manohar and Iain~W. Stewart.
\newblock {The Zero-Bin and Mode Factorization in Quantum Field Theory}.
\newblock {\em Phys.Rev.}, D76:074002, 2007.

\bibitem{Kelley:2011ng}
Randall Kelley, Matthew~D. Schwartz, Robert~M. Schabinger, and Hua~Xing Zhu.
\newblock {The two-loop hemisphere soft function}.
\newblock {\em Phys.Rev.}, D84:045022, 2011.

\bibitem{Khelifa:2011zu}
Kamel Khelifa-Kerfa.
\newblock {Non-global logs and clustering impact on jet mass with a jet veto
  distribution}.
\newblock {\em JHEP}, 1202:072, 2012.

\bibitem{Dasgupta:2001sh}
M.~Dasgupta and G.P. Salam.
\newblock {Resummation of nonglobal QCD observables}.
\newblock {\em Phys.Lett.}, B512:323--330, 2001.

\bibitem{Dasgupta:2002bw}
Mrinal Dasgupta and Gavin~P. Salam.
\newblock {Accounting for coherence in interjet $E_t$ flow: A Case study}.
\newblock {\em JHEP}, 0203:017, 2002.

\bibitem{Banfi:2002hw}
A.~Banfi, G.~Marchesini, and G.~Smye.
\newblock {Away from jet energy flow}.
\newblock {\em JHEP}, 0208:006, 2002.

\bibitem{Hornig:2011iu}
Andrew Hornig, Christopher Lee, Iain~W. Stewart, Jonathan~R. Walsh, and Saba
  Zuberi.
\newblock {Non-global Structure of the $O({\alpha}_s^2)$ Dijet Soft Function}.
\newblock {\em JHEP}, 1108:054, 2011.

\bibitem{Plehn:2000be}
T.~Plehn.
\newblock {Single stop production at hadron colliders}.
\newblock {\em Phys.Lett.}, B488:359--366, 2000.

\bibitem{Han:2009ya}
Tao Han, Ian Lewis, and Thomas McElmurry.
\newblock {QCD Corrections to Scalar Diquark Production at Hadron Colliders}.
\newblock {\em JHEP}, 1001:123, 2010.

\bibitem{Martin:2009bu}
A.D. Martin, W.J. Stirling, R.S. Thorne, and G.~Watt.
\newblock {Uncertainties on $\alpha_s$ in global PDF analyses and implications
  for predicted hadronic cross sections}.
\newblock {\em Eur.Phys.J.}, C64:653--680, 2009.

\bibitem{Cacciari:2011ma}
Matteo Cacciari, Gavin~P. Salam, and Gregory Soyez.
\newblock {FastJet User Manual}.
\newblock {\em Eur.Phys.J.}, C72:1896, 2012.

\bibitem{Ellis:1985er}
R.~Keith Ellis and J.C. Sexton.
\newblock {QCD Radiative Corrections to Parton Parton Scattering}.
\newblock {\em Nucl.Phys.}, B269:445, 1986.

\bibitem{Becher:2007ty}
Thomas Becher, Matthias Neubert, and Gang Xu.
\newblock {Dynamical Threshold Enhancement and Resummation in Drell-Yan
  Production}.
\newblock {\em JHEP}, 0807:030, 2008.

\bibitem{Becher:2011fc}
Thomas Becher, Christian Lorentzen, and Matthew~D. Schwartz.
\newblock {Resummation for W and Z production at large $p_T$}.
\newblock {\em Phys.Rev.Lett.}, 108:012001, 2012.

\bibitem{Becher:2012xr}
Thomas Becher, Christian Lorentzen, and Matthew~D. Schwartz.
\newblock {Precision Direct Photon and W-Boson Spectra at High $p_T$ and
  Comparison to LHC Data}.
\newblock {\em Phys.Rev.}, D86:054026, 2012.

\bibitem{Dasgupta:2007wa}
Mrinal Dasgupta, Lorenzo Magnea, and Gavin~P. Salam.
\newblock {Non-perturbative QCD effects in jets at hadron colliders}.
\newblock {\em JHEP}, 0802:055, 2008.

\end{thebibliography}
\end{document}